%% file: main.tex
\ifdefined\pdfinfoomitdate\pdfinfoomitdate=1\fi
\ifdefined\pdfsuppressptexinfo\pdfsuppressptexinfo=-1\fi
\ifdefined\pdftrailerid\pdftrailerid{}\fi
\documentclass[conference,compsoc]{IEEEtran}
\usepackage[utf8]{inputenc}
\usepackage[T1]{fontenc}
\usepackage{amsmath,amssymb,amsthm}
\usepackage{booktabs}
\usepackage{graphicx}
\usepackage{xcolor}
\usepackage{tikz}
\usetikzlibrary{arrows.meta,calc}
\usepackage{fontawesome5}
\usepackage[hyphens]{url}
\usepackage{xspace}
\usepackage[hidelinks]{hyperref}
\usepackage{cleveref}
\usepackage{placeins}
\hypersetup{pdfauthor={Xin Jin, Maximillian Liao, Handong Chen, Yi Nancy Li, Junyu Liu},pdftitle={Correctable Enrollment and One-Transaction Seed Proofs for Ethereum's Post-Quantum Migration},pdfcreator={},pdfproducer={}}
\crefname{section}{Section}{Sections}
\crefname{figure}{Fig.}{Figs.}
\crefname{table}{Table}{Tables}
\crefname{appendix}{Appendix}{Appendices}
\newtheorem{theorem}{Theorem}
\newtheorem{proposition}{Proposition}
\newtheorem{corollary}{Corollary}
\theoremstyle{definition}
\newtheorem{definition}{Definition}
\crefname{theorem}{Theorem}{Theorems}
\crefname{proposition}{Proposition}{Propositions}
\crefname{corollary}{Corollary}{Corollaries}
\crefname{definition}{Definition}{Definitions}
\newcommand{\runsub}[1]{\par\smallskip\noindent\refstepcounter{subsection}\textbf{\thesubsection\enskip #1.}\enskip\ignorespaces}
\newcommand{\runpar}[1]{\par\smallskip\noindent\textbf{#1.}\enskip\ignorespaces}
\def\extendedversion{}
\newif\ifextended
\ifdefined\extendedversion\extendedtrue\fi
\newcommand{\keynum}[1]{\ifextended#1\else\textbf{#1}\fi}
\DeclareRobustCommand{\proofsref}{\ifextended\cref{app:proofs}\else the extended version\fi}
\DeclareRobustCommand{\extref}[1]{\ifextended\cref{#1}\else the extended version\fi}
\DeclareRobustCommand{\Extref}[1]{\ifextended\Cref{#1}\else The extended version\fi}

\DeclareRobustCommand{\yes}{\tikz[baseline=-0.6ex]{\fill (0,0) circle (0.55ex);}}
\DeclareRobustCommand{\no}{\tikz[baseline=-0.6ex]{\draw[line width=0.3pt] (0,0) circle (0.55ex);}}
\DeclareRobustCommand{\partly}{\tikz[baseline=-0.6ex]{\draw[line width=0.3pt] (0,0) circle (0.55ex); \fill (0,-0.55ex) arc (-90:90:0.55ex) -- cycle;}}

\input{generated/bill_macros}

\input{generated/bill_gpu_macros}

\input{generated/exchange_macros}

\input{generated/backlog_macros}

\input{generated/sens_macros}

\begin{document}

\title{Correctable Enrollment and One-Transaction Seed Proofs for Ethereum's Post-Quantum Migration}
\author{\IEEEauthorblockN{Xin Jin\textsuperscript{1}, Maximillian Liao\textsuperscript{1}, Handong Chen\textsuperscript{2}, Yi Nancy Li\textsuperscript{2} and Junyu Liu\textsuperscript{1,*}}
\IEEEauthorblockA{\textsuperscript{1}University of Pittsburgh, Pittsburgh, PA, USA\\
\textsuperscript{2}SeQure Ltd, Chicago, IL, USA\\
\textsuperscript{*}Corresponding author: \href{mailto:junyuliu@pitt.edu}{junyuliu@pitt.edu}}}
\maketitle

\begin{abstract}
A quantum computer running Shor's algorithm can extract the private key of every Ethereum account whose public key is published, and accounts exposed by their own transactions hold 54.6\,\% of execution-layer ETH. As the first such machine may stay secret, a migration cannot wait for visible thefts, and once keys are extractable, legacy signatures no longer tell owners from attackers. We design a migration in which owners enroll post-quantum keys cheaply with their existing signatures before ECDSA is retired at a scheduled time, while a zero-knowledge proof of the wallet's BIP-39 seed can correct any such enrollment until a common deadline afterward; an account nobody enrolled stays frozen and recoverable by seed proof without a deadline. Each proof binds the new key, a nonce and a fee cap, so neither a paying sponsor nor a reordering block builder can redirect or replay it. Validators switch to hash-based keys first, at a finalized checkpoint and under a stated bound on the attacker's voting power. We prove the account guarantees under explicit assumptions, model-check the rules, and implement the design in go-ethereum and a Lean consensus client. A three-layer recursive hash-based STARK yields a seed proof of about 190\,KB that fits one transaction, verifies in 38\,ms and takes a median of 111\,s to generate on a laptop; four proofs aggregate into one. In a ten-node simulation with mainnet-derived accounts and fee demand, block space rather than proving sets the duration: the 4.45 million accounts with at least 0.1\,ETH would migrate in 29.4 days if all enroll before ECDSA is retired and in 235 days if all recover after it in batches of four. In a mass enrollment, an enrollment waits a median of 19.1 minutes in the public mempool, longer than published estimates for a fast quantum computer to extract its key, so owners with a seed and an unexposed key should migrate by seed proof.
\end{abstract}

\section{Introduction}\label{sec:intro}

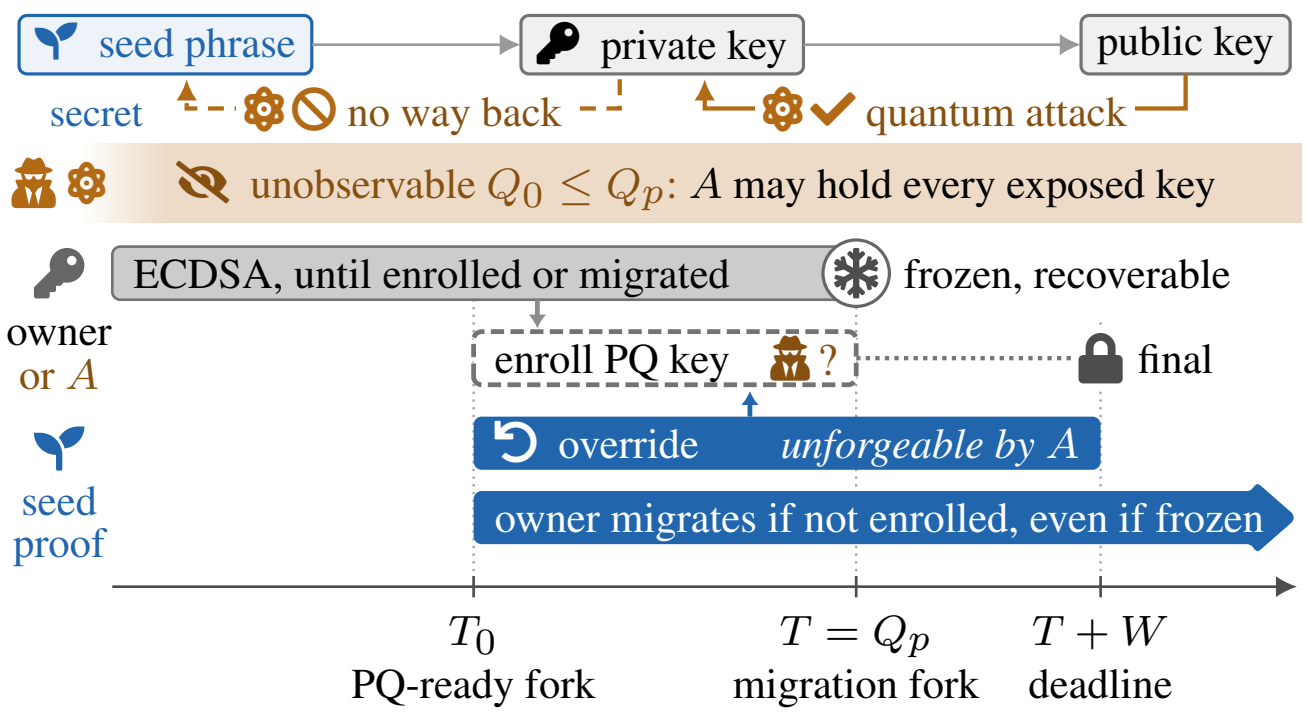
\begin{figure}[b]
\centering
\input{figures/fig1}
\caption{Overview of the migration. The adversary $A$ has a quantum computer from an unobservable $Q_0\le Q_p$, possibly before $T_0$, within the capacity bound of \cref{sec:adversary}. Until $T+W$, a zero-knowledge seed proof overrides any enrollment, even $A$'s, and restores the owner's authority (\cref{sec:nongoals}).}
\label{fig:timeline}
\end{figure}

Every Ethereum account that has sent a transaction has exposed its public key, because a transaction's sender address is derived by recovering that key from its ECDSA signature over the secp256k1 curve~\cite{Wood14}. Even without sending a transaction, an account exposes its public key whenever it signs an off-chain message such as a token permit~\cite{ERC2612} or a relayed EIP-7702 delegation~\cite{EIP7702}, while every validator's BLS public key is recorded openly in the consensus state. Shor's algorithm~\cite{Shor97} enables a cryptographically relevant quantum computer (CRQC) to extract the private keys from these public keys (\cref{fig:timeline}). On Ethereum mainnet in November 2026, transaction-exposed externally owned accounts (EOAs) hold 54.6\,\% of execution-layer ETH in wallets and contracts, and 61\,\% of staked ETH returns to such EOAs when withdrawn (\cref{fig:exposure}). As quantum hardware advances rapidly and algorithms for breaking classical public--private key pairs grow increasingly efficient, the estimated number of gates needed to recover a single 256-bit key has fallen by three orders of magnitude since 2017~\cite{RNSL17,Litinski23,BZG26}, to $7 \times 10^7$ Toffoli gates on fewer than 1,450 logical qubits. In physical error-corrected hardware, this puts key extraction within reach of either a trapped-ion machine with about 20,000 physical qubits running for 26 days~\cite{HTY26} or a faster superconducting machine with under half a million physical qubits running in minutes~\cite{BZG26}.

Once a quantum adversary extracts an account's private key, signatures under that key can no longer distinguish the legitimate owner from the adversary. To rescue compromised accounts, Buterin proposed an emergency hard fork~\cite{Buterin24} that relies on a deeper secret held by modern wallets: most wallets derive their private keys from a BIP-39 mnemonic seed through a chain of one-way hashes~\cite{BIP39,BIP32,BIP44}. Because quantum computers can invert these hashes only by generic search, extracting a private key does not reveal the underlying seed, provided the mnemonic has sufficient entropy and was never leaked (\cref{sec:adversary}). The fork therefore disables ECDSA-authorized transactions and lets an owner reclaim an account under post-quantum authentication with a zero-knowledge STARK proof of knowing a seed that derives the account's key. Yet turning this idea into a live migration protocol raises four questions.

\textbf{When does the adversary appear?} The moment $Q_0$ at which a CRQC first exists is unobservable because an adversary may keep it secret and delay visible theft. A migration fork that disables ECDSA at a public time $Q_p$ must therefore stay secure when $Q_0 \le Q_p$, because the adversary can then sign for every exposed account until $Q_p$ exactly as its owner can. After $Q_0$ only an owner with a second secret like a mnemonic seed can still be distinguished from the adversary, whereas an owner without a seed must finish migrating before $Q_0$ to stay safe. Existing voluntary ``off-ramp'' proposals~\cite{EIP8141,BIP360} let owners move to post-quantum keys or quantum-resistant outputs via legacy signatures. That approach fails once $Q_0$ passes, because the adversary can bind any exposed account to its own post-quantum key (a \emph{stealth enrollment}), and even a previously unexposed owner exposes its public key when signing the migration transaction. The emergency hard fork~\cite{Buterin24} instead waits until visible thefts occur after $Q_0$, rolls the chain back to the first theft and requires every EOA to recover with a seed proof. This reactive approach is costly on a live network: rolling back finalized history reverts honest transactions and risks splitting the chain (as Ethereum's 2016 DAO fork did~\cite{KLM17}), while waiting for an emergency forces every account at once to compute an expensive zero-knowledge proof.

\textbf{Who orders transactions and who pays their fees?} Under proposer--builder separation, a slot's winning builder chooses which transactions enter the block and in what order~\cite{HKFW23}\ifextended, a power long used for front-running~\cite{DGK20}\fi. Without controlling consensus, the adversary can therefore insert transactions signed with exposed keys ahead of or between an owner's recovery steps to make them fail. Before the fork it can also drain an account's liquid ETH while stake withdrawals, tokens or contract permissions remain tied to its address. Reclaiming that address in place then requires a third-party sponsor to pay the recovery gas without being able to change the new key. Even when ETH remains, after the fork the retired legacy key can no longer authorize a fee before the new key is installed.

\textbf{Does a seed proof fit in one transaction?} Compact elliptic-curve proofs break under quantum attacks, whereas post-quantum hash-based proofs have been prohibitively large and costly to verify. A recent seed proof for the simpler EdDSA key derivation already measures 5.4\,MB and takes 2.3\,s to verify~\cite{BCRS26}. That size far exceeds what one Ethereum transaction can carry under its $2^{24}$ gas limit~\cite{EIP7825}, even before accounting for Ethereum's heavier derivation: five BIP-32 levels and three secp256k1 scalar multiplications.

\textbf{What does migration cost at mainnet scale?} Pausing Ethereum for a dedicated migration, which takes months on Bitcoin~\cite{Pont24}, would freeze its millions of accounts, institutional custodians and hundreds of thousands of validators. Migrating on a live chain instead stresses every system layer: proofs are generated, propagated, verified and stored across all nodes, while migration transactions compete with ordinary traffic for block space under EIP-1559~\cite{EIP1559}. Most transactions are included within a minute, before a fast quantum computer could extract the key that they reveal~\cite{BZG26}, but during a mass enrollment an ECDSA-signed enrollment can wait in the public mempool, its public key exposed, long enough for such a machine to extract its private key. Existing estimates fix the migration's share of block space in advance~\cite{Pont24,KK26}, leaving throughput, fees and mempool exposure under live traffic unmeasured (\cref{sec:related}).

\textbf{Correctable binding.} We present a two-stage migration protocol that combines lightweight ECDSA-signed key enrollment before a scheduled cutoff with zero-knowledge seed proofs that can both recover unmigrated accounts and override fraudulent enrollments (\cref{fig:timeline}). First, a PQ-ready fork at $T_0$ enables post-quantum transaction types while ECDSA remains active, opening an enrollment period $[T_0, T)$ before a second fork at the public cutoff time $T = Q_p$ permanently disables protocol-level secp256k1 authority. During $[T_0, T)$, active owners and custodians without a seed can enroll a post-quantum key such as ML-DSA~\cite{FIPS204} cheaply with an ECDSA-signed transaction and no zero-knowledge proof. Because a quantum adversary who appears before $T$ could forge such an ECDSA enrollment to hijack an exposed account, every ECDSA-signed enrollment remains correctable until a common deadline $T + W$ one year after $T$ (\emph{correctable binding}). An account not enrolled by $T$ is frozen, not destroyed: its owner can reclaim the address and its bound stake or permissions at any later time with a seed proof, though liquid assets moved before $T$ are not recovered. Each seed proof binds the owner's new key, a migration nonce and a fee cap, so even if an untrusted sponsor pays the gas or a malicious builder reorders the block, nobody can substitute the key, replay the proof or overcharge the account; a recovery that such ordering interrupts fails without effect and can be resubmitted.

\textbf{Validators switch first.} Protected accounts can recover after compromise under the stated conditions, but validators collectively control the finality those recoveries need. As \cref{prop:minimal} shows, a migration that only the old chain authenticates must finalize the registrations of the new validator set before the adversary can suppress finality past the recovery deadlines, which it can by keeping more than a third of the active stake slashed. The 7,647 largest validators hold more than a third of mainnet stake; recovering their BLS keys one by one is estimated at 48 to 122 days on one superconducting quantum computer, but one discrete logarithm of the sum of their public keys forges the slashing evidence for all of them (\cref{sec:n1-bound}). The protocol therefore switches consensus first: before $T_0$, each validator uses its BLS key to register a hash-based leanXMSS key~\cite{DKKW25,KKW25,DKKW25b}, and at $T_0$ the chain moves to Lean consensus from a finalized checkpoint. This rests on an explicit capacity bound: before $T_0$ the adversary cannot forge BLS authorizations, individually or in aggregate, for a consensus threshold, and validators whose keys it registered hold less than a third of the new set. If the adversary breaks some BLS keys before $T_0$, the checkpoint excludes every validator with conflicting registrations, so a conflicting registration cannot displace an owner's registration finalized by $T_0$, and the owner of an excluded or hijacked validator with a protected withdrawal account can still recover that account with a seed proof, exit and withdraw what remains of its stake.

\textbf{A seed proof in one transaction.} To fit Ethereum's full seed derivation\ifextended\else{} (\cref{sec:relation})\fi{} into a single transaction, we build a three-layer recursive zero-knowledge STARK~\cite{BBHR18,BCS16} from hash functions alone.\ifextended{} Its relation covers the complete derivation of an account from a 64-byte BIP-39 seed through six HMAC-SHA512 evaluations along a five-level BIP-44 path, three secp256k1 scalar multiplications and Keccak-256 (\cref{sec:relation}).\fi{} Two layers of recursive compression shrink the proof to about 190\,KB, which one transaction can carry, and add hiding commitments at the outer layer to achieve zero knowledge. The proof verifies in 38\,ms inside go-ethereum and is nearly $30\times$ smaller and $60\times$ faster to verify than the recent EdDSA seed proof~\cite{BCRS26}. A two-layer recursive aggregator combines four owners' final proofs into one aggregate proof of about 190\,KB, so one batch transaction can migrate four accounts at once. Each layer has 128 to 130 bits of modeled round-by-round IOP soundness~\cite{soundcalc}, while quantum zero knowledge and weak simulation extractability of the composed proof are explicit assumptions (\cref{sec:proof-security}).

\textbf{Cost at mainnet scale.} We implement the protocol in go-ethereum~\cite{GoEthereum} paired with the Ream Lean consensus client~\cite{Ream}, test that no adversarial ordering redirects or replays a recovery, and evaluate the full system on ten nodes with mainnet's account population and ordinary EIP-1559 demand. A seed proof, which an enrollment does not need, takes a median of \simOwnerProofSeconds\,s and 31\,GiB on a laptop CPU; as a cost reference, a rented A100 machine proves for \$\gpuBestDollarsPerThousand{} per thousand owners. A self-paid recovery consumes 8.74\,M gas (about half the $2^{24}$ per-transaction gas cap). Bidding at most 1\,gwei, migration transactions take only the block space that ordinary users leave, \simEnrollMigShare\,\% of the 30\,M gas target. With this share, during a mass enrollment an ECDSA enrollment waits a median of \simEnrollWaitMedianMin{} minutes in the public mempool, longer than the 9 to 12 minutes estimated for a fast superconducting quantum computer with precomputation to extract its private key~\cite{BZG26}, so unexposed accounts with a seed should migrate by seed proof instead. For the 4.45\,M EOAs holding at least 0.1\,ETH, migration takes \simDaysBEnroll{} days if all enroll before $T$, \simDaysBBatched{} days if all recover after $T$ in batches of four, and \simDaysBSplitBatched{} days on a mixed path where accounts active since November 2023 enroll before $T$ and the rest recover in batches (\cref{sec:eval-audit}). At the 1\,gwei cap, an enrollment costs \simEnrollFeeEth\,ETH and a batched recovery \simBatchedFeeEth\,ETH per account. Finally, centralized exchanges hold at least \exShareEl\,\% of execution-layer ETH, and none of the \exExamined{} we examined documents a mnemonic seed. The exchanges' \exKeys{} EOAs holding at least 0.1\,ETH can finish enrolling at $T_0$ in \exEnrollMin{} minutes of block space for \exFeeEth\,ETH in total fees. In summary, we make five contributions:
\begin{itemize}[\setlength{\IEEElabelindent}{0pt}]
\item \textbf{A security audit (\cref{sec:threat,sec:security,app:surface}).} We map what a quantum computer breaks in Ethereum (\cref{tab:surface}) and prove that any migration of the kind we study needs two upgrades: an owner check by an unexposed secret from which the account derives, and timely finalization of new validator keys (\cref{prop:minimal}). Our protocol provides both.
\item \textbf{A correctable migration (\cref{sec:threat,sec:design,sec:security}).} Owners enroll new keys cheaply, and a seed proof can correct any enrollment until $T+W$, so the protocol need not know when the first quantum computer appears. We define authenticated migration as a game, prove the account guarantees with concrete bounds (\cref{thm:auth,thm:account,cor:recovery}) and model-check them.
\item \textbf{A seed proof that fits one transaction (\cref{sec:proof}).} Our hash-based zero-knowledge proof of Ethereum's key derivation takes about 190\,KB and verifies in 38\,ms.
\item \textbf{Validators switch first (\cref{sec:n1,sec:security}).} Validators move to hash-based keys first, so recoveries still reach finality under a stated bound on the attacker. We prove the switch and run it on a Lean consensus client.
\item \textbf{Cost at mainnet scale (\cref{sec:eval}).} Running the protocol at mainnet scale shows that block space, not proving, sets how long the migration takes, and that at the 1\,gwei cap the 4.45\,M EOAs with at least 0.1\,ETH pay \simFeeEthBSplitBatchedRound\,ETH in fees on the mixed path, while proving costs \$\gpuBestDollarsPerThousand{} per thousand owners on a rented A100 machine, a cost reference, since an owner proves on a machine that it trusts with the seed.
\end{itemize}

\section{Background, Threat Model and Goals}\label{sec:background}

\begin{figure}[t]
\centering
\includegraphics[width=\columnwidth]{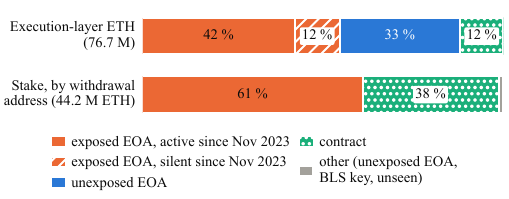}
\caption{Value behind exposed keys on mainnet, 3--4~November 2026 (data: \cref{sec:eval-audit}). Top: the 76.7\,M ETH on the execution layer. Bottom: the 44.2\,M ETH of stake, by the kind of its withdrawal address (orange: every exposed EOA). Shares are rounded to whole percent.}
\label{fig:exposure}
\end{figure}

\runpar{Accounts, Keys and Exposure}\label{sec:bg-accounts}
An externally owned account (EOA) is controlled by a secp256k1 private key $k$ and public key $K = kG$, and its address is the last 20 bytes of $\mathsf{keccak256}(K)$~\cite{Wood14}. Every ECDSA signature from an EOA in a transaction or elsewhere publishes $K$, whereas an account that has never signed exposes only its address, a 160-bit truncated hash. Contract accounts such as multisignature wallets hold no key pair of their own and are usually controlled by EOAs.\ifextended{} On the consensus layer, validators sign with BLS over BLS12-381, and the consensus state holds their public keys.\fi\ifextended{} \Cref{fig:exposure} shows the value behind exposed keys on mainnet.\fi

\runpar{Hierarchical Deterministic Wallets}\label{sec:bg-hd}
Wallets usually derive keys from a BIP-39 mnemonic~\cite{BIP39}, whose 12 to 24 words encode 128 to 256 bits of entropy; PBKDF2-HMAC-SHA512 with 2,048 iterations, salted with an optional passphrase, stretches them into a 64-byte $\mathit{seed}$. BIP-32~\cite{BIP32} then computes a master key and chain code $(k_0, c_0) = \mathsf{HMAC\text{-}SHA512}(\text{``Bitcoin seed''},\allowbreak\ \mathit{seed})$. To reach an Ethereum account's private key $k$, the wallet steps down the five-level BIP-44 path $m/44'/60'/j'/c/i$~\cite{BIP44} (for account $j$, change level $c$ and address index $i$), deriving each child key and chain code $(k_\ell, c_\ell)$ from its parent's with one more HMAC-SHA512\ifextended{} whose left half is added to the parent private key modulo the group order\fi: the three primed (\emph{hardened}) levels hash the parent private key directly, whereas the two unprimed (\emph{non-hardened}) levels hash the parent public key and thus require a secp256k1 scalar multiplication. Because a CRQC recovers any private key from its public key, exposing any level's public key and chain code $(K_\ell, c_\ell)$ reveals $(k_\ell, c_\ell)$ and hence every descendant key, hardened or not. Yet even given $(k_0, c_0)$, inverting the initial HMAC-SHA512 to recover $\mathit{seed}$ admits only generic quantum search, so knowing $\mathit{seed}$ distinguishes an owner from an adversary who already holds the account's private key~\cite{Buterin24,BCRS26,SW23}.

\runpar{Validators and Lean Consensus}\label{sec:bg-lean}
In Ethereum's proof-of-stake consensus, validators attest to blocks with aggregated BLS signatures~\cite{ConsensusSpecs}, and votes from two thirds of the stake finalize a checkpoint~\cite{BG17,BHK20}. A CRQC that breaks validator keys can sign in their name to get an honest validator penalized and stripped of voting power (\emph{slashed}) for signing conflicting votes (\emph{equivocation}), stall finality with more than a third of the stake, or finalize a conflicting chain with two thirds, so validators also need post-quantum keys~\cite{BZG26}. Lean consensus is a proposed redesign of the consensus layer~\cite{Ream} in which validators sign with leanXMSS, a stateful hash-based signature scheme of the XMSS family, and hash-based succinct arguments aggregate their signatures~\cite{DKKW25,KKW25,DKKW25b}\ifextended; Ream~\cite{Ream} is a Lean consensus client in Rust\fi. A leanXMSS key signs a bounded number of messages, one per leaf, and never reuses a one-time leaf key, so its signing state must never roll back.

\runsub{Threat Model}\label{sec:threat}\label{sec:adversary}\label{sec:bg-quantum}
Governance fixes the forks $T_0$ and $T = Q_p$ and the window $W$ in advance (\cref{fig:timeline}). An \emph{owner} controls an externally owned account, and a \emph{sponsor} is any account that pays another account's transaction fee. Because an exposed ECDSA key can no longer distinguish its owner from a quantum attacker, owners fall into two groups depending on whether they hold a deeper derivation secret:
\textbf{Protected owners (with a seed):} we protect owners whose account derives from a BIP-39 mnemonic along a five-level BIP-44 path, whose mnemonic has at least 128 bits of entropy and was never exposed, and whose wallet keeps the seed on a device the owner trusts and computes proofs there; the account's public key may well be exposed.
\textbf{Owners without a seed:} an owner whose account derives from no seed along such a path has no witness for the seed proof (\cref{sec:relation}) and is an \emph{owner without a seed}, whose only checkable secret is the account's ECDSA key. This group includes the centralized exchanges examined in \cref{sec:eval-audit}, as far as their documentation shows, and the owners of keys generated without a seed, of keys on paths of another shape than $m/44'/60'/j'/c/i$ (such as the four-level path $m/44'/60'/0'/i$ of Ledger's legacy Ethereum app), and of keys that hardware security modules (HSMs) or threshold multi-party computation (MPC) generate without one.

\emph{Key recovery.} The adversary $A$ is external to the protocol and has a CRQC from an unknown time $Q_0 \le Q_p$ (\cref{fig:timeline}). Shor's algorithm~\cite{Shor97} solves the discrete-logarithm problem in polynomial time, breaking both ECDSA (over secp256k1) and BLS signatures. From $Q_0$, $A$ knows the secp256k1 private key of every account whose public key has been published (in a transaction, a typed-data signature, an EIP-7702 tuple, a multisignature co-signature or off chain); with a fast machine it can also mount \emph{on-spend} attacks that recover the private key from a transaction waiting in the mempool and replace it before inclusion~\cite{BZG26}. On the consensus layer, $A$ can recover any validator's BLS private key, or the secret of a sum of validators' public keys, enabling aggregate forgery on a common message. Because $Q_0$ is unobservable, we assume $A$ may hold every exposed account key at any time before $T$, while the capacity bound limits the validator signatures that $A$ can forge, individually or in aggregate.

\emph{Ordering and observation.} Without a validator key or a majority, $A$ can manipulate transaction ordering by acquiring the building rights of any slot (\cref{sec:intro}). It may then include, omit, reorder and delay any valid transactions that it has seen or can create, split bundles and replace transactions in the public mempool; natural reorganizations occur as well. $A$ sees every transaction and proof in the mempool and in blocks, holds funds and may act as a sponsor.

\emph{Quantum adversary limits (A1--A8).} Beyond the discrete logarithms that Shor's algorithm solves, we assume that $A$ breaks nothing that the protocol relies on: it knows no protected owner's mnemonic or seed (A1) and has only generic quantum search~\cite{Grover96,BHT98} against the hash functions Keccak-256, SHA-512 and Poseidon2 (A2). Against these hash functions, Grover's search~\cite{Grover96,BHT98} requires about $2^{n/2}$ sequential quantum evaluations to find an input matching a known $n$-bit hash output, so finding a preimage of a known 160-bit address takes about $2^{80}$ quantum work for a single target and $2^{80}/\sqrt{N}$ across $N$ targets ($\sqrt{N} \approx 2^{14}$ for the 345\,M mainnet addresses in \cref{sec:eval-audit}), which remains infeasible under realistic fault-tolerance and parallelization costs~\cite{BZG26}. We also assume that $A$ cannot forge post-quantum ML-DSA~\cite{FIPS204} or hash-based leanXMSS~\cite{DKKW25,KKW25} signatures (A3), and that the seed proof is zero knowledge and weakly simulation extractable against quantum adversaries (A4, \cref{sec:proof-security}).

We do not assume that $A$ is slow on the execution layer: our account guarantees hold even if $A$ instantly holds all exposed account keys from $Q_0$ on. For validators, however, we add a \emph{capacity bound} (A5): until the registration deadline $D_{\mathrm{reg}} = T_0$, $A$ cannot forge BLS authorizations (individually or in aggregate) for a consensus threshold, and the validators that enter the new set with keys registered by $A$ hold less than a third of that set (\cref{sec:n1-bound}). \ifextended Unlike an account, a validator carries consensus weight; without the bound, the validators that $A$ took over could withhold finality from the owners' corrections. \fi The protocol thus needs no knowledge of $Q_0$, only this bound before $T_0$. Finally, we assume that $A$ cannot alter the last pre-$T_0$ finalized checkpoint $B_{\mathrm{ck}}$ that nodes fix out of band (A6, \cref{sec:n1-reg}), cannot violate the safety or liveness of Lean consensus under its usual network assumptions while faulty weight stays below a third (A7), and cannot delay the finalization of a recovery that stays valid, fees included, past time $\Delta$ from broadcast (A8).

\begin{figure*}[t]
\centering
\input{figures/fig_overview}
\caption{Where the protocol changes Ethereum. The Account, Proof and Validator columns map the three flows on the right to the system layers audited in \cref{tab:audit}; numbers identify steps within each flow. Only the seed proof is zero knowledge (ZK), and it touches the most layers. Validators switch first, since recovery and override count only once finalized (\cref{sec:design,sec:n1,sec:proof}).}
\label{fig:overview}
\end{figure*}
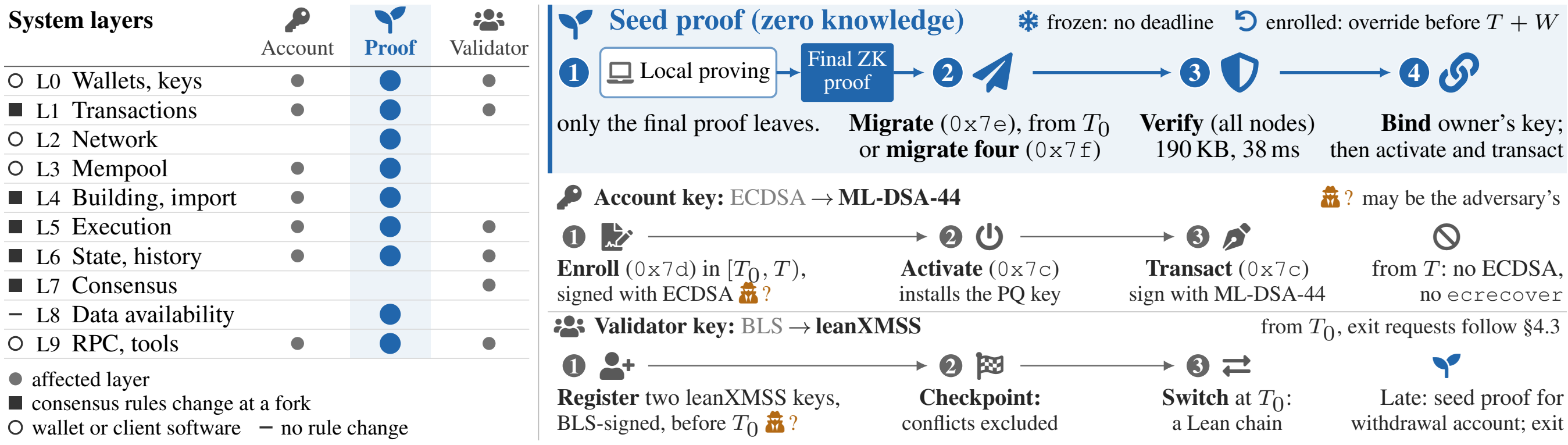

\runsub{Goals}\label{sec:goals}
G1--G4 and G6 concern accounts and need no bound on $A$'s capacity, given a chain whose finality $A$ cannot control; G5 concerns validators and provides that chain under the capacity bound (\cref{sec:n1}).

\textbf{G1 (Retirement).} From $T$ on, no secp256k1 key authorizes a state change through the protocol\ifextended: transaction envelopes, EIP-7702 tuples and the \texttt{ecrecover} precompile accept no secp256k1 authority\else{} (\cref{sec:retire})\fi, and an account that has not enrolled a post-quantum key by $T$ is frozen, not destroyed.

\textbf{G2 (Authenticated migration).} Only the owner can bind an account to a post-quantum key with a seed proof: an accepted recovery installs exactly the owner's key, once, at a fee within the owner's cap, and the proof reveals nothing about the seed; \cref{sec:security} states G2 as a game and bounds $A$'s success under A1, A2 and A4.

\textbf{G3 (Correctable binding).} Until $T + W$, the owner's seed proof replaces an enrolled key, even one from $A$'s stealth enrollment, and no transaction authorized by that key is valid afterwards.\ifextended{} Every enrollment precedes $T$, so the owner has at least $W$ after $A$'s last chance to enroll.\fi

\textbf{G4 (Ordering-robust recovery).} Under every ordering, split and delay of valid transactions by $A$, an owner's recovery either binds the owner's key within the owner's fee cap or fails without effect and can be resubmitted, by a sponsor if $A$ has drained the account. For its part, the owner's wallet \ifextended sends a transaction only if every state that $A$ can reach by ordering the account's transactions in other hands (\cref{sec:wallet}) is acceptable, and it\else follows \cref{sec:wallet} and\fi{} detects a stealth enrollment in time to override it.

\textbf{G5 (Validator transition).} An eligible validator whose valid registration of a leanXMSS key is finalized by $D_{\mathrm{reg}}$ moves to that key or, if $A$ registered a conflicting one, is excluded; if its withdrawal credentials name a protected owner's account, it can then request its exit through that account (\cref{sec:n1-late}). Under the capacity bound, $A$'s validators start below a third of the new set; while they stay below it, $A$ can neither finalize conflicting checkpoints nor stall finality.

\textbf{G6 (Owners without a seed).} An enrollment by an owner without a seed that is final before $Q_0$ is the owner's and binds the account to the owner's keys for good; once $A$ holds the account's key and the chain records no key of the owner, no protocol that checks only that key can guarantee such a binding (\cref{prop:noseed}).

\runsub{Non-Goals}\label{sec:nongoals}
\emph{Assets.} With an exposed key, $A$ can spend an account's assets before $T$, which only an earlier $T$ prevents; an override restores authority over the address but neither returns what $A$ moved nor revokes permissions that $A$ granted in third-party contracts, which wallets must revoke (\cref{sec:wallet}).

\emph{Other threats.} We exclude attacks that need no CRQC, such as theft with a leaked mnemonic, denial of service, including censoring a recovery, and fee or ordering games against unexposed accounts; other adversaries, such as malicious RPC providers and wallet vendors, or hosted provers, which receive the seed; implementation defects; and systems beyond Ethereum's base layer, such as layer-2 chains and bridges. Proving in a trusted execution environment, which needs trust in the hardware and post-quantum attestation, is future work.

\section{Protocol Design: Accounts}\label{sec:design}

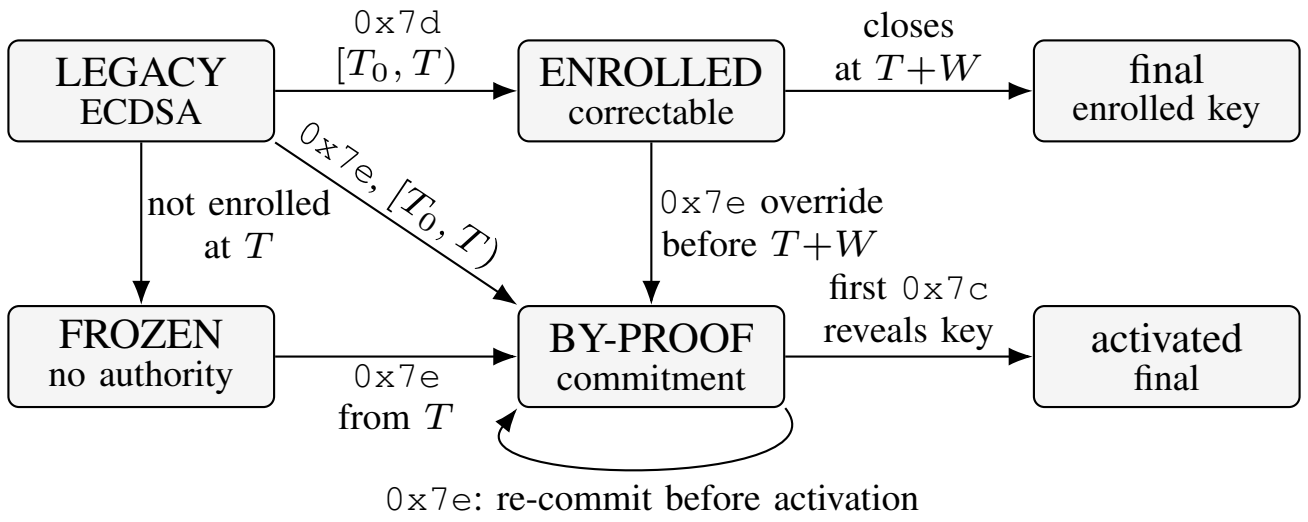
\begin{figure}[t]
\centering
\begin{tikzpicture}[x=1pt, y=1pt, font=\footnotesize, >={Latex[length=1.6mm]},
  st/.style={draw, rounded corners=2pt, minimum width=50pt, minimum height=19pt, inner sep=1.5pt, align=center, fill=black!4},
  lab/.style={font=\scriptsize, align=center, inner sep=1pt}]
  \node[st] (leg) at (25,50) {LEGACY\\[-2pt]{\scriptsize ECDSA}};
  \node[st] (enr) at (121,50) {ENROLLED\\[-2pt]{\scriptsize correctable}};
  \node[st] (fin) at (218,50) {final\\[-2pt]{\scriptsize enrolled key}};
  \node[st] (fro) at (25,0) {FROZEN\\[-2pt]{\scriptsize no authority}};
  \node[st] (prf) at (121,0) {BY-PROOF\\[-2pt]{\scriptsize commitment}};
  \node[st] (act) at (218,0) {activated\\[-2pt]{\scriptsize final}};
  \draw[->] (leg) -- node[lab, above] {\texttt{0x7d}\\$[T_0, T)$} (enr);
  \draw[->] (enr) -- node[lab, above] {closes\\at $T{+}W$} (fin);
  \draw[->] (leg) -- node[lab, right, pos=0.5] {not enrolled\\at $T$} (fro);
  \draw[->] (fro) -- node[lab, below] {\texttt{0x7e}\\from $T$} (prf);
  \draw[->] (leg.south east) -- node[lab, above, sloped, pos=0.45] {\texttt{0x7e}, $[T_0, T)$} (prf.north west);
  \draw[->] (enr) -- node[lab, right] {\texttt{0x7e} override\\before $T{+}W$} (prf);
  \draw[->] (prf) -- node[lab, above] {first \texttt{0x7c}\\reveals key} (act);
  \draw[->] (prf.south east) .. controls +(12,-15) and +(-12,-15) .. node[lab, below=2pt] {\texttt{0x7e}: re-commit before activation} (prf.south west);
\end{tikzpicture}
\caption{Authority states of an account under the protocol. ECDSA authority ends at the enrollment, at a migration by seed proof or at $T$, whichever comes first. An enrolled account, even once activated, stays correctable until $T + W$. In a final state, no seed proof can replace the key.}
\label{fig:states}
\end{figure}

The protocol changes the execution layer for accounts (\cref{sec:n1} treats validators): a registry of per-account authority records, four transaction types (\cref{tab:tx}) and consensus rules R1--R7 (\cref{sec:retire,sec:fees,sec:pool}), activated by two forks (\cref{fig:timeline}). \Cref{fig:overview} shows where every change of \cref{sec:design,sec:n1,sec:proof} falls in the ten layers that \cref{sec:eval-audit} audits, and the order of each flow; \cref{fig:states} shows an account's life.

\runsub{Account State}\label{sec:state}
A system contract, the \emph{registry}, holds a record $\mathit{Rec}(a) = (\mathit{mode}, \mathit{alg}, \mathit{enrolled}, \mathit{mnonce}, \mathit{commit})$ for each account $a$, written only by protocol transitions. The $\mathit{mode}$ is LEGACY, ENROLLED (correctable until $T + W$), BY-PROOF (by a seed proof) or NATIVE (an account whose address is a hash of a post-quantum key); from $T$ on, an externally owned account still in LEGACY is \emph{frozen}, with no deadline. Here $\mathit{alg}$ identifies the post-quantum signature scheme, the commitment $\mathit{commit} = \mathsf{keccak256}(\mathit{alg} \,\|\, \mathit{pk})$ fixes the post-quantum public key $\mathit{pk}$, and the migration nonce $\mathit{mnonce}$ grows with every change of authority. The enrollment's time, $\mathit{enrolled}$,\ifextended{} shares a storage slot with the mode and migration nonce and\fi{} tells the wallet from when to review the account's permissions after an override.

\begin{table}[t]
\caption{Transactions of an account under the protocol. PoP: proof of possession of the new key. Gas measured on our client: for legacy a plain transfer, for a self-paid \texttt{0x7e} and for \texttt{0x7f} the mean over ten fresh proofs or batches.}
\label{tab:tx}
\centering
\footnotesize
\setlength{\tabcolsep}{3pt}
\begin{tabular}{@{}llll@{}}
\toprule
Transaction & Authorized by & Valid & Gas \\
\midrule
\texttt{0x00}--\texttt{04} legacy & ECDSA & before $T$, LEGACY & 21\,k \\
\texttt{0x7d} enroll & ECDSA + PoP & $[T_0, T)$, LEGACY & 175\,k \\
\texttt{0x7e} migrate & seed proof & $\ge T_0$ & 8.74\,M \\
\texttt{0x7f} migrate four & seed proofs, sponsor & $\ge T_0$ & 15.07\,M \\
\texttt{0x7c} activate & ML-DSA-44 & $\ge T_0$ & 393\,k \\
\texttt{0x7c} transfer & ML-DSA-44 & activated & 117\,k \\
\texttt{0x7c} \texttt{rotate} & ML-DSA-44 + PoP & final & 493\,k \\
\bottomrule
\end{tabular}
\end{table}

\runsub{Transactions}\label{sec:tx}
\textbf{Enrollment} (\texttt{0x7d}) is valid in $[T_0, T)$. Its ECDSA-signed envelope carries a post-quantum public key and a proof of possession (PoP) of it, an ML-DSA signature over the chain, the sender, its nonce and the key, so a mistyped or injected key fails at once. The enrollment sets the mode to ENROLLED, clears any EIP-7702 delegation and voids the account's ECDSA authority, which confines the race for an exposed account to the enrollment itself.

\textbf{Migration} (\texttt{0x7e}, valid from $T_0$) carries no signature: the seed proof is its authorization. Its public input is a 198-byte string (\cref{app:pi}), which a node recomputes from the transaction's fields; it fixes the account, its migration nonce, the key commitment, an expiry, the fee cap and the fee terms of \cref{sec:fees}, among other fields, so the proof applies at most once and installs only the owner's key. A node accepts a \texttt{0x7e} only through one of three gates: (i)~migrating a LEGACY account, which recovers it once frozen and, before $T$, lets an owner whose key was never published migrate without signing (\cref{sec:wallet}); (ii)~overriding an enrollment before its deadline; or (iii)~re-committing a BY-PROOF account that has not yet activated, which lets an owner replace a wrong key. The transition sets the mode to BY-PROOF with the new commitment, increments the migration nonce and clears any installed key or EIP-7702 delegation, moving no asset apart from its fee.

A \textbf{post-quantum transaction} (\texttt{0x7c}) is an ordinary transaction of an enrolled, migrated or native account, signed with ML-DSA-44~\cite{FIPS204} under the registered key.\ifextended{} A node cannot recover the signer's address from a post-quantum signature, so the sender is an explicit field.\fi{} The first such transaction carries the public key, which the node checks against the commitment and installs as the account's code, marking the account \emph{activated}. A final, activated account changes its key with a \texttt{0x7c} that calls \texttt{rotate} on the registry, again with a proof of possession of the new key.

An \textbf{aggregated migration} (\texttt{0x7f}) migrates several accounts under one proof. A sponsor collects the owners' public proofs and proves that all of them verify (\cref{sec:aggregation}); each account still gets its own migration nonce, commitment and fee terms. The node checks each account's expiry, state and nonce on its own: an account that fails is skipped and logged, and the others apply. Only a fault of the whole batch, such as an invalid aggregate proof or sponsor signature, makes its block invalid. A low balance never causes a skip, since an account reimburses the sponsor the smaller of its cap and its balance.

\runsub{Correctable Binding}\label{sec:override}
An enrollment authorized by ECDSA may be $A$'s, so the protocol keeps it correctable (G3): until $T + W$, a \texttt{0x7e} replaces the enrolled commitment with the owner's, and from that transaction on the enrolled key authorizes nothing. Three choices protect the window. First, nothing that the enrolled key authorizes can close or shorten it, since that key may be $A$'s: only time closes it, and the seed holder may end it early by overriding the enrollment. Second, one deadline, $T + W$, holds for every account, so each owner has at least $W$ after $A$'s last chance to enroll. We set $W$ to 365 days, enough for every exposed EOA with at least 0.1\,ETH to override in batches at the simulated rates (\cref{sec:eval-audit}), and a release may lengthen it. Third, if a quantum computer becomes evident before $T$, an emergency fork may move $T$ forward to $T' < T$, but it must extend every running window to $W' \ge T + W - T'$, so that no deadline moves earlier (the \emph{window lemma}). Every release that schedules or moves $T$, or changes $W$, is published before the earliest moment it changes, so that no node applies it retroactively.

\runsub{Retiring ECDSA}\label{sec:retire}
Three rules void secp256k1 authority (G1). \textbf{R1} rejects ECDSA-signed transactions of non-LEGACY senders before $T$ and every such transaction from $T$ on, decided by type and block timestamp before any sender is recovered. \textbf{R2} skips the EIP-7702 tuples of non-LEGACY accounts and, from $T$ on, every tuple; from $T$ on it also executes a call into an account that delegates through EIP-7702 as a call into an empty account, because the delegated code may accept the account's own ECDSA signature. \textbf{R3} makes the \texttt{ecrecover} precompile return its failure value for non-LEGACY accounts, which costs a cold read of the registry (2,100 gas) before $T$, and for every input from $T$ on.\ifextended{} An exemption from R3 would keep open exactly the signatures that $A$ forges: contracts take secp256k1 signatures as consent, as in permits~\cite{ERC2612}, $A$ forges them for every exposed key, including those of oracles and other signers that are not accounts, and a key that authorized an exemption might itself be $A$'s.\fi{} In return, \textbf{R5} gives activated accounts a signature inside the EVM: they report non-empty code, and a call of the ERC-1271 method \texttt{isValidSignature}~\cite{ERC1271} on such an account verifies an ML-DSA signature natively, at a provisional 50,000 gas, so permits of contracts that support ERC-1271 keep working. Since activated accounts thus report code, \textbf{R6} exempts post-quantum accounts from the rule that a sender has no code~\cite{EIP3607}.

R3 ends every verification that a contract performs through \texttt{ecrecover}. Contracts whose authority is their caller keep working once their owners migrate; a Safe, for instance, accepts a migrated owner's signature as a contract signature, which R5 checks. A contract that accepts only recovered secp256k1 signatures and cannot be upgraded freezes what it holds at $T$, so its operators must replace it before $T$; such contracts hold at least 4.8\,\% of the stake that withdraws to contracts (\cref{app:surface}).

\runsub{Fees Without a Signature}\label{sec:fees}
Because a \texttt{0x7e} has no signer who could authorize a fee, and from $T$ on the account's ECDSA key authorizes nothing, the proof authorizes the maximum charge as well: its public input fixes a fee cap (\textbf{R4}). A self-paid \texttt{0x7e} is valid only if its gas limit times its maximum gas price stays within the cap and the account's balance covers it. In a sponsored one, the sponsor, an activated account other than the accounts the transaction carries, signs it with its post-quantum key, pays the gas and is reimbursed at most the cap. A failing \texttt{0x7e} invalidates its block and so charges no one. Since anyone could wrap a public proof in a sponsored transaction and collect the cap, the public input also fixes one of four fee modes (\cref{tab:pi}). In \textbf{mode~0}, the account pays the actual cost of its own transaction. The other three modes govern sponsors: \textbf{mode~1} reimburses only the owner's named sponsor up to the cap until a deadline, \textbf{mode~2} lets any sponsor carry the proof unreimbursed, and \textbf{mode~3} is an explicit opt-in public bounty that reimburses any sponsor as in mode~1. For overrides, $A$ can use the enrolled key to drain the account before the override, so the owner's wallet never tops up such an account and submits the override in mode~1 or~2 (G4).

\runsub{Mempool, Builder and Network Rules}\label{sec:pool}
Each rule is enforced at four sites. Nodes check transactions at mempool admission (reinjection after a reorganization included), at each new head when evicting stale transactions, and at block import, which alone decides validity. The fourth site is block building: because a failing \texttt{0x7e} invalidates its block (\cref{sec:fees}), builders re-validate every transaction against the block's timestamp and the state left by those already placed, popping any \texttt{0x7e} that fails at its position (\textbf{R7}). \ifextended The mempool refuses a \texttt{0x7f} in which an account already fails. After cheap checks, it verifies a \texttt{0x7e}'s proof and caches the outcome under a hash of public input and proof, so a rejected proof takes no slot. A \texttt{0x7e} has no signer, so a single pending slot per account would let anyone replace the owner's transaction with a copy whose sponsor later backs out; the mempool therefore keys pending migrations by account and payer, so such a copy expires beside the owner's transaction. Like other large transactions in go-ethereum, a migration of about 190\,KB is announced to peers rather than pushed; a peer fetches it, verifies its proof, and only then admits and announces it.\else The mempool refuses a \texttt{0x7f} in which an account already fails, verifies each \texttt{0x7e}'s proof once and keys pending migrations by account and payer, so that a copy whose sponsor backs out cannot replace the owner's transaction.\fi

\runsub{Wallet Rules}\label{sec:wallet}
The owner's wallet derives the post-quantum key from the same seed: $\mathsf{HMAC\text{-}SHA512}$ keyed with the seed over a label that names the algorithm, the account's path and a key epoch yields the 32-byte seed of ML-DSA key generation, so the mnemonic stays the only backup; the epoch is the migration nonce at which the key is bound, so a rotation derives a fresh key. The wallet proves on the owner's device and lets only the final proof leave it, because the inner proofs are not zero knowledge. For G4, the wallet considers the account's transactions in other hands: those that left the owner's devices or appeared in a block or mempool, and those that $A$ could create with the keys it holds. It sends a new transaction only if, whichever of these are included with the new one and in whatever order, the account does not end up bound to a key that $A$ knows. It also resubmits a recovery that $A$'s ordering made fail. It reports at once an enrollment that it did not make, so the owner can override it in time, and after an override it revokes with the new key the permissions $A$ granted (\cref{sec:nongoals}). By default, a wallet migrates by seed proof an account whose key was never published, since an enrollment would expose the key in the mempool (\cref{sec:eval-audit}); the owner may still enroll, which is safe only before $Q_0$.

\section{Protocol Design: Validators}\label{sec:n1}

Accounts cannot be migrated on a chain whose finality $A$ controls, because every override and frozen-account recovery of \cref{sec:design}, like every validator exit below, takes effect only once finalized, and most come after $T$. If validators still signed with BLS, $A$ would decide which chain finalizes (\cref{sec:adversary}). The protocol therefore switches consensus to leanXMSS keys at the PQ-ready fork, $S = T_0$, before the first transaction that can bind an account. Instead of proving a BLS key from a seed, which would need a proof of a derivation other than BIP-32, it uses the address in a validator's withdrawal credentials, to which the stake is paid. Validators thus migrate as accounts do: before a deadline, the old key authorizes the registration of a leanXMSS key, as in the registry of EIP-8450~\cite{EIP8450}; afterwards, the owner exits through the withdrawal address (\cref{sec:n1-late}).

\runsub{Registration and Checkpoint}\label{sec:n1-reg}
\emph{Registration.} Before a deadline $D_{\mathrm{reg}}$, \ifextended each validator generates leanXMSS keys from fresh randomness (in Ream, one for attestations and one for proposals) and registers them with a 312-byte message signed by its BLS key, under a dedicated signing domain that separates it from every other consensus message. The message names the chain, the validator's index and BLS key, the new keys, the signature scheme and the slots in which the keys will sign.\else each validator generates leanXMSS keys from fresh randomness and registers them with a message signed by its BLS key under a dedicated signing domain.\fi{} A registry contract accepts registrations only while the block's timestamp precedes $D_{\mathrm{reg}}$ and folds each into a hash accumulator; the contract checks no signature. On chain, the inclusion, order and time of registrations are consensus facts that anyone can recompute, whereas an off-chain collector would be a second trusted party whose censorship would stay invisible.

\emph{Checkpoint.} A registration authorized by BLS alone does not authenticate its key against $A$, and after $D_{\mathrm{reg}}$, $A$ may break every BLS key, including those of validators that have left, and sign an alternative history. The transition therefore rests on a trusted checkpoint: $B_{\mathrm{ck}}$, the last block finalized by $D_{\mathrm{reg}}$, which nodes fix out of band, as they already obtain a recent finalized checkpoint when they sync~\cite{ConsensusSpecs}. Our claims for validators are conditional on this checkpoint and on the capacity bound. \ifextended A checkpoint tool reads the registrations up to $B_{\mathrm{ck}}$, checks them against the accumulator, and verifies each BLS signature against the validator set at $B_{\mathrm{ck}}$, together with the encoding of the new keys and their slots.\else A checkpoint tool checks the registrations up to $B_{\mathrm{ck}}$ against the accumulator and the validator set there.\fi{} Its output, the new validator set and the anchor $B_{\mathrm{ck}}$, defines the genesis of a Lean chain whose first block extends $B_{\mathrm{ck}}$. Unfinalized blocks after $B_{\mathrm{ck}}$ are dropped with their registrations, and the execution layer keeps its state at $B_{\mathrm{ck}}$. Registration closes at the switch, $D_{\mathrm{reg}} = S$, because the capacity bound covers only the time before $D_{\mathrm{reg}}$.

\runsub{Conflicts and the Capacity Bound}\label{sec:n1-bound}
\emph{Conflict-as-exclusion defeats ordering attacks.}
$A$ can forge a registration only for a validator whose BLS key it has broken, and because $A$ controls transaction ordering, any rule that keeps the first or the last registration lets $A$ win (by front-running the owner, or by registering last and censoring re-registrations near $D_{\mathrm{reg}}$). The checkpoint tool therefore groups valid registrations by validator and resolves conflicts by exclusion:
(i)~one distinct key tuple (however often submitted) admits the validator;
(ii)~two or more distinct tuples \emph{exclude} the validator regardless of order; and
(iii)~zero registrations mark it \emph{missed}.
Invalid signatures are ignored, so $A$ cannot exclude unbroken validators, while any owner whose registration finalizes by $D_{\mathrm{reg}}$ leaves $A$ no race to win: $A$ can at worst force an exclusion, after which the owner exits via \cref{sec:n1-late}. Censoring the owner's registration until $D_{\mathrm{reg}}$, like censoring an override for $W$, is a denial of service outside our model (\cref{sec:nongoals}). Moreover, registrations omit leanXMSS proofs of possession (which would prematurely spend a stateful one-time leaf key) and permit duplicate keys across different validators: copying a key grants $A$ no signing power, whereas forbidding duplicates would let $A$ exclude honest validators by copying their registrations.

\emph{Capacity bound over the shrunk validator set.}
Every excluded or missed validator, together with any pre-$T_0$ exits forced by $A$ (\cref{sec:n1-late}), shrinks the new validator set $V$, which alone runs the Lean chain, so we measure $A$'s share in $V$ rather than in the old set, since shrinking $V$ amplifies $A$'s relative weight (A5, \cref{sec:adversary}): $f_A = w(U)/w(V) < 1/3$ (G5), where $U \subseteq V$ holds the validators that enter $V$ with keys that $A$ registered and $w$ is the weight that the new consensus gives validators. Before $D_{\mathrm{reg}} = T_0$, A5 additionally requires that the validators whose BLS authorizations $A$ can forge, individually or in aggregate, hold $< 1/3$ of active effective balance (preserving the honesty of $B_{\mathrm{ck}}$) and fewer than two thirds of the 512 positions in any light-client sync committee~\cite{ConsensusSpecs}. A5 thus bounds $A$'s capacity, but only before $D_{\mathrm{reg}} = T_0$.

\emph{Why aggregate forgery matters: one-shot mass slashing.}
Although breaking a BLS signing key cannot steal stake (which follows withdrawal credentials), a consensus attack need not recover keys one by one. In finalized slot 15,575,968, the 7,647 largest validators hold over $1/3$ of all stake; breaking their keys individually takes an estimated 48--122 days on one superconducting CRQC (\extref{app:keys}), \emph{but a single discrete logarithm of their aggregate public key} suffices to forge both signatures of a double-vote slashing proof. Because Ethereum's slashing check verifies the aggregate signature without checking committee membership~\cite{ConsensusSpecs}, that single proof, once included, slashes all of them that are still slashable.

\runsub{Late Path and Exit Requests}\label{sec:n1-late}
A validator that missed the deadline, was excluded or was taken over by $A$ can leave only through its withdrawal address, in one of four ways. (i)~\emph{An account derived from the owner's seed:} the owner migrates it with a seed proof (\cref{sec:tx}); an enrollment does not suffice, because it may be $A$'s. A \texttt{0x7c} from the address then sends an exit request to the EIP-7002 contract~\cite{EIP7002}, and the owner re-stakes with new leanXMSS keys. The protocol's part ends with the request; processing the exit and a later deposit under leanXMSS keys is Lean consensus's part. (ii)~\emph{An account of an owner without a seed:} the validator can exit late only through an enrollment, from $T + W$ on (\cref{prop:noseed}), so it should have its registration final by $D_{\mathrm{reg}}$. (iii)~\emph{BLS withdrawal credentials (type \texttt{0x00}):} with no such path, the change to an execution address~\cite{ConsensusSpecs} must be finalized by $D_{\mathrm{reg}}$ or the checkpoint excludes the validator. The change exposes the BLS withdrawal public key, so after $Q_0$, $A$ may race it with a forged change. (iv)~\emph{A contract,} as in staking pools: it follows its own rules, and if it authorizes exits only by recovered secp256k1 signatures, R3 stops them from $T$ on (\cref{sec:retire}).

\emph{Forced exits.} Exposed withdrawal addresses need one more rule. Until $T$, $A$ holds the key of every exposed address (\cref{sec:adversary}), and from such addresses it could exit honest validators until the honest part of the new set shrank and $f_A$ exceeded a third. From $T_0$ on, the request contracts of EIP-7002 and of EIP-7251, which merges validators~\cite{EIP7251}, therefore accept a request only from an account migrated by a seed proof, a native post-quantum account, an enrollment whose window has closed, or a contract, whose own checks its operator must make post-quantum (\cref{tab:surface}). Neither an activated account, whose code is only its key, nor an account with an EIP-7702 delegation counts as a contract: otherwise $A$ could enroll and activate an exposed withdrawal address, or delegate it, and exit through it. This is the transition's only rule in the execution layer. A validator that $A$ took over works for $A$ until its owner's exit, and $A$ can have it slashed; the exit returns what is left, not what was lost.

\section{Seed Proof System}\label{sec:proof}

The seed proof must resist quantum provers, be zero knowledge, require no trusted setup, fit one transaction (\cref{sec:tx}) and express SHA-512, secp256k1 and Keccak arithmetic.\ifextended{} Pairing-based SNARKs~\cite{Groth16,GWC19,KZG10} fall to Shor's algorithm, as does any proof wrapped in one; lattice-based SNARKs~\cite{BS23} are young, and proofs from MPC or VOLE in the head~\cite{IKOS07,AHIV17,BBDGKORS23} grow with this large relation.\fi{} We therefore build a recursive hash-based STARK~\cite{BBHR18} in Plonky3~\cite{Plonky3}, where secp256k1 appears solely as circuit arithmetic.

\runsub{Relation}\label{sec:relation}
Given the 198-byte public input $x$ (\cref{app:pi}) and witness $(\mathit{seed}, j, c, i)$ with 64-byte $\mathit{seed}$, $j, i < 2^{31}$ and $c \in \{0, 1\}$ (hiding the indices and covering MetaMask at $j = c = 0$ and Ledger Live at $c = i = 0$), the relation checks that the derivation of \cref{sec:bg-hd} along $m/44'/60'/j'/c/i$~\cite{BIP32,BIP44} yields a valid master key and a valid key $k$ at each step, and that the last 20 bytes of $\mathsf{keccak256}(kG)$ match the address in $x$. The relation starts at the 64-byte seed, not at the mnemonic: every key of the wallet derives from the seed, and A1 keeps the seed as secret as the mnemonic (\cref{sec:adversary}), so proving PBKDF2's 2,048 rounds, which only stretch the mnemonic into the seed, would enlarge the relation without strengthening it. This takes six HMAC-SHA512 evaluations, three secp256k1 scalar multiplications and one Keccak-256 permutation; each proof layer binds all 198 bytes of $x$ into its Fiat--Shamir transcript before drawing challenges. The leaf proof $\mathsf{L}$ arithmetizes this relation as an algebraic intermediate representation (AIR) of four KoalaBear-field tables joined by lookup arguments: field operations, SHA-512, Keccak and byte pairs\ifextended{} (\cref{app:pi})\fi.

\begin{figure}[t]
\centering
\includegraphics[width=\columnwidth]{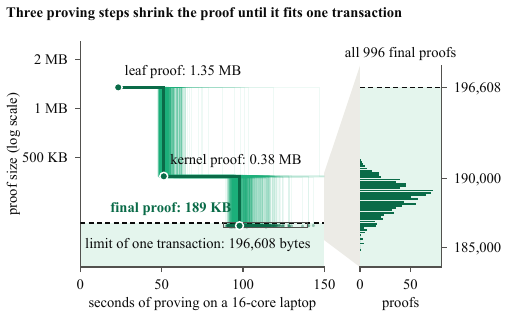}
\caption{The size of the proof during proving, for the \simOwnerCount{} seed proofs of \cref{app:bill}: one thin line each, the thick line through their medians. The leaf proof $\mathsf{L}$ of the key derivation gives way to the kernel $\mathsf{B}$, which verifies $\mathsf{L}$, and to the final proof $\mathsf{F}$, which verifies $\mathsf{B}$. Right: the final proofs, enlarged. Seven proofs end beyond the time axis.}
\label{fig:seedproof}
\end{figure}

\runsub{Recursion and Zero Knowledge}\label{sec:recursion}
Because $\mathsf{L}$ is 1.35\,MB, two recursion layers compress it (\cref{fig:seedproof,tab:layers}): the kernel AIR $\mathsf{B}$ verifies $\mathsf{L}$, and the final AIR $\mathsf{F}$ verifies $\mathsf{B}$ (we wrote both; they constrain every inner check, Merkle path and transcript). Using a larger blowup and fewer queries to trade proving time for size, $\mathsf{F}$ averages 189\,KB and reaches at most 192,501 bytes across all query positions, within our 196,608-byte transaction bound. While $\mathsf{L}$ and $\mathsf{B}$ stay in wallet memory without hiding (\cref{sec:wallet}), $\mathsf{F}$ achieves zero knowledge by randomizing its trace and quotient~\cite{HK24} and salting its Merkle leaves with fresh OS randomness, exposing at most 45 values per private column against 16,384 random degrees of freedom so that two proofs of one witness differ in every opened value.

\runsub{Aggregation}\label{sec:aggregation}
A sponsor aggregates four owners' public final proofs $\mathsf{F}_1, \dots, \mathsf{F}_4$ in two more layers: the first runs on $\mathsf{B}$'s AIR to verify all four $\mathsf{F}_k$, and the second runs on $\mathsf{F}$'s AIR to compress the first over the four 198-byte public inputs. Because evaluating each $\mathsf{F}_k$'s periodic columns at its out-of-domain point $\zeta_k$ in-circuit would exceed $\mathsf{B}$'s row limit, the aggregator takes those values as advice and outputs each $\zeta_k$ with a Poseidon2 digest of the values while keeping every other check of $\mathsf{F}_k$ in-circuit. The node's verifier recomputes the periodic values from $\mathsf{F}$'s fixed key and compares the digests, so an accepted aggregate with false advice yields a Poseidon2 collision.

\runsub{Soundness and Assumptions}\label{sec:proof-security}
Because some seed derives every address, G2 requires \emph{knowledge soundness} and \emph{weak simulation extractability} (after seeing the owner's proofs for public inputs of its choice, $A$ cannot prove a new public input of the owner's address without knowing a seed). \ifextended Classically, the Fiat--Shamir transform of FRI and Plonk-like FRI-based arguments is knowledge sound~\cite{BGKTTZ23}, BCS arguments are UC secure given an honest-verifier zero-knowledge IOP with state-restoration knowledge soundness and a straightline extractor~\cite{BCS16,CF24}, and recursion from straight-line extractors loses little~\cite{CGSY24,CCGOS23}; in the quantum random-oracle model (QROM), the BCS transform has soundness and extraction results under round-by-round IOP premises~\cite{CMS19,COS20}. \Cref{prop:zk} proves that the IOP underlying $\mathsf{F}$ is honest-verifier zero knowledge for uniform masks; the step from it to $\mathsf{F}$'s proof remains part of A4. \else The extended version reviews classical and quantum~\cite{CMS19} random-oracle results. \fi Each layer uses FRI in the Johnson-bound list-decoding regime~\cite{BBHR18FRI,BCIKS20} without grinding, achieving 128, 128 and 130 bits of modeled round-by-round IOP soundness in soundcalc~\cite{soundcalc} (\ifextended\cref{tab:params} in \cref{app:pi}\else parameters in the extended version\fi), and commits with 8-element Poseidon2~\cite{GKS23} Merkle digests (about 124 bits of classical and 82.6 bits of quantum collision resistance~\cite{BHT98}, where realistic cost models find no quantum advantage~\cite{Bernstein09}). Because these figures cover neither multi-height leaf roll-ins nor non-interactive compilation and recursive composition, quantum zero knowledge and weak simulation extractability of the composed proof and its aggregation (quantifying $\varepsilon_{\mathrm{ZK}}$ and $\varepsilon_{\mathrm{SE}}$ in \cref{sec:security}) remain an explicit assumption (A4, \cref{sec:adversary}), without a concrete quantum security level for the composed system.

\section{Security Analysis}\label{sec:security}
\ifextended We state the goals of \cref{sec:goals} as theorems, in three steps: a reduction for the seed proof (\cref{thm:auth}), invariants of an account's authority record (\cref{thm:account}) and recovery under an inclusion assumption (\cref{cor:recovery}); \cref{prop:noseed} gives the guarantee and the limit for owners without a seed (G6), and \cref{prop:minimal} shows which upgrades are necessary. \Cref{sec:adversary} states the assumptions A1--A8. G1 needs no assumption. The proof system enters only through A4 (\cref{sec:proof-security}). The reductions and the induction are proved by hand, in \cref{app:proofs}; TLC checks the model to five blocks and Apalache the window lemma and the authority record for all integers (\cref{sec:eval-model}); A4 is assumed, and the adversarial runs test that the client follows the model (\cref{tab:adv}).\else We state the goals of \cref{sec:goals} as theorems under A1--A8 (\cref{sec:adversary}). G1 needs none, and the proof system enters only through A4 (\cref{sec:proof-security}). The hand proofs are in the extended version, part of the artifact (\cref{sec:impl}).\fi

\begin{definition}[Game $\mathrm{AUTH}$]\label{def:auth}
For a quantum adversary $A$ and a path $p^\ast$:

\smallskip\noindent\fbox{\small\begin{tabular}{@{}r@{\ \,}p{0.80\linewidth}@{}}
1: & draw a seed $s$ as the owner's wallet draws it;\ \ $a^\ast \gets \mathsf{Der}(s, p^\ast)$;\ \ $Q \gets \emptyset$ \\
2: & $\mathit{tx} \gets A^{H, \mathcal{P}}(p^\ast, \mathsf{mk}(s))$ \\
3: & $A$ wins if $\mathit{tx}$ is an accepted \texttt{0x7e} or \texttt{0x7f} that carries an $x \notin Q$ with $a_x = a^\ast$ \\[1pt]
\multicolumn{2}{@{}p{0.85\linewidth}@{}}{$\mathcal{P}(x, p)$: if $\mathsf{Der}(s, p) = a_x$, add $x$ to $Q$ and return the owner's proof of $x$} \\
\end{tabular}}\smallskip

\noindent Here $\mathsf{Der}(s, p)$ is the address that the derivation of \cref{sec:relation} yields from $s$ along $p$, $a_x$ is the account of a public input $x$, and $\mathsf{mk}(s) = \mathsf{HMAC\text{-}SHA512}(\text{``Bitcoin seed''}, s)$ is the wallet's master key and chain code, hence every BIP-32 private key of the wallet, exposed or not. $A$ queries the proof system's random oracle $H$ in superposition and $\mathcal{P}$ classically. \emph{Accepted} means passing the node's proof verification, which does not depend on the chain's state.
\end{definition}

\begin{theorem}[Authenticated migration, G2]\label{thm:auth}
Under A4, every quantum adversary $A$ satisfies
\[ \mathsf{Adv}_A := \Pr[A \text{ wins } \mathrm{AUTH}] \;\le\; \varepsilon_{\mathrm{ZK}} + \varepsilon_{\mathrm{SE}} + \varepsilon_{\mathrm{wit}}: \]
the advantages against the zero knowledge and the weak simulation extractability of the proof, and the probability of computing, from $\mathsf{mk}(s)$, a seed and a path that derive $a^\ast$, each at the resources of $A$ with the simulator and the extractor.
\end{theorem}

\emph{Proof sketch.} Let $\mathsf{Adv}^{\mathsf{S}}_A$ be the probability that $A$ wins when the simulator answers $H$ and, after the same test, $\mathcal{P}$; its proofs do not depend on $s$. A win there yields, by weak simulation extractability, a seed and a path that derive $a^\ast$, computed from $\mathsf{mk}(s)$ and $p^\ast$:
\[ \mathsf{Adv}_A \;\le\; \mathsf{Adv}^{\mathsf{S}}_A + \varepsilon_{\mathrm{ZK}} \;\le\; \varepsilon_{\mathrm{SE}} + \varepsilon_{\mathrm{wit}} + \varepsilon_{\mathrm{ZK}}. \]
Under A1 and A2, only generic search over the mnemonic or the 160-bit address computes such a seed (\cref{sec:bg-quantum}). A second proof of an $x \in Q$ is no win and gains nothing: the migration nonce lets a public input apply only once.

\ifextended\else \Cref{thm:account} adds $\varepsilon_{\mathrm{sig}}$ and $\varepsilon_{\mathrm{hash}}$, the advantages against ML-DSA and the hash functions. In $\varepsilon_{\mathrm{hash}}$ alone, A2 asks more than preimage resistance, namely that the key derivation of \cref{sec:wallet} is pseudorandom even given $\mathsf{mk}(s)$.\fi

\begin{theorem}[Account safety: G1, G3, G4]\label{thm:account}
Let $a$ be an account of a protected owner whose wallet follows \cref{sec:wallet}, and let a quantum adversary $A$ hold its secp256k1 key from the start and order every block. The execution of the rules of \cref{sec:design} that $A$ produces is \emph{ideal}, that is, no seed proof, signature or hash preimage in it is a forgery (\proofsref), except with probability
\[ \Pr[\text{not ideal}] \;\le\; \varepsilon_{\mathrm{auth}} + \varepsilon_{\mathrm{sig}} + \varepsilon_{\mathrm{hash}} \]
over all that $A$ does, where $\varepsilon_{\mathrm{auth}}$ is the bound of \cref{thm:auth}. Write $(m, c, n)$ for the mode, key commitment and migration nonce of $a$'s record at time $t$, with primes after a step. An ideal execution, of any length, has four properties.
\emph{(a)~Retirement.} A secp256k1 authority of $a$ takes effect only if $m = \text{LEGACY}$ and $t < T$, and $m \ne \text{LEGACY}$ implies $m' \ne \text{LEGACY}$.
\emph{(b)~Proof authority.} A migration applies only a public input that the owner proved, at most once, with its key commitment and a charge under the proof of at most its fee cap; $m = \text{BY-PROOF}$ implies $m' = \text{BY-PROOF}$, and every key installed in this mode is the owner's.
\emph{(c)~Correctable binding.} If $m = \text{ENROLLED}$ and $t < T + W$, the override gate is open, and a step under the enrolled key has $(m', c', n') = (m, c, n)$; an override gives $m' = \text{BY-PROOF}$, so by (b) only keys of the owner are installed afterwards. An enrollment needs $t < T$, so the gate is open for more than $W$. If $m = \text{LEGACY}$ and $t \ge T$, the account is frozen, and its record changes only by a seed proof, without a deadline.
\emph{(d)~Ordering.} A recovery of $a$ applies with the effect that its public input fixes (the commitment, the next migration nonce and a charge under the proof within its fee cap), or it makes no write to the record of $a$ and no charge of its own. While $a$ is not final, $n' \ne n$ only for a migration that the owner proved or for an enrollment, possible once and only at $t < T$.
\end{theorem}

\ifextended \emph{Proof sketch.} After hops to simulated proofs ($\varepsilon_{\mathrm{ZK}}$) and uniform seeds for the wallet's post-quantum keys ($\varepsilon_{\mathrm{KDF}}$: here alone A2 asks more than preimage resistance, namely that the key derivation of \cref{sec:wallet} is pseudorandom even given $\mathsf{mk}(s)$), an execution is ideal unless (i)~a migration of $a$ applies a public input the owner did not prove ($\varepsilon_{\mathrm{SE}} + \varepsilon_{\mathrm{wit}}$, as in \cref{thm:auth}), (ii)~a message the owner did not sign is accepted under a post-quantum key of the wallet (an ML-DSA forgery, $\varepsilon_{\mathrm{sig}}$ over all the wallet's keys), (iii)~a key that is not the wallet's is installed against the commitment of one that is (a second preimage), or (iv)~a native account or a contract arises at $a$ (a preimage of the address); $\varepsilon_{\mathrm{hash}}$ sums $\varepsilon_{\mathrm{KDF}}$ and the two preimage advantages. Properties (a)--(d) follow by induction over the seven transitions that change or use an account's authority (\cref{app:proofs}); the window lemma (\cref{sec:eval-model}) extends (c) to releases that move $T$ or change $W$.\fi

\begin{corollary}[Recovery]\label{cor:recovery}
Assume A8, and consider an ideal execution in which a sponsor keeps every condition of each recovery of the owner, other than the account's record, valid for $\Delta$: its gas, the payer's authorization, the proof's expiry and, in a fee mode that has one, the fee deadline. Let the owner send a seed proof for $a$ at time $t$.
(i)~If $a$ is frozen, this one proof binds $a$ to a key of the owner: it is finalized by $t + \Delta$, unless another migration that the owner proved applies first and binds the account.
(ii)~If $a$ is enrolled, the same holds provided that $t + \Delta < T + W$.
(iii)~If $a$ is in mode LEGACY and $t < T$, at most two proofs are needed, the second sent at a time $t_2$ with $t_2 + \Delta < T + W$.
\end{corollary}

By (c) and (d), only by enrolling $a$, once, can $A$ make a pending recovery fail without binding $a$ to a key of the owner; one more proof then overrides the enrollment. No rule forces inclusion on a builder (\cref{sec:nongoals}): without A8, $A$ can delay a recovery but neither change its effect nor charge $a$ more under the proof than its fee cap.

\begin{proposition}[Owners without a seed, G6]\label{prop:noseed}
Let $a$ be the account of an owner without a seed whose post-quantum keys come from fresh randomness.
(a)~Before $Q_0$ only the owner signs for $a$, so an enrollment of $a$ that is final before $Q_0$ is the owner's. Once an enrollment by the owner is final, no secp256k1 authority of $a$ takes effect, and, except with the probability $\varepsilon_{\mathrm{auth}} + \varepsilon_{\mathrm{sig}} + \varepsilon_{\mathrm{hash}}$ of \cref{thm:account}, $a$ stays bound to keys of the owner: no seed proof for $a$ is accepted, and only a rotation that the owner signs changes the key. Here the $\varepsilon_{\mathrm{wit}}$ in $\varepsilon_{\mathrm{auth}}$ is the probability that $A$ finds a seed and a path that derive $a$.
(b)~Let $A$ hold the secp256k1 key of $a$ while the chain records nothing that tells a key of the owner from a key of $A$'s. If the holder of that key can produce whatever a protocol accepts from the owner, the protocol cannot guarantee that $a$ ends bound to keys of the owner alone: of two runs that differ only in who is the owner, it fails in one with probability at least $1/2$.
\end{proposition}

\ifextended For (a), the reduction and transitions of \cref{thm:account} apply from the owner's enrollment on. \fi For (b), from $Q_0$ on $A$ holds every exposed key and runs the owner's program for a key of its own. In two runs that differ only in which of the two is the owner, the chain sees the same, so each of the events $S_0$ and $S_1$, that $a$ ends bound to keys of the one alone or of the other alone, has the same probability in either run; they exclude each other, so $\Pr[S_0] + \Pr[S_1] \le 1$, and in one of them the owner succeeds with probability at most $1/2$\ifextended{} (\cref{app:proofs})\fi.

\emph{Validator transition (G5).} \ifextended\Cref{prop:n1} in \cref{app:proofs} states G5\else The extended version proves G5\fi{} in three parts. A validator's outcome depends only on the set of its valid registrations up to $B_{\mathrm{ck}}$, not on their order. Under A5--A7, the chain starting from the new set is safe and live while $A$'s share of its weight stays below a third. From $T_0$ on, an exit that a protected owner did not authorize needs an enrollment that stays uncorrected until $T + W$.\ifextended{} A withdrawal contract whose own checks a CRQC breaks remains its operator's responsibility; requests that $A$ makes through it can raise $A$'s share of the new set. An adversary controlling finality can censor overrides, so \cref{thm:account}, \cref{cor:recovery} and \cref{prop:noseed} speak of the chain that the validator transition provides.\fi

\begin{proposition}[Necessary upgrades]\label{prop:minimal}
Let a protocol migrate accounts and validators against $A$ without knowing $Q_0$. (a)~Let it tell an owner from $A$ only by the account's own records and the owner's wallet secrets, and let no record final before $Q_0$ tie an exposed account to a secret from which it does not derive. The protocol can guarantee that the account ends bound to keys of the owner alone only by checking a secret from which the account derives and that $A$ cannot compute from what it holds; even if $A$ may hold every BIP-32 key of the wallet (\cref{def:auth}), the seed is one, and G2 requires the check to reveal nothing about it. (b)~Let the keys that replace BLS count only through registrations that the old chain finalizes. The protocol can guarantee that corrections of accounts are finalized by their deadlines only if the old chain finalizes the registrations that determine the new validator set before $A$ can suppress its finality until those deadlines pass.
\end{proposition}

\ifextended For (a), the two runs of \cref{prop:noseed}(b) leave only a check of a secret that $A$ cannot compute, and a secret from which the account does not derive, $A$ can replace by its own; the seed stays unavailable to $A$ under A1 and A2, and revealing it would let $A$ prove for the owner's other accounts and derive the wallet's post-quantum keys. For (b), registrations and corrections count only once finalized, and without a finalized new set only the old chain can finalize them. $A$ can suppress its finality with forged equivocations: once the slashed validators hold $w_{\mathrm{sl}} > w/3$ of the active stake $w$, unslashed votes weigh at most $w - w_{\mathrm{sl}} < 2w/3$, so no checkpoint gains two thirds, and since $A$ may break every BLS key, it can slash more as the inactivity leak~\cite{ConsensusSpecs} drains them (\cref{app:proofs}). \fi Within these classes, a zero-knowledge check of a derivation secret and timely finalization of replacement registrations are thus necessary; the protocol provides them with the seed proof and the validator switch, whose registrations the old chain finalizes by $D_{\mathrm{reg}}$ while A5 bounds $A$. \Cref{thm:auth,thm:account}, \cref{cor:recovery}, \cref{prop:noseed} and the validator transition give G1--G6 under A1--A8 and the conditions they state, among them the wallet rules, timely recovery and $A$'s share of the new set staying below a third. \Cref{tab:surface} maps the other components a CRQC breaks to these goals and assumptions.

\section{Implementation and Evaluation}\label{sec:eval}\label{sec:impl}
\looseness=-1 We implement the protocol in one execution and one consensus client; the code is public at \url{https://github.com/DaseinDa/ethereum-pq-transition}, whose README describes its parts and how to build them.
A modified go-ethereum implements the registry, the transactions of \cref{tab:tx}, rules R1--R7 at all four sites of \cref{sec:pool} and the request rule, and verifies ML-DSA-44 signatures natively and $\mathsf{F}$ (\cref{app:pi}) with a linked Rust library. The three layers of \cref{sec:proof} are Rust programs on a pinned Plonky3, and the wallet derives the post-quantum key (\cref{sec:wallet}) and builds the transactions. A modified Ream~\cite{Ream} drives each go-ethereum node through the Engine API, and our tools and the registry contract of \cref{sec:n1-reg} carry out the validator switch.

\ifextended Unless noted, measurements run on one laptop with an Intel i9-12950HX CPU (16 cores, 24 threads), an RTX A5500 GPU (\cref{app:gpu}) and 64\,GB of memory, of which Linux (WSL~2) sees 50\,GiB. Two go-ethereum nodes, each driven by a Ream node, form a local chain with 12-second slots; the audit of \cref{sec:eval-audit} adds mainnet data and ten-node runs, each on one Amazon EC2 m7i.4xlarge instance with emulated links (\cref{app:bill}). Every key and seed is a public test value. We give the number of repetitions $n$ where a measurement was repeated; the end-to-end and adversarial runs are pass/fail checks, run once each. \Cref{sec:eval-proof,sec:eval-gas} show that a seed proof fits one transaction and what batching saves, \cref{sec:eval-audit} what migration costs at mainnet scale and how long an enrollment waits in the mempool, and \cref{sec:eval-runs,sec:eval-n1,sec:eval-model} report the end-to-end, adversarial and validator-switch runs, which all pass, and the TLA+ models, whose properties all hold.\else \looseness=-1 Unless noted, measurements run on a laptop with an Intel i9-12950HX CPU (16 cores), an RTX A5500 GPU (\cref{app:gpu}) and 64\,GB of memory (50\,GiB under WSL~2), on a local chain of two go-ethereum nodes, each driven by a Ream node, with 12-second slots; the audit of \cref{sec:eval-audit} adds mainnet data and ten-node runs, each on one Amazon EC2 m7i.4xlarge instance (\cref{app:bill}). Keys and seeds are public test values; $n$ counts repetitions. \Cref{app:model} reports the end-to-end, adversarial and validator-switch runs, which all pass, and the TLA+ models, whose properties all hold. \Cref{sec:eval-proof,sec:eval-gas} show that a seed proof fits one transaction, and \cref{sec:eval-audit} what migration costs at mainnet scale.\fi{}

\runsub{Proofs and Verification}\label{sec:eval-proof}
\looseness=-1 A seed proof takes its owner about \keynum{two minutes}, once per account (\cref{tab:layers}): in a build for this CPU, \simProveAll\,s of proving, \simOwnerProofSeconds\,s from start to final proof (\gpuLaptopOwnerSeconds\,s with the GPU), and 31\,GiB at peak, the medians over the \simOwnerCount{} proofs of \cref{app:bill}. Proving thus needs a machine with at least 32\,GiB that the owner trusts with the seed, so a hardware wallet's owner enrolls before $T$ or enters the mnemonic on such a machine. As a cost reference, a rented A100 machine proves and aggregates \gpuBestOwnersPerHour{} owners an hour ($n = \gpuBestOwners$), \keynum{\$\gpuBestDollarsPerThousand} per thousand (\cref{app:gpu}). \ifextended Over ten proofs, $\mathsf{F}$ has $188{,}709 \pm 1{,}154$ bytes. Inside go-ethereum, built for portable x86-64, it verifies in $37.9 \pm 0.4$\,ms ($n = 10$).\else Inside go-ethereum, built for portable x86-64, $\mathsf{F}$ verifies in $37.9 \pm 0.4$\,ms ($n = 10$).\fi{} Each recursion layer rejects the corrupted inner proofs and wrong public inputs of our negative tests, 73 in all.

\runsub{Gas and Throughput}\label{sec:eval-gas}
\looseness=-1 A seed proof costs as much gas as \keynum{50} enrollments. No precompile prices a STARK, so we price $\mathsf{F}$ at the median gas per nanosecond (0.143) of three already priced precompiles (\texttt{ecrecover} and the BN254 and BLS12-381 pairings~\cite{Wood14,EIP2537}), timed in one portable binary on one CPU core ($n = 10$ each). Verifying $\mathsf{F}$ on that core takes a median of 39.7\,ms\ifextended{} (37.9 in go-ethereum)\fi, a charge of 5,666,000 gas. A self-paid recovery (\texttt{0x7e}) then costs $8{,}738{,}837 \pm 18{,}210$ gas over ten fresh proofs (at most 8,770,396), most of it for verification and for 189\,KB of calldata. The three rates span 0.104 to 0.261 gas per nanosecond: at the lowest a recovery would cost 7.2\,M gas and at the highest 13.4\,M, where the 21.5\,M of a batch's verification would not fit in one transaction. The recovery paths' days scale with the verification charge.

\ifextended Batching doubles the recoveries in a block. Mainnet blocks have a 60\,M gas limit (\cref{sec:eval-audit}). A builder adds a transaction only if the gas left covers its declared limit, for a recovery its use plus 10\,\% (9.75\,M for the sponsored recoveries of our runs, which use 8.8\,M). Such a block thus holds six recoveries but 343 enrollments, and one day of blocks at most 43,200 recoveries or 2.47\,M enrollments. Over ten batches of four fresh proofs, the aggregate proof of a \texttt{0x7f} has $188{,}965 \pm 1{,}026$ bytes (at most 192,757 over all query positions) and verifies, with the recomputation of the periodic values, in 82.4\,ms ($n = 10$), a charge of 11,771,000 gas. In go-ethereum a batch of four recoveries then uses $15{,}073{,}675 \pm 16{,}443$ gas, 3.77\,M per account, so a block holds three batches, twelve recoveries or 86,400 a day, twice as many as one by one. The sponsor proves a batch in \simBatchSeconds\,s at 31\,GiB in a build for this CPU (\simBatchSecondsPortable\,s portable), the median over \simBatchCount{} batches (\cref{app:bill}).\else \looseness=-1 Batching doubles the recoveries in a block. A builder adds a transaction only if the gas left covers its declared limit, for a recovery its use plus 10\,\% (9.75\,M for the sponsored recoveries of our runs, which use 8.8\,M), so a block at mainnet's 60\,M gas limit holds \keynum{six} recoveries but \keynum{343} enrollments. Over ten batches of four fresh proofs, the aggregate proof of a \texttt{0x7f} has $188{,}965 \pm 1{,}026$ bytes (at most 192,757 over all query positions) and verifies, with the recomputation of the periodic values, in 82.4\,ms ($n = 10$), a charge of 11,771,000 gas. In go-ethereum a batch of four recoveries then uses $15{,}073{,}675 \pm 16{,}443$ gas, 3.77\,M per account, so a block holds three batches, or \keynum{twelve} recoveries. The sponsor proves a batch in \simBatchSeconds\,s at 31\,GiB in a build for this CPU (\simBatchSecondsPortable\,s portable), the median over \simBatchCount{} batches (\cref{app:bill}).\fi{}

\runsub{Cost on Mainnet, Simulated}\label{sec:eval-audit}
\looseness=-1 We audit what the migration adds to each layer of \cref{fig:overview} (\cref{tab:audit}) by running the protocol, not by multiplying unit costs. The finalized beacon state of slot 15,575,968 (3~November 2026) lists 814,177 live validators
staking 44.2\,M ETH. At block 26,344,089 of the next day (BigQuery's public Ethereum data~\cite{BigQueryEthereum}, 345\,M addresses), EOAs that have sent
a transaction, exposing their public key, hold \keynum{54.6\,\%} of the 76.7\,M ETH on the execution layer without the
deposit contract and two burn addresses (a lower bound: keys exposed only by permits or EIP-7702 authorizations are
not counted), unexposed EOAs 33.3\,\% and contracts, which migrate through their code or owners (\cref{sec:retire}), 12.0\,\%.
4.45\,M EOAs hold at least 0.1\,ETH: 1.51\,M have sent since November 2023, 1.24\,M are exposed but have sent nothing since, and 1.70\,M are unexposed.

\ifextended \looseness=-1 \emph{Simulation setup.} We run the implementation on a population with mainnet's composition: ten geth/Ream nodes, each in its own network
namespace on a random 4-regular graph (20--80\,ms latency, 50\,Mbit/s links), with mainnet's 60\,M gas limit
and unchanged EIP-1559. Ordinary traffic reproduces mainnet's demand: each slot draws the gas used of one of the 30 consecutive
mainnet blocks 26,345,047 to 26,345,076 of 5~November 2026 (8,381 transactions; 30.2\,M gas on average, the target) and fills it with calls that burn the gas of sampled mainnet transactions at their fee bids; a
user whose bid is below the base fee does not send. These bids are set against that day's base fee of about
0.05\,gwei and understate what users would pay: in 2025, at median base fees of 0.4--2\,gwei, blocks still carried
18\,M gas. In the \emph{demand of record}, half of the ordinary users re-price to the base fee; at \emph{today's bids}, all
keep that day's bids, which bounds the migration's speed from above. Migration transactions bid at most 1\,gwei. Enrollment
runs move 10,000 accounts and recovery runs 1,000 owners with fresh seed proofs. A time on mainnet is a population
divided by the steady-state accounts per slot (migration gas per slot over the gas of one account, activation included), at 7,200 slots a
day; the steady state runs from the block where the base fee reaches 0.8\,gwei to the one where 90\,\% of the accounts are done.\else \looseness=-1 \emph{Simulation setup.} We run the implementation on a population with mainnet's composition: ten geth/Ream nodes on emulated links, with mainnet's 60\,M gas limit and unchanged EIP-1559. Ordinary traffic reproduces mainnet's demand: each slot draws the gas used of one of 30 consecutive mainnet blocks of 5~November 2026 (30.2\,M gas on average, the target) and fills it with calls that burn the gas of sampled mainnet transactions at their fee bids; a user whose bid is below the base fee does not send. These bids are set against that day's base fee of about 0.05\,gwei and understate what users would pay: in 2025, at median base fees of 0.4--2\,gwei, blocks still carried 18\,M gas. In the \emph{demand of record}, half of the ordinary users re-price to the base fee; at \emph{today's bids}, all keep that day's bids, which bounds the migration's speed from above. Migration transactions bid at most 1\,gwei. Enrollment runs move 10,000 accounts and recovery runs 1,000 owners with fresh seed proofs. A time on mainnet is a population divided by the steady-state accounts per slot, at 7,200 slots a day (\cref{app:bill}).\fi{}

\begin{figure}[t]
\centering
\includegraphics[width=\columnwidth]{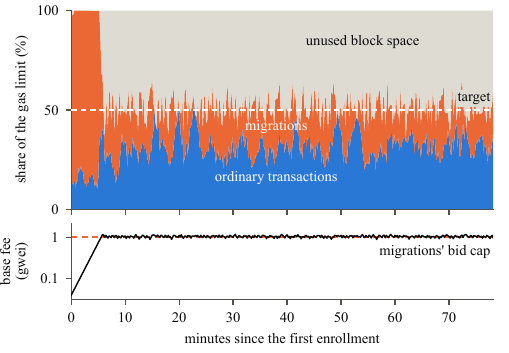}
\caption{Block space during the enrollment run at the demand of record: the share of each block's 60\,M gas limit used by ordinary transactions and by migrations or left unused (one-minute moving average), above the block's base fee.}
\label{fig:blockspace}
\end{figure}

\ifextended
\begin{figure}[t]
\centering
\includegraphics[width=\columnwidth]{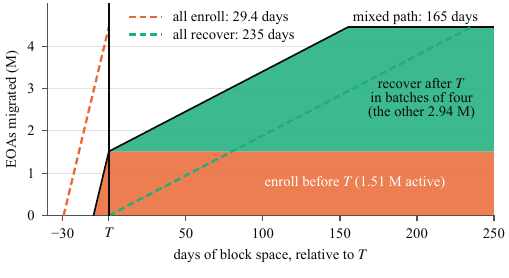}
\caption{Migration of the 4.45\,M EOAs holding at least 0.1\,ETH at the simulated rates of record, in days of block space relative to $T$. Filled: the mixed path, in which the 1.51\,M accounts active since November 2023 enroll before $T$ (\simDaysBActiveEnroll{} days, orange) and the other 2.94\,M recover in batches of four after it (\simDaysBRestBatched{} days, green), \simDaysBSplitBatched{} days in all. Dashed: all enroll before $T$ (\simDaysBEnroll{} days) or all recover after it in batches (\simDaysBBatched{} days).}
\label{fig:progress}
\end{figure}
\fi

\input{generated/bill_table}

\begin{figure}[t]
\centering
\includegraphics[width=\columnwidth]{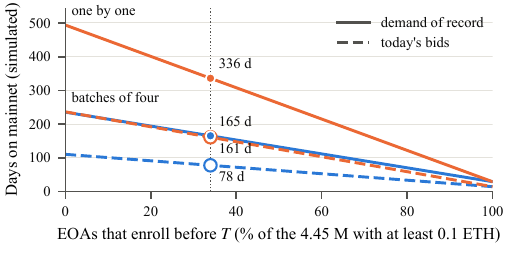}
\caption{Days on mainnet at the simulated rates to migrate the 4.45\,M EOAs with at least 0.1\,ETH, against the share that enrolls before~$T$; the others recover after~$T$, in batches of four (blue) or one by one (orange). Solid: half of the ordinary users raise their bids to the base fee (demand of record); dashed: all keep their current bids (today's bids). The dotted line marks the owners who have sent a transaction since November 2023 (34\,\%).}
\label{fig:uptake}
\end{figure}

\looseness=-1 \emph{Block space sets the days.} At 1\,gwei, EIP-1559 holds the average block at its target and the migration gets what ordinary
users leave: below 1\,gwei a block fills with migration transactions, above it only ordinary transactions enter (\cref{fig:blockspace}). At the
demand of record the migration takes \simEnrollMigShare\,\% of the target, about a fifth of the limit, and
\simEnrollServed\,\% of the ordinary demand is still included, 19.0\,M gas a block and as much as users bought at such fees in 2025 (\simBaselineServed\,\% without the migration); at
today's bids it takes \simEnrollMigShareUpper\,\% and leaves \simEnrollServedUpper\,\%. Enrolling the 4.45\,M EOAs
before~$T$ takes \keynum{\simDaysBEnroll{} days}, the 1.51\,M active ones \simDaysBActiveEnroll, and recovering the 4.45\,M
after~$T$ in batches of four \keynum{\simDaysBBatched} (\ifextended\cref{tab:sim,fig:progress}\else\cref{tab:sim}\fi), so $T$ must be announced early enough for the active owners to enroll. Overrides take this path too and may cluster near the single deadline $T + W$ (\cref{sec:override}): $W$ admits \simOverridesInWBatchedM\,M of them in batches, and if $A$ enrolled all \simEoasBExposedM\,M exposed EOAs with at least 0.1\,ETH, their overrides would take \simDaysBExposedBatched{} days (\simDaysAExposedBatched{} for the \simEoasAExposedM\,M with at least 0.01\,ETH, for which a release would lengthen $W$). \ifextended On the mixed path of \cref{fig:progress}, the
other 2.94\,M, which recover in batches, include\else On the mixed path, in which the 1.51\,M active accounts enroll before~$T$ and the other 2.94\,M recover in batches (\keynum{\simDaysBSplitBatched{} days} in all, \cref{fig:uptake}), the latter include\fi{} the 1.70\,M unexposed EOAs that the wallet rule of \cref{sec:wallet} sends to seed proofs. \ifextended With the 1.33\,M EOAs that hold
at least 100 USDT, USDC or DAI but less than 0.1\,ETH, all \simTokenEoasM\,M recovering after~$T$ take \simDaysTokenBatched{} days in batches and
\simDaysTokenSingle{} one by one. \Cref{tab:sim} gives other balances, and \cref{fig:uptake} the time against the early
share. \fi \ifextended The rate depends neither on the population's size (in batches at today's bids, 500 owners recover at \simLinHalf{} accounts per slot,
1,000 at \simLinFull) nor on the machine or the number of sponsors (\cref{app:bill}). Across random draws, enrollment takes \simDaysBEnroll{} to \simDaysBEnrollSeedTwo{} days and batched recovery
\simDaysBBatchedFastest{} to \simDaysBBatchedSlowest; February 2025's demand moves both by at most \sensDemandMaxPct\,\%, and if three quarters of users re-price, enrollment takes \sensDaysEnrollRepriceHigh{} days and batched recovery \sensDaysBatchedRepriceHigh{} (\extref{tab:sens}).\else The rate depends neither on the population's size nor on the machine or the number of sponsors (\cref{app:bill}); if three quarters of users re-price, enrollment takes \sensDaysEnrollRepriceHigh{} days and batched recovery \sensDaysBatchedRepriceHigh{} (\extref{tab:sens}).\fi{}

\input{generated/exchange_table}

\emph{Exchanges enroll at $T_0$, in minutes.} Public labels~\cite{Spellbook,EthLabels,DefiLlamaAdapters} and the reserve lists of five exchanges~\cite{PoRBinance,PoRBybit,PoRBitstamp,PoRHTX,PoRBackpack} attribute \exAddresses{} addresses to \exNumExchanges{} centralized exchanges and custodians. These hold \exEthM\,M ETH, \exShareEl\,\% of the execution layer (a lower bound), \exKeysEthM\,M of it in \exKeys{} EOAs with at least 0.1\,ETH and \exExposedValueShare\,\% of that behind exposed keys. Of the \exTopExchanges{} exchanges among the 1,000 largest balances, the \exExamined{} whose custody we examined, with \exExaminedShare\,\% of their ETH, document no seed that the relation of \cref{sec:relation} accepts (\cref{tab:exchanges}); Fireblocks and Fordefi derive addresses from a master key, not a seed~\cite{CustodyFireblocksPaths,CustodyFordefiPaths,CustodyFireblocksRecovery}. As far as their documentation shows, they are owners without a seed and cannot migrate by seed proof: an account still in LEGACY at $T$ stays frozen, and from $Q_0$ on no enrollment can be told apart from $A$'s (\cref{prop:noseed}). By G6 an exchange's enrollment is surely its own only if final before $Q_0$, so it enrolls all its keys at $T_0$. \Cref{tab:exchanges} gives what this costs each exchange: all \exKeys{} keys enroll in \keynum{\exEnrollMin{} minutes} of block space at the demand of record (\exEnrollMinUpper{} at today's bids), for \keynum{\exFeeEth\,ETH} in fees at 1\,gwei, with no proving.\ifextended{} The \exEthContractM\,M ETH in their contracts follows the contracts' code (\cref{sec:retire}).\fi{} \exValidators{} validators with \exStakeM\,M ETH, \exStakeShare\,\% of the stake, withdraw to exchange addresses and register before $D_{\mathrm{reg}}$ (\cref{sec:n1-late}). Exchanges thus need $T_0$ to precede $Q_0$, and minutes after it; $T$ and $W$ serve the other accounts, which need days.

\begin{figure}[t]
\centering
\includegraphics[width=\columnwidth]{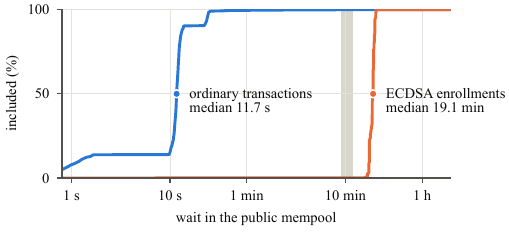}
\caption{Share of the transactions sent in the steady state of the same run that are included within a given wait in the public mempool. Every enrollment waits longer than the 9 to 12 minutes in which a fast quantum computer extracts its key (shaded)~\cite{BZG26}, while ordinary transactions are included within about one slot.}
\label{fig:wait}
\end{figure}

\ifextended \emph{A mass enrollment waits minutes in the mempool.} At the demand of record, an enrollment sent in the steady state waits a median of \simEnrollWaitMedian\,s, or \simEnrollWaitMedianMin{} minutes (10th and 90th percentiles \simEnrollWaitPten{} and \simEnrollWaitPninety\,s; \cref{fig:wait}), while the ordinary transactions of the same blocks wait \simOrdWaitMedian\,s, 99\,\% of them less than \simOrdWaitPninetyNine\,s; at today's bids an enrollment waits \simEnrollWaitMedianUpper\,s. The nodes' logs give each wait, from a transaction's first acceptance by a mempool to the moment the proposer hands over the block that carries it. The queue sets this wait, not the protocol: our driver lets 2,000 owners enroll at once, \simEnrollBacklog{} were pending on average, and Little's law, the pending count over the \simEnrollLittleSealedPerSlot{} enrollments sealed per slot, gives the mean wait within \simEnrollLittleGap\,\%. More owners sending at once lengthen the wait in proportion: in runs with 500, 1,000 and 4,000 pending, an enrollment waits a median of \backlogWaitMinHalfK, \backlogWaitMinOneK{} and \backlogWaitMinFourK{} minutes (\cref{tab:backlog}).

\looseness=-1 \emph{The wait opens on-spend races.} Babbush et al.~\cite{BZG26} estimate that a first fast-clock CRQC needs about 9 or 12 minutes from a published key to its private key, and conclude that such a machine can hardly race Ethereum transactions, most of which are included within a minute. A mass enrollment is different: with 1,000 enrollments pending, about three full blocks, \backlogOverNineOneK\,\% of them waited longer than 9 minutes, and with 500 none did (\extref{tab:backlog}). \else \looseness=-1 \emph{The mempool wait opens on-spend races.} At the demand of record, an enrollment sent in the steady state waits a median of \keynum{\simEnrollWaitMedianMin{} minutes} (\cref{fig:wait}; \simEnrollWaitMedianUpper\,s at today's bids), while the ordinary transactions of the same blocks wait \simOrdWaitMedian\,s, 99\,\% of them less than \simOrdWaitPninetyNine\,s. The nodes' logs give each wait, from a transaction's first acceptance by a mempool to the proposer's hand-over of its block. The queue, not the protocol, sets this wait: our driver lets 2,000 owners enroll at once, and Little's law, the \simEnrollBacklog{} pending on average over the \simEnrollLittleSealedPerSlot{} enrollments sealed per slot, gives the mean wait within \simEnrollLittleGap\,\%. Babbush et al.~\cite{BZG26} estimate that a first fast-clock CRQC needs about 9 or 12 minutes from a published key to its private key, and conclude that such a machine can hardly race Ethereum transactions, most of which are included within a minute. A mass enrollment is different: with 500, 1,000 and 4,000 enrollments pending, the median wait is \backlogWaitMinHalfK, \backlogWaitMinOneK{} and \backlogWaitMinFourK{} minutes, and with 1,000, about three full blocks, \backlogOverNineOneK\,\% waited longer than 9 minutes, with 500 none (\extref{tab:backlog}). \fi An owner who bids above the others is included sooner and lengthens the wait of the others, and $A$ can delay an enrollment in the slots that it builds (\cref{sec:adversary}); a machine that needs 9 minutes per key races about 160 enrollments a day and chooses the most valuable, since a pending enrollment names its account. \ifextended For an account whose key was never published, the \texttt{0x7d} thus hands the key to $A$ after $Q_0$: $A$ may enroll the account first, which the override corrects (G3), or sweep it before the enrollment lands, which nothing undoes (\cref{sec:nongoals}). The seed-proof migration of a LEGACY account before $T$ (\cref{sec:tx,sec:wallet}) removes this exposure, since its owner never signs with secp256k1; it costs a seed proof instead of an enrollment. If $A$ already holds the key and enrolls first, the owner's migration fails on its nonce, and a new seed proof overrides the enrollment.\else A pending \texttt{0x7d} thus hands a key that was never published to $A$ after $Q_0$, for a stealth enrollment, which the override corrects (G3), or a sweep, which nothing undoes (\cref{sec:nongoals}); the seed-proof migration of a LEGACY account before $T$ (\cref{sec:tx,sec:wallet}) avoids this, since its owner never signs with secp256k1.\fi

\ifextended
Without the migration, the median node on EC2 downloads \simNodeBaselineDownGB\,GB a day, \simNodeBaselineClDownGB\,GB of
it consensus traffic (aggregated leanXMSS signatures), and uses \simNodeBaselineGethCpu\,s (geth) and
\simNodeBaselineReamCpu\,s (Ream) of CPU per slot; the migration adds \simNodeEnrollAddedDownGB\,GB a day during the
enrollment and \simNodeBatchedAddedDownGB\,GB during the batched recovery. Ordinary traffic burns gas without
mainnet's calldata, so only these additions carry over to mainnet. \Cref{tab:audit} lists what each layer bears and
\cref{app:bill} the full bill.
\fi

\looseness=-1 \emph{Lower gas per account, not faster proving, shortens the migration.} Block space, not proving or propagation, sets how long the migration takes\ifextended{}
(\cref{fig:layers})\fi. Verification takes \keynum{65\,\%} of a recovery's gas and \keynum{78\,\%} of a batch's, and calldata most of the rest; the calldata and 37\,\% of a batch's verification are paid once per proof, and the rest, 1.86\,M gas per account in a batch of four, once per account. A prototype verifier that evaluates its own periodic columns exactly and sparsely, from a table fixed by the verifying key, takes 33\,\% of the batch verifier's time and 40\,\% of the single one's, which scaled to our records gives 27.3 instead of 82.4\,ms (3.90 instead of 11.77\,M gas) and 15.9 instead of 39.7\,ms (2.27 instead of 5.67\,M gas). Faster proving shortens only how long a proof takes (\cref{app:gpu}).

\ifextended
\runsub{End-to-End and Adversarial Runs}\label{sec:eval-runs}
On the two-node chain, an owner recovers a frozen account, activates the new key and spends with ML-DSA-44; the recovery survives a reorganization, which removes it and includes it again, and a restart of both nodes, and it is finalized. Key rotation and sponsored recoveries in fee modes 1--3 execute and agree on both nodes. On a line of three go-ethereum nodes under Ream, ten recoveries of 190--193\,KB, each submitted to the first node, cross both hops by announcement and fetch (\cref{sec:pool}); a hop takes 0.55--0.56\,s ($n = 10$ per hop), of which 0.5\,s is go-ethereum's wait before it fetches an announced transaction and 39\,ms the verification of $\mathsf{F}$. After a restart of all three nodes, they reach the same new head and keep every receipt. On four nodes, eight owners migrate by seed proof 15 minutes before $T$, singly and in batches, and every mempool then refuses each owner's ECDSA transaction. \ifextended\Cref{tab:adv} in \cref{app:model}\else The extended version\fi{} lists the runs in which $A$ holds the account's exposed key and builds blocks; each ends as \cref{sec:design} predicts, and in particular, when $A$ front-runs one owner of a four-owner batch of real proofs, the batch loses only that owner's entry.

\runsub{Validator Switch}\label{sec:eval-n1}
We run the transition of \cref{sec:n1} on the same two nodes, from a snapshot of ten validators. A stand-in chain carries the registrations until $D_{\mathrm{reg}}$, and a Lean chain takes over from $S$. Node~H holds six honest validators and node~X the key that $A$ registered for a seventh. The registrations cover a validator with BLS withdrawal credentials and every case of the conflict rule: honest registrations, one of them sent twice; invalid signatures; $A$'s registration for a validator whose owner missed the checkpoint; and conflicts in both orders. The checkpoint includes seven validators and excludes the conflicts and the BLS credential, so $f_A = 1/7$.

\emph{The switch.} The first Lean block extends $B_{\mathrm{ck}}$. In the final run, both execution clients drop their unfinalized tail 7.3\,s after $S$; its only transaction, the registration of the owner who missed the checkpoint, returns to the mempool and reverts in the first Lean block, past $D_{\mathrm{reg}}$. Finality passes $B_{\mathrm{ck}}$ 76\,s after $S$. \emph{$A$ silent.} With $A$'s validator online but silent from slot 31 of the Lean chain, the six honest validators advance finality from slot 25 to slot 36 within twelve slots. \emph{Late path.} After $T$, the withdrawal address of a validator excluded for a conflict is recovered, activated and sends an EIP-7002 exit request, which its block carries and Ream hands to the execution client; on the Lean chain the validator set stays the one fixed at the checkpoint, and no exit, withdrawal or deposit is processed. \emph{Request rule.} From $T_0$ on, EIP-7002 and EIP-7251 requests from a legacy address, or from an enrolled and activated address until $T + W$, revert.

\emph{Key lifetime.} Each validator signs with a leanXMSS key valid for $2^{22}$ slots (1.6 years at 12\,s), with a 52-byte public key; a signature in the last slot verifies.

\emph{Repetitions.} All checks pass in the final run, whose $D_{\mathrm{reg}}$ falls 108\,s before $S$ and whose $T_0$ 12\,s after it, and in a run of the released package with $D_{\mathrm{reg}} = S = T_0$; the conflict-rule and switch checks pass in six runs with independent keys.

\runsub{Model Checking}\label{sec:eval-model}
We model the account rules and the validator switch in TLA+ with ideal signatures and proofs (\ifextended\cref{tab:model} in \fi\cref{app:model}). TLC explores every ordering by $A$ of up to five blocks, Apalache proves the window lemma of \cref{sec:override} and an inductive invariant of an account's idealized authority record over unbounded integers (\proofsref), and every modeled property holds.
\fi

\section{Related Work}\label{sec:related}
\ifextended \Cref{tab:compare} compares our protocol with other approaches; the text gives what the table cannot.\fi

\ifextended
\begin{table}[t]
\caption{Our protocol and related approaches (\yes: provided; \partly: in part, or left open; \no: absent; --: does not apply). Before legacy signatures end: SM, a signed move (a migration that the legacy signature authorizes while it is still valid); Cor, correctable (the owner can undo such a move by the adversary); NR, no rollback. After they end: Rec, exposed keys recover; Fee, a fee without the legacy key; Ord, the adversary's ordering; Val, validator keys. Last columns: proof size and verification; Sc, client and chain scale.}
\label{tab:compare}
\centering
\scriptsize
\setlength{\tabcolsep}{1.8pt}
\begin{tabular}{@{}lccccccclc@{}}
\toprule
 & \multicolumn{3}{c}{Before} & \multicolumn{4}{c}{After} & & \\
\cmidrule(lr){2-4}\cmidrule(lr){5-8}
Approach & SM & Cor & NR & Rec & Fee & Ord & Val & Proof, verify & Sc \\
\midrule
Signed off-ramps~\cite{EIP8141,BIP360} & \yes & \no & \yes & \no & -- & \no & \no & -- & \no \\
Vault contract~\cite{KBF26} & \yes & \no & \yes & \no & -- & \partly & \no & -- & \partly \\
Commit, delay, reveal~\cite{Stewart18} & -- & -- & \yes & \no & \no & \partly & -- & -- & \no \\
Emergency fork~\cite{Buterin24} & \no & -- & \no & \yes & \no & \no & \no & not given & \no \\
Freeze, recovery~\cite{EIP8425} & \partly & \partly & \yes & \yes & \no & \no & \no & not given & \no \\
Legacy sunset~\cite{BIP361} & \yes & \no & \yes & \partly & \no & \no & -- & not given & \no \\
Signature lifting~\cite{SW23} & -- & -- & \yes & \yes & \yes & \yes & -- & 100s of KB & \no \\
Seed proofs, EdDSA~\cite{BCRS26} & -- & -- & \yes & \yes & \no & \no & \no & 5.4\,MB, 2.3\,s & \no \\
Hidden-key proofs~\cite{KK26} & \yes & \partly & \yes & \partly & \partly & \partly & \no & not given & \partly \\
Key registry~\cite{EIP8450} & -- & -- & -- & -- & -- & -- & \partly & -- & \partly \\
\midrule
This paper & \yes & \yes & \yes & \yes & \yes & \yes & \yes & 189\,KB, 38\,ms & \yes \\
\bottomrule
\end{tabular}
\end{table}
\fi

\emph{Emergency forks and Bitcoin.} Buterin's emergency fork~\cite{Buterin24} shares our core primitive, an aggregated STARK proof that an address derives from a secret preimage, but rolls the chain back after visible thefts (\cref{sec:intro}). Our protocol instead retires ECDSA at a scheduled time $T$, with cheap enrollment before it, correctable binding until $T + W$ and validators moved too. EIP-8425~\cite{EIP8425}, a concurrent draft, proposes a similar freeze with recovery by a commitment registered before a trusted cutoff or by a seed proof, but leaves the mechanism, proof system and schedule open. Our protocol needs no trusted cutoff for accounts because seed proofs override every ECDSA-signed enrollment until $T + W$, and it instantiates and benchmarks the proof. BIP-361~\cite{BIP361} could rescue frozen Bitcoin outputs by proving knowledge of a hardened BIP-32 parent key, whereas we prove knowledge of the BIP-39 seed and handle account-chain delegations, \texttt{ecrecover} contract checks and validators. PQ-Vault~\cite{KBF26} leaves Ethereum unchanged and protects only funds moved into a vault contract secured by a quantum-safe sidechain, whereas our protocol migrates account and validator authority in place, so unmigrated accounts stay frozen and recoverable.

\emph{Seed proofs.} \ifextended Sattath and Wyborski~\cite{SW23} lift a signature scheme whose keys derive through a post-quantum one-way function, as in HD wallets, to a post-quantum scheme over the same keys; we instantiate such a lift for Ethereum's derivation, from the BIP-39 seed through BIP-32 to a secp256k1 key and its Keccak-256 address, as a transparent proof small enough for one transaction. Baldimtsi et al.~\cite{BCRS26} prove knowledge of an RFC~8032~\cite{RFC8032} seed for EdDSA chains in a zkVM, in 6.2\,s per proof. Deegan~\cite{Deegan25} sketches the statement for ECDSA wallets, PBKDF2 included, and leaves its performance open; we instantiate and measure it from the 64-byte seed.\else Sattath and Wyborski~\cite{SW23} lift HD-wallet signature schemes to post-quantum schemes over the same keys, which our proof instantiates for Ethereum; Baldimtsi et al.~\cite{BCRS26} prove an EdDSA seed in a zkVM, and Deegan~\cite{Deegan25} sketches the ECDSA statement without measuring it.\fi{} Kiraz and Kardas~\cite{KK26} bind a still-hidden public key to a post-quantum key, and let an exposed key enroll under both signatures while it is uncompromised. \ifextended Ray et al.~\cite{RGR26} prove knowledge of the private key behind a hash-committed address, for keys never exposed, and quote proofs of 2 to 10\,MB from others' benchmarks. Once $A$ can compute the key, these works require a separate rescue policy or a recovery credential committed in advance; our seed proof is such a rescue for every account on a five-level BIP-44 path. Kiraz and Kardas's capacity scenarios divide an assumed quarter of the gas limit by one enrollment's gas, about 50 days per million accounts, and Pont et al.~\cite{Pont24} count at least 76 days of full blocks for Bitcoin's outputs; we measure what the fee market leaves the migration under mainnet's demand (\cref{sec:eval-audit}).\else Ray et al.~\cite{RGR26} prove knowledge of the private key behind a hash-committed address, for keys never exposed. Once $A$ can compute the key, these works require a separate rescue policy or a recovery credential committed in advance; our seed proof is such a rescue for every account on a five-level BIP-44 path. Kiraz and Kardas assume a quarter of the gas limit, and Pont et al.~\cite{Pont24} full Bitcoin blocks; we measure what the fee market leaves the migration under mainnet's demand (\cref{sec:eval-audit}).\fi

\emph{Validator keys.} EIP-8450~\cite{EIP8450}, a concurrent draft, lets each validator register a dormant leanXMSS key~\cite{DKKW25} under its BLS key, which can also replace the registered key, and requires that the switch, or a fork that stops key changes, come while BLS is believed secure. We specify the switch and admit an $A$ that may already hold BLS keys, through registration under BLS before a checkpoint, the conflict rule, a capacity bound measured in the new set, and the request rule (\cref{sec:n1}).

\section{Conclusion}\label{sec:conclusion}
Our protocol keeps every ECDSA-signed enrollment correctable by a seed proof until $T + W$ and moves validators to leanXMSS keys first, under a stated capacity bound. An owner without a seed is bound for good by an enrollment final before $Q_0$. An account that no one enrolled and whose owner never returns stays frozen, as unspendable as one whose key is lost today. The protocol migrates no account on its owner's behalf and neither burns nor moves its balance. Its seed proof, a recursive hash-based STARK of about 190\,KB, verifies in go-ethereum in under 40\,ms. \ifextended The protocol thus lets Ethereum retire ECDSA on a date fixed in advance, without knowing $Q_0$, and shows that lower gas per account, not faster proving, shortens the migration.\else The protocol thus lets Ethereum retire ECDSA on a date fixed in advance, without knowing $Q_0$. Block space, not proving, limits the migration, so lower gas per account shortens it.\fi

\FloatBarrier
\bibliographystyle{IEEEtranLink}
\bibliography{refs}

\appendices
\crefalias{section}{appendix}
\raggedbottom

\section{Public Input Layout}\label{app:pi}
A node recomputes the public input of a \texttt{0x7e} from the transaction's fields and the chain (\cref{tab:pi}). Integers are big-endian with fixed widths, the domain tag is $\mathsf{keccak256}(\mathit{tag} \,\|\, \mathit{chain\_id})$ for a fixed ASCII version string $\mathit{tag}$, and the two fee fields are zero in the modes that do not use them.

\ifextended
\emph{Leaf AIR.} The leaf proof $\mathsf{L}$ arithmetizes the relation as an algebraic intermediate representation (AIR) of four tables over the KoalaBear field ($p = 2^{31} - 2^{24} + 1$), with a degree-8 extension, joined by lookup arguments: a table of 256-bit field operations that runs a fixed, committed program for the curve and the key additions ($2^{14}$ rows, 296 columns), a SHA-512 table ($2^{11}$ rows), a Keccak table ($2^{8}$ rows) and a byte-pair lookup table ($2^{16}$ rows).
\fi

\ifextended\begin{table}[t]\else\begin{table}[b]\fi
\caption{The 198-byte public input of a migration.}
\label{tab:pi}
\centering
\footnotesize
\begin{tabular}{@{}lrrl@{}}
\toprule
Field & Offset & Bytes & Content \\
\midrule
chain id & 0 & 32 & the chain \\
domain tag & 32 & 32 & constant of the chain \\
account & 64 & 20 & the address being migrated \\
migration nonce & 84 & 8 & must equal the registry's \\
algorithm & 92 & 1 & post-quantum scheme \\
key commitment & 93 & 32 & $\mathsf{keccak256}(\mathit{alg} \,\|\, \mathit{pk})$ \\
verifier version & 125 & 4 & proof format and parameters \\
expiry & 129 & 8 & last valid block timestamp \\
fee cap & 137 & 32 & maximum charge to the account \\
fee mode & 169 & 1 & 0--3 (\cref{sec:fees}) \\
named sponsor & 170 & 20 & mode 1 only \\
fee deadline & 190 & 8 & modes 1 and 3 only \\
\bottomrule
\end{tabular}
\end{table}

\begin{table}[t]
\caption{The three proof layers on a 16-core laptop CPU: median proving time over the \simOwnerCount{} proofs of \cref{app:bill} (\gpuLaptopOwnerCount{} with the GPU), and peak memory and mean proof size over ten fresh proofs. Rows: height of the tallest AIR table.}
\label{tab:layers}
\centering
\footnotesize
\setlength{\tabcolsep}{3.5pt}
\begin{tabular}{@{}llrrrrr@{}}
\toprule
Layer & Proves & Rows & CPU & GPU & Memory & Size \\
\midrule
$\mathsf{L}$ & the relation & $2^{16}$ & \simProveL\,s & \gpuLaptopProveL\,s & 18.4\,GiB & 1.35\,MB \\
$\mathsf{B}$ & $\mathsf{L}$ verifies & $2^{16}$ & \simProveB\,s & \gpuLaptopProveB\,s & 24.3\,GiB & 382\,KB \\
$\mathsf{F}$ & $\mathsf{B}$ verifies (hiding) & $2^{14}$ & \simProveF\,s & \gpuLaptopProveF\,s & 31.0\,GiB & 189\,KB \\
\bottomrule
\end{tabular}
\end{table}

\ifextended
\begin{table}[t]
\caption{Parameters of the three proof layers; none uses grinding. Soundness is soundcalc's modeled round-by-round IOP figure~\cite{soundcalc}, not a security level for the composed non-interactive proof. The leaf calculation uses its tallest committed height; see \cref{sec:proof-security}.}
\label{tab:params}
\centering\footnotesize
\setlength{\tabcolsep}{4pt}
\begin{tabular}{@{}lrrr@{}}
\toprule
 & $\mathsf{L}$ & $\mathsf{B}$ & $\mathsf{F}$ \\
\midrule
Rows of the tallest table & $2^{16}$ & $2^{16}$ & $2^{14}$ \\
Blowup & $2^{6}$ & $2^{6}$ & $2^{9}$ \\
Queries & 43 & 43 & 29 \\
FRI arities (rounds) & $\le 4$ (10) & $\le 4$ (8) & 16, 8 (2) \\
Transcript rounds & 15 & 13 & 7 \\
\midrule
Modeled IOP soundness (bits) & 128 & 128 & 130 \\
\bottomrule
\end{tabular}
\end{table}
\fi

\section{Model Checking and Adversarial Runs}\label{app:model}\ifextended\else\label{sec:eval-model}\label{sec:eval-runs}\label{sec:eval-n1}\fi
We model the account rules in TLA+~\cite{Lamport02} with ideal signatures and proofs. TLC explores an owner whose key $A$ holds: ECDSA transactions, EIP-7702 delegations, enrollments by both parties, override, frozen recovery, re-commit, activation and rotation, over five blocks ordered by $A$, in up to 1.3\,M states, with and without the seed-proof migration of a LEGACY account before $T$; with it, an owner who never enrolled reaches BY-PROOF before $T$, and without it, never. Apalache~\cite{KKT19} proves the window lemma of \cref{sec:override}, including its publication rule, and an inductive invariant of an account's idealized authority record (\proofsref) for all integer values. Every property of retirement, correctable binding, fee authorization and single application holds. Two models of \cref{sec:n1} add two more results. The conflict rule holds for every inclusion order and cut point (1.96\,M states). Under the request rule, $A$ obtains an exit only from an enrollment that the owner left uncorrected until $T + W$; the property fails without the rule. \ifextended\Cref{tab:model} lists the models and their bounds, and \cref{tab:adv} the adversarial runs of \cref{sec:eval-runs}.\else The extended version lists the models with their bounds and each adversarial run.

\emph{Runs.} The end-to-end and adversarial runs are pass/fail checks, run once each. On the two-node chain, an owner recovers a frozen account, activates the new key and spends; the recovery survives a reorganization and a restart and is finalized, and key rotation and sponsored recoveries in fee modes 1--3 agree on both nodes. On three nodes a recovery crosses a hop in 0.55--0.56\,s, 0.5\,s of it go-ethereum's delay before fetching an announced transaction; on four, eight owners migrate by seed proof before $T$, after which every mempool refuses their ECDSA transactions. Every run in which $A$ holds the exposed key and builds blocks ends as \cref{sec:design} predicts; front-running one owner of a four-owner batch removes only that owner's entry. In the final run of the validator switch (\cref{sec:n1}), with registrations for every case of the conflict rule, the checkpoint includes seven of ten validators, one with a key that $A$ registered, and excludes the conflicts ($f_A = 1/7$); finality passes $B_{\mathrm{ck}}$ 76\,s after $S$ and advances while $A$'s validator is online but silent. After $T$, an excluded validator's withdrawal address is recovered, activated and sends an EIP-7002 exit request, while the Lean chain keeps the checkpoint's validator set; from $T_0$ on, EIP-7002 and EIP-7251 requests from a legacy address, or from an enrolled and activated one until $T + W$, revert. All checks pass in the final run and in a run of the released package, and the conflict-rule and switch checks in six runs with independent keys; the extended version reports each run in full.\fi

\ifextended
\begin{table}[t]
\caption{Model checking with ideal signatures and proofs. In every model $A$ holds the owner's exposed key and orders all transactions. TLC checks each model up to the stated bound, and Apalache proves the window lemma and an invariant of the authority record (\proofsref) for all integers; every property holds. A control must end in a violation (a rule removed or weakened, a property the design does not promise, or a vacuity test); a scenario is a behavior that the model must be able to reach. All 108 jobs end as expected.}
\label{tab:model}
\centering\footnotesize
\setlength{\tabcolsep}{3pt}
\begin{tabular}{@{}p{0.40\columnwidth}p{0.32\columnwidth}r@{}}
\toprule
Model & Bound & States \\
\midrule
\raggedright Account rules (G1, G3), six variants & \raggedright 28 transactions, 5 blocks of 3 & 0.21\,M--1.26\,M \\
\raggedright Fees and single application (G2, G4) & \raggedright 41 transactions, 3 blocks of 3 & 9.76\,M \\
\raggedright Batches, adversarial sponsor & \raggedright 36 transactions, 4 blocks of 3 & 0.21\,M \\
\raggedright Window lemma & \raggedright up to 3 releases & 0.21\,M \\
\raggedright Window lemma, inductive & \raggedright all integers & -- \\
\raggedright Authority record, inductive (\cref{thm:account}) & \raggedright all integers & -- \\
\raggedright Wallet's key epochs & \raggedright 6 writes, any reorganization & 812 \\
\raggedright Checkpoint's conflict rule (G5) & \raggedright 3 validators, 5 registrations & 1.96\,M \\
\raggedright Request rule (G5) & \raggedright times 0 to 7 & 4,468 \\
\midrule
\multicolumn{3}{@{}l@{}}{Controls violated as they must be: 17; scenarios reached: 74} \\
\bottomrule
\end{tabular}
\end{table}

\begin{table}[t]
\caption{Runs in which $A$ holds the exposed key and orders blocks, on two-node chains (\cref{sec:eval-runs}), the last two with real four-owner batches; each ran once.}
\label{tab:adv}
\centering
\footnotesize
\setlength{\tabcolsep}{3pt}
\begin{tabular}{@{}p{0.43\columnwidth}p{0.53\columnwidth}@{}}
\toprule
$A$'s action & Outcome \\
\midrule
Places its transfer and a token approval before the owner's override in one block & Both valid: the transfer and approval stand (\cref{sec:nongoals}); the account ends bound to the owner's key \\
Sends a transaction with the enrolled key after the override & Refused by both mempools, removed by the builder; a block carrying it is rejected at import \\
Signs a \texttt{0x7c} with the enrolled key before the override and releases it after (component test) & Rejected at mempool admission, reinjection, builder and import \\
Drains the balance that would pay the override & The self-paid override fails at its position and is removed; a sponsored override succeeds \\
Sends ECDSA transactions after $T$ & Both nodes reject every block carrying one \\
Places an owner's own seed proof, as a \texttt{0x7e} that it pays for, before the owner's four-owner \texttt{0x7f} & Both valid: the owner migrates to its own key, its batch entry is skipped (stale nonce), the other three migrate \\
Enrolls its own key for an owner, with the owner's exposed key, before the owner's four-owner \texttt{0x7f} & Both valid: the entry is skipped (stale nonce), the other three migrate; in the next block the owner's override binds the owner's key \\
\bottomrule
\end{tabular}
\end{table}
\fi

\section{The Simulated Bill}\label{app:bill}
\begin{table*}[t]
\caption{What the migration adds per layer, per unit and at mainnet scale (60\,M gas limit, 7,200 slots a day): times from the simulated runs on EC2 (\cref{sec:eval-audit}) at the demand of record, today's bids in parentheses; laptop hours on the laptop of \cref{sec:eval}; per-unit L2 and L3 figures from ten nodes on that laptop (20--80\,ms, 50\,Mbit/s), L9 from four (one run).}
\label{tab:audit}
\centering
\footnotesize
\setlength{\tabcolsep}{3pt}
\begin{tabular}{@{}p{0.10\textwidth}p{0.24\textwidth}p{0.29\textwidth}p{0.31\textwidth}@{}}
\toprule
Layer & Migration adds & Measured per unit & At mainnet scale \\
\midrule
\raggedright L0 Wallets, keys & a seed proof per account recovered after~$T$; two leanXMSS keys per validator & \simOwnerProofSeconds\,s (\simProveAll\,s proving), 31\,GiB per proof; \simValKeySeconds\,s, \simValSecretMB\,MB per validator & \simOwnerLaptopHoursBRound{} laptop hours if all 4.45\,M EOAs recover; \simValKeyLaptopHours{} laptop hours, \simValSecretTB\,TB of validator keys \\
\raggedright L1 Transactions & \texttt{0x7d}, \texttt{0x7e}, \texttt{0x7f}, \texttt{0x7c} & \texttt{0x7d}: 175\,k gas, 3.8\,KB; \texttt{0x7e}: 8.8\,M, 191\,KB; \texttt{0x7f}: 3.77\,M, 48\,KB per account; \texttt{0x7c}: 117\,k, 2.5\,KB (393\,k to activate) & \simEnrollMigShare\,\% of the target (\simEnrollMigShareUpper\,\%); 4.45\,M EOAs: \simDaysBEnroll{} days before~$T$ (\simDaysBEnrollUpper), \simDaysBBatched--\simDaysBSingle{} after, \simDaysBSplitBatched{} if 1.51\,M active enroll \\
\raggedright L2 Network & gossip of 191\,KB recoveries & 1.75\,s per hop under load (0.55\,s idle, \cref{sec:eval-runs}), 3.5\,s to ten nodes; 0.86\,MB per node and recovery & \simNodeSingleAddedDownGB\,GB a day more per node for single recoveries, \simNodeBatchedAddedDownGB\,GB for batches \\
\raggedright L3 Mempool & a proof check before admission & 42--44\,ms per recovery; at most 16 pending per sponsor & at most 16 recoveries per sponsor, or 64 in batches \\
\raggedright L4 Building, import & blocks of six recoveries & up to \simSingleBlockKBMax\,KB per block; \simSingleSixProcessSeconds\,s to process, \simSingleSixImportDelay\,s later import (medians) & six fill \simSingleSixFill\,\% of the 60\,M limit; no slot missed (phase guard, \cref{app:bill}) \\
\raggedright L5 Execution & the verification charge & 5.67\,M gas per proof, 11.77\,M per aggregate & EIP-7976's proposed calldata floor of 64 gas a byte~\cite{EIP7976}: 12.3\,M a recovery, four per block \\
\raggedright L6 State, history & two registry slots, the key as code; proofs in bodies & 1.4\,KB of state per account; 191\,KB of history per recovery & 6.4\,GB of state for 4.45\,M EOAs; 231--868\,GB of history after~$T$ \\
\raggedright L7 Consensus & registration, leanXMSS votes & 46\,k gas per registration; 2.7\,KB per vote, 188\,KB per aggregate & \simValRegHours\,h (\simValRegHoursUpper) for 814\,k validators; 8{,}705 with BLS withdrawal keys (278\,k ETH); 61\,\% of stake in 14{,}115 exposed EOAs \\
\raggedright L8 Data availability & recoveries in payloads & up to \simSingleBlockKBMax\,KB per block against mainnet's 155\,KB median & up to \simBatchedBlockKBMax--\simSingleBlockKBMax\,KB per block for \simDaysBBatched--\simDaysBSingle{} days after~$T$ \\
\raggedright L9 RPC, tools & new types, account queries & 0.53\,s to submit, 0.44\,s to fetch (382, 388\,KB JSON); 68\,ms smallest call; 4.1\,ms to build a sponsored recovery & about a second of wallet time per recovery, off the critical path \\
\bottomrule
\end{tabular}
\end{table*}

\Cref{tab:sim} gives every path at both demands, \cref{fig:uptake} the days against the share that enrolls before~$T$, and \extref{tab:backlog} an enrollment's wait at other numbers of enrollments pending at once; the authors provide the runs' evidence and the tables' generators on request. \ifextended \Cref{fig:layers} puts the times of \cref{tab:audit} on one axis. Each machine's work for one recovery takes under two minutes, at most the \simOwnerProofSeconds\,s of the seed proof; everything longer is set by block space: in a mass enrollment, an enrollment waits a median of \simEnrollWaitMedianMin{} minutes in the mempool, ten times as long as a proof takes, and the 4.45\,M EOAs need \simDaysBEnroll{} to \simDaysBSingle{} days. A wallet's own calls to its node add about a second per recovery, most of it the link carrying the 191\,KB transaction as 382\,KB of JSON.\else Each machine's work for one recovery takes under two minutes, at most the \simOwnerProofSeconds\,s of the seed proof, and everything longer is set by block space; the extended version puts the times of \cref{tab:audit} on one axis.\fi{}

\ifextended
\begin{figure}[t]
\centering
\includegraphics[width=\columnwidth]{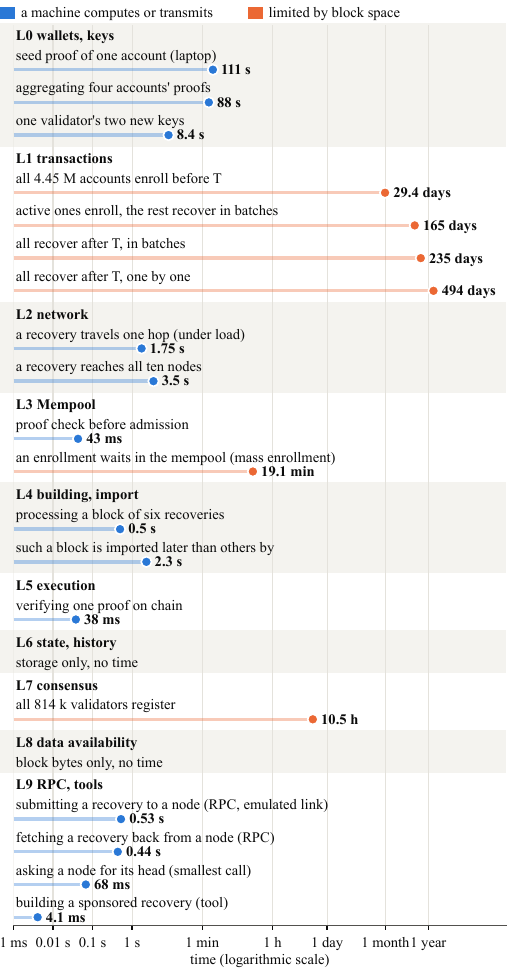}
\caption{Time per layer of \cref{tab:audit}, on a logarithmic axis. Blue: what a machine computes or transmits for one account, one transaction or one block. Orange: what block space limits at the demand of record: an enrollment's wait in the mempool during the simulated mass enrollment, the calendar time for the 4.45\,M EOAs with at least 0.1\,ETH, and the registration of all validators.}
\label{fig:layers}
\end{figure}
\fi

\emph{Machines.} Each run of record puts its ten nodes on one Amazon EC2 m7i.4xlarge instance (Intel Xeon Platinum 8488C, 16 vCPU, 64\,GiB)\ifextended\else, each in its own network namespace on a random 4-regular graph (20--80\,ms latency, 50\,Mbit/s links)\fi, one run per instance at a time; the recovery runs submit seed proofs and batches that the laptop of \cref{sec:eval} proved beforehand, and the validators' run, whose registration gas alone enters the bill, ran on the laptop. \ifextended A node's CPU per slot depends on the machine, so \extref{tab:sim} sets each run's node
load against the run without migration on the same instance type and names the runs that shared its instance: Ream's CPU per slot also differs between instances of that type, by more than the migration adds. \fi The laptop's proof campaign proved 1,000 owners and 250 batches; the medians of \cref{sec:eval-proof,sec:eval-gas} leave out four owners and one batch whose proving overlapped other work, and the first twelve batches, which a portable build proved.

\emph{Rate.} A path's accounts per slot are the migration gas that its steady state's blocks include per slot, a missed slot counting as an empty block, over one account's gas.\ifextended\else{} The steady state runs from the block where the base fee reaches 0.8\,gwei to the one where 90\,\% of the accounts are done, and the ordinary demand draws from the mainnet blocks 26,345,047 to 26,345,076 (8,381 transactions).\fi{} In the steady state of the enrollment of record, blocks sealed \simEnrollSealedPerSlot{} enrollments (\simEnrollLittleSealedPerSlot{} while the driver fed the queue, \cref{sec:eval-audit}) and \simEnrollPerSlotCounted{} activations per slot, which, weighted by their gas, 175\,k and 393\,k, give \simEnrollPerSlot{} accounts per slot, since the steady state's activations include accounts enrolled before it. A recovery run's 48 sponsors each keep one batch, or up to twelve single recoveries, pending, below the mempool's limit of 16 per sponsor (\cref{tab:audit}). The batched recovery of record pools the steady states of \simBatchedRuns{} runs with other random draws, which migrated \simBatchedPerSlotMin{} to \simBatchedPerSlotMax{} accounts per slot, so its time on mainnet is the mean of theirs; \simBatchedRoomShare\,\% of their blocks at or below 1\,gwei ended with room for another batch (\simEnrollRoomShare\,\% for another enrollment in the enrollment of record). \ifextended With 12
sponsors, \simSponsorCheckBatchedRoomTwelve\,\% of these blocks did, and the batched recovery migrated
\simSponsorCheckBatchedTwelve{} accounts per slot against \simSponsorCheckBatchedFortyEight{} with 48 (single
recoveries at today's bids: \simSponsorCheckSingleTwelve{} against \simSponsorCheckSingleFortyEight). More sponsors
leave fewer of these blocks with room, but the migration still gets what ordinary users leave of the target (\cref{sec:eval-audit}): with
12 sponsors the batched recovery's rate lies within the range of its runs with 48.\else With 12 sponsors instead of 48, the batched recovery's rate stays within the range of its runs with 48.\fi{}

\emph{Machines compared.} With the same seed and 12 sponsors, the batched recovery at the demand of record migrates \simMachineCheckRecord{} accounts per slot on EC2 and \simMachineCheckOther{} on the laptop, \simMachineCheckGap\,\% apart: rates, and so times on mainnet, carry over between machines; node CPU, propagation and import times and the mempool's
proof checks do not, and \cref{tab:audit,tab:sim} give each with its machine.

\ifextended
\emph{Phase guard.} Our port of the Ream client cancels a proposal whenever another message waits in its queue, and
its status exchanges with peers, on a 4\,s timer whose phase within the slot is set at start-up, supply such
messages. Where they fall in the first two seconds of a slot, a proposal carrying recoveries slips past its slot and
the lateness cascades: on EC2, where the nodes started more slowly, a single-recovery run without this control missed 33
of 410 slots. Every recovery run and the enrollment at today's bids therefore start each node so that its timer
falls 2.2--2.8\,s into the slot, restarting it before genesis if needed, and record every node's phase. The
enrollment of record, its replication and the run without migration started without this control and missed no
slot; only in the enrollment of record were proposals deferred, and these still went out within their slot. The
laptop's batched recovery had its ten nodes in the band; the validators' run, at 2.8--3.2\,s, missed none of its ten
slots.
\else
\emph{Phase guard.} Our port of the Ream client cancels a proposal whenever another message waits in its queue, so every recovery run and the enrollment at today's bids start each node with its 4\,s status timer 2.2--2.8\,s into the slot (details in the extended version).
\fi

\ifextended
\emph{Sensitivity.} \Cref{tab:sens} changes one parameter of the bill at a time. The demand of February 2025 draws from the 30 blocks 21,848,090 to 21,848,119: of the months from February to July 2025, February's median base fee, 0.76\,gwei, came closest to the migrations' cap. The enrollment runs repeat the enrollment of record, with its 10,000 accounts and random draws, on the laptop; the recovery runs migrate 500 owners of the campaign and are set against a control with the settings of record, which takes \sensDaysControl{} days in batches (\sensDaysRecordSeedOne{} for the run of record with the same draws). In every run the blocks average the target and the base fee of the steady state stays near the migrations' bid cap, so the migration gets what ordinary users leave at that fee. Users who re-price always send, so their share bounds the migration's: in enrollment it takes \sensShareNone\,\% of the target if none re-price, \sensShareHalfLo{} to \sensShareHalfHi\,\% if half do and \sensShareThreeQ\,\% if three quarters do. Migration bids of at most 0.5 or 2\,gwei move the base fee with them and leave the migration \sensShareCapLo{} to \sensShareCapHi\,\%. Verification priced at the lowest or the highest rate of \cref{sec:eval-gas} changes the gas of a recovery but not the migration's share; at the highest, the client rejects every batch, whose verification exceeds the $2^{24}$ gas of one transaction, and owners recover one by one in \sensDaysSingleHighRate{} days. Every node of these runs started with the phase guard, and no run missed a slot.

\input{generated/sens_table}
\fi

\ifextended
\input{generated/backlog_table}
\fi

\section{Proving with a GPU}\label{app:gpu}
The GPU path computes the coset low-degree extensions and the hiding Merkle commitments of all five provers (the three layers and the aggregation's two) on the GPU and leaves the rest on the CPU. For the same salts its proofs are bit-identical to the CPU path's, so the verifiers accept them unchanged; the laptop's verifier builds, the client's Rust library among them, accepted all \gpuReverifiedOwnerProofs{} owner proofs and \gpuReverifiedCarriers{} batch proofs of the A100 machines. Without the GPU, the hiding Merkle commitments take half to two thirds of every proof step, 49 to 66\,\% on the three machines \ifextended of \cref{tab:gpu}\else measured\fi. With the GPU they run 1.9 to 8.4 times faster and a step 1.3 to 2.5 times faster, and no stage then takes more than 42\,\% of a step; on the A100 machine with 20.4 CPUs, the stages left on the CPU take as long on both paths (\cref{fig:stages}). \Extref{tab:gpu} gives the throughput and the number of owners proved in each configuration. \ifextended On the laptop's 16\,GB GPU, a pipeline peaks at \gpuLaptopDeviceOwnerGiB\,GiB of device memory for an owner and \gpuLaptopDeviceBatchGiB\,GiB for a batch, with \gpuLaptopHostGiB\,GiB of host memory. One GPU serves several pipelines: with three on the machine with 13.6 CPUs it is busy 27\,\% of the time, so CPU cores and host memory, not the GPU, limit a machine. \fi For the 4.45\,M EOAs with at least 0.1\,ETH, the cheapest configuration, \gpuBestPipelines{} pipelines on the A100 with \gpuBestCpus{} CPUs, needs \gpuMachineHoursBRound{} machine-hours, \$\gpuDollarsBk\,k (the cheapest configuration\ifextended{} of \cref{tab:gpu}\fi{} without the GPU, on the A100 with 20.4 CPUs: \$\gpuCpuOnlyDollarsBk\,k), and \gpuMachinesToKeepUp{} such machines keep pace with the chain's batched recovery; a sponsor that only aggregates needs \gpuSponsorGpusToKeepUp{} A100s. The GPU shortens the proving, not the days, which block space sets.

\begin{figure}[t]
\centering
\includegraphics[width=\columnwidth]{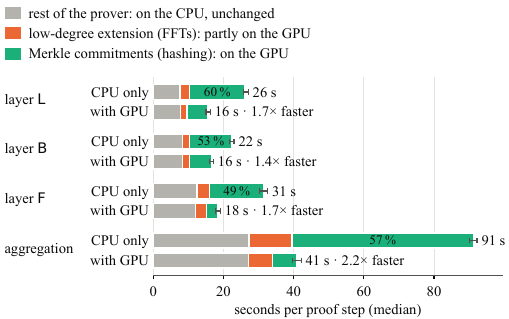}
\caption{Proving time on the A100 machine with 20.4 CPUs, one pipeline, for the three proof layers and the aggregation of four owners' proofs, without and with the GPU. Gray is the rest of the prover (trace generation, constraint evaluation, opening), which stays on the CPU. Each bar ends in its median and, with the GPU, the step's speed-up; inside the CPU-only bars stands the Merkle commitments' share; whiskers span the step's 10th to 90th percentile. Without the GPU 8 owner proofs and 2 batches, with it 50 and 12.}
\label{fig:stages}
\end{figure}

\ifextended
\input{generated/bill_gpu_table}
\fi

\section{What a Quantum Computer Breaks}\label{app:surface}
\Cref{tab:surface} lists the components of Ethereum's base layer, wallets and contracts that rest on a problem that Shor's algorithm solves and, for contrast, the hash functions.

\ifextended\else\looseness=-1 \fi\emph{Contracts that hold stake.} Of the stake that withdraws to contracts (\cref{fig:exposure}), 78\,\% sits in upgradeable contracts (57\,\% in Lido's withdrawal vault alone) and 10\,\% in immutable or owner-governed contracts with a caller- or role-based path, mostly Safe and classic Gnosis multisig wallets. At least 4.8\,\% (4.9\,\% with one probable case; 0.81\,M ETH, 1.8\,\% of all stake, chiefly BitGo-type 2-of-3 wallets~\cite{CustodyBitstampWalletSimple}) sits in contracts that can neither be upgraded nor move funds without an \texttt{ecrecover} signature. We classified the 200 largest addresses by their bytecode; they hold 93\,\% of that stake.

\begin{table}[t]
\caption{What a CRQC breaks in Ethereum's base layer, wallets and contracts, and the goals and rules that each component concerns.}
\label{tab:surface}
\centering
\footnotesize
\setlength{\tabcolsep}{2.5pt}
\begin{tabular}{@{}p{0.36\columnwidth}p{0.27\columnwidth}p{0.33\columnwidth}@{}}
\toprule
\raggedright Component (primitive) & \raggedright A CRQC can & \raggedright Goals; in our protocol \tabularnewline
\midrule
\raggedright Transactions, EIP-7702 tuples, \texttt{ecrecover} (ECDSA) & \raggedright sign for every exposed key & \raggedright G1--G6; R1--R3, seed proof, \cref{sec:n1-late} \tabularnewline
\raggedright Contracts' own secp256k1 or RSA (\texttt{MODEXP}) checks, P-256 precompile~\cite{EIP7951} & \raggedright forge their authorizations & \raggedright outside G1, but withdrawal contracts touch G5; operators replace them \tabularnewline
\raggedright Validator keys and duties, RANDAO included; sync committees (BLS) & \raggedright forge duties and aggregate slashing evidence; attack finality and light clients & \raggedright G3, G5; switch at $T_0$; A5--A7 \tabularnewline
\raggedright BLS withdrawal credentials (\texttt{0x00}) & \raggedright forge a change once the public key is exposed & \raggedright G5; change finalized by $D_{\mathrm{reg}}$, safe only before $Q_0$ \tabularnewline
\raggedright Deposit signatures (BLS)~\cite{ConsensusSpecs} & \raggedright front-run a new validator's deposit & \raggedright G5; safe only before $Q_0$ \tabularnewline
\raggedright BIP-32 derivation (secp256k1) & \raggedright derive every descendant key from $(K_\ell, c_\ell)$ & \raggedright none beyond \cref{def:auth}: the seed stays secret \tabularnewline
\raggedright KZG blob commitments and proofs~\cite{EIP4844} & \raggedright open them to other data & \raggedright no direct use; blob availability outside our goals; A7, A8 \tabularnewline
\raggedright Pairing precompiles~\cite{EIP2537,EIP197} & \raggedright forge pairing-based proofs and BLS signatures & \raggedright as contracts' own checks \tabularnewline
\raggedright Node keys: RLPx~\cite{RLPx}, discv5, libp2p (secp256k1, X25519) & \raggedright impersonate peers, read their traffic & \raggedright via A6--A8 only; outside our scope \tabularnewline
\raggedright Hash functions & \raggedright generic speed-ups & \raggedright G2--G6 rest on A2--A4 \tabularnewline
\bottomrule
\end{tabular}
\end{table}

\ifextended
\section{Keys for a Consensus Threshold}\label{app:keys}
We count keys in the finalized beacon state of slot 15,575,968 (3~November 2026, 19:53~UTC) of \cref{sec:eval-audit}, which a public checkpoint-sync server provided. Its hash tree root, which we recomputed from the file, equals the state root in the slot's header on two other providers' beacon nodes. A validator counts if it is active at the state's epoch: its activation epoch has passed and its exit epoch has not. These 814,041 validators, the 814,177 of \cref{sec:eval-audit} without the 136 that await activation, hold 44,238,469\,ETH of effective balance, the weight to which the finality thresholds refer~\cite{BG17,ConsensusSpecs}. Sorted by effective balance, largest first, the shortest prefix with more than a third of this weight has 7,647 validators and 14,747,223\,ETH. Each of them holds at least 1,598\,ETH and has compounding (\texttt{0x02}) withdrawal credentials, which let a validator hold up to 2,048\,ETH~\cite{EIP7251}. No smaller set of keys holds more than a third of the stake at this slot. The count measures how concentrated the stake is, not $A$'s effort: one discrete logarithm of their aggregate public key enables common-message aggregate forgeries (\cref{sec:n1-bound}). In validators of 32\,ETH, more than a third of the stake would take 460,818 keys. More than half of the weight takes 122,795 keys, and two thirds take 353,204. We estimate serial recovery by multiplying the key count by the published time for one 256-bit secp256k1 key: 9 or 12~minutes from a precomputed state and 18 or 23~minutes for a whole run on a superconducting machine~\cite{BZG26}, 10~minutes per key with non-local connections~\cite{Litinski23}, and 25.7~days on a trapped-ion machine~\cite{HTY26}. The 7,647 keys thus take one superconducting machine 48 to 122 days, and a trapped-ion one 538 years without counting its success probability of 63\,\%. Following~\cite{BZG26}, these illustrative serial estimates use secp256k1 timings as a proxy for BLS12-381. The authors provide the script and its aggregate output on request.
\fi

\ifextended
\section{Security Proofs}\label{app:proofs}

\emph{Notions.} The seed proof is a non-interactive argument in the random-oracle model for the relation of \cref{sec:relation}, $R = \{(x, (s, p)) : \mathsf{Der}(s, p) \text{ is the account of } x\}$; the transcript binds the other fields of $x$. A \emph{carrier} is a \texttt{0x7e} with its public input or a \texttt{0x7f} with its four; it is \emph{accepted} if it passes the node's proof verification, for a \texttt{0x7f} with the checks of \cref{sec:aggregation}, whatever the chain's state. Adversaries are quantum: they query the oracle $H$ in superposition and request proofs classically. Each $\varepsilon$ is an advantage at the resources of the adversary at hand; for $A$, these include the simulator and the extractor. \emph{Zero knowledge:} a polynomial-time simulator answers the oracle queries and outputs proofs $\mathsf{S}(x)$ without a witness, and no adversary that adaptively chooses pairs $(x, w) \in R$ distinguishes the prover's proofs under $H$ from the simulator's with advantage above $\varepsilon_{\mathrm{ZK}}$. \emph{Weak simulation extractability:} there is a polynomial-time extractor $\mathsf{E}$ such that, for every adversary that runs with the simulator, obtains simulated proofs for a set $Q$ of public inputs and outputs a carrier, the following event has probability at most $\varepsilon_{\mathrm{SE}}$: the carrier is accepted and carries some $x \notin Q$ for which $\mathsf{E}$, run with that adversary, outputs no $w$ with $(x, w) \in R$. The notion is weak in that only a new public input counts, not a new proof of an old one. A \texttt{0x7f} may aggregate simulated proofs beside the adversary's own. A4 assumes both notions for the composed proof and the second also for its aggregation (\cref{sec:proof-security}); \cref{prop:zk} proves honest-verifier zero knowledge of the IOP underlying $\mathsf{F}$, a premise of the first.

$\varepsilon_{\mathrm{wit}}$ is the largest probability with which an algorithm with these resources, given $\mathsf{mk}(s)$ and $p^\ast$ as in \cref{def:auth}, outputs a seed $s'$ and a path $p'$ with $\mathsf{Der}(s', p') = \mathsf{Der}(s, p^\ast)$: the owner's seed, or another seed that reaches the same address. Generic search~\cite{Grover96} finds the first with a probability on the order of $\min\{1, q_m^2 2^{-\kappa}\}$, for a mnemonic that encodes $\kappa \ge 128$ uniform bits (A1), and the second with $\min\{1, q_a^2 2^{-160}\}$, where $q_m$ and $q_a$ count the candidates that each search tests; a mnemonic costs more to test, since its test includes PBKDF2. A2 assumes that nothing does better.

\begin{proof}[Proof of \cref{thm:auth}]
Game~0 is $\mathrm{AUTH}$. Game~1 answers the oracle queries and the requests for proofs with the simulator, after the same test that $\mathsf{Der}(s, p)$ is the account of $x$. An adversary that draws $s$ itself, runs $A$ and outputs whether $A$ wins distinguishes real from simulated proofs, so the two games differ by at most $\varepsilon_{\mathrm{ZK}}$. In Game~1, a winning output is an accepted carrier that carries a public input $x^\ast$ for $a^\ast$ outside the set $Q$ of public inputs of which $A$ obtained a proof. By weak simulation extractability, $\mathsf{E}$ then outputs a witness $(s', p')$ of $x^\ast$, so $\mathsf{Der}(s', p') = a^\ast$, except with probability $\varepsilon_{\mathrm{SE}}$. Game~1 uses $s$ only for $\mathsf{mk}(s)$ and for the tests, and $\mathsf{mk}(s)$ suffices for both, since the address of every path follows from it. $A$, the simulator and $\mathsf{E}$ thus form an algorithm that computes $(s', p')$ from $\mathsf{mk}(s)$ and $p^\ast$, which succeeds with probability at most $\varepsilon_{\mathrm{wit}}$. For $N$ target accounts of the wallet, a union bound in the same game gives at most $N$ times the bound.
\end{proof}

\begin{table*}[t]
\caption{The authority transitions of an account: its record $(m, c, n)$ of mode, key commitment and migration nonce, and its installed key $k$ ($\bot$ if none), at time $t$. The account is final if $m = \text{BY-PROOF}$ and $k \ne \bot$, or if $m = \text{ENROLLED}$ and $t \ge T + W$. Checks that only restrict a transition, such as proofs of possession and fee rules, are left out.}
\label{tab:machine}
\centering
\footnotesize
\setlength{\tabcolsep}{4pt}
\begin{tabular}{@{}p{0.25\textwidth}p{0.12\textwidth}p{0.36\textwidth}p{0.21\textwidth}@{}}
\toprule
Transition & Authorized by & Valid if & Effect \\
\midrule
ECDSA use (\texttt{0x00}--\texttt{04}, EIP-7702 tuple, \texttt{ecrecover}) & secp256k1 key & $m = \text{LEGACY}$ and $t < T$ & record unchanged \\
Enrollment of $c'$ (\texttt{0x7d}) & secp256k1 key & $m = \text{LEGACY}$ and $T_0 \le t < T$ & ENROLLED, $c'$, $n + 1$; $k = \bot$ \\
Migration of $x$ (\texttt{0x7e}, entry of a \texttt{0x7f}) & seed proof of $x$ & $t \ge T_0$; the nonce of $x$ is $n$; $x$ has not expired; and $m = \text{LEGACY}$, or $m = \text{ENROLLED}$ and $t < T + W$, or $m = \text{BY-PROOF}$ and $k = \bot$ & BY-PROOF, the commitment of $x$, $n + 1$; $k = \bot$; a charge within the fee cap of $x$ \\
Activation with $\mathit{pk}$ (first \texttt{0x7c}) & signature under $\mathit{pk}$ & $m \ne \text{LEGACY}$, $k = \bot$, and $c$ commits to $\mathit{pk}$ & $k = \mathit{pk}$ \\
Use (\texttt{0x7c}, or a signature that R5 verifies) & signature under $k$ & $k \ne \bot$ & record unchanged \\
Rotation to $\mathit{pk}'$ & signature under $k$ & $k \ne \bot$ and the account is final & the commitment to $\mathit{pk}'$, $n + 1$; $k = \mathit{pk}'$ \\
Time & & & $t$ grows \\
\bottomrule
\end{tabular}
\end{table*}

\begin{proof}[Proof of \cref{thm:account}]
\Cref{tab:machine} abstracts every rule of \cref{sec:design} that writes the authority over an existing account or acts under it. A transaction may take several of its steps, as a first \texttt{0x7c} activates and then acts, and the checks that it leaves out only remove executions. An EIP-7702 delegation changes no condition of the table and is cleared by an enrollment and by a migration. The migration nonce starts at zero and is a 64-bit counter whose increment the node refuses at its largest value, so it never wraps. A native account arises only at the address that the hash of its post-quantum key fixes, and a contract only at an address that a hash of its creator's data fixes. Let $K_O$ be the post-quantum keys of the owner's wallet, $C_O$ their commitments and $\mathrm{Auth}$ the public inputs for $a$ that the owner proved; the wallet proves only commitments in $C_O$, signs only its own transactions and rotates only to keys in $K_O$ (\cref{sec:wallet}).

\emph{Ideal executions.} Call an execution \emph{ideal} if (i) every migration of $a$ applies a public input in $\mathrm{Auth}$, (ii) the owner signed every message that is accepted under a key in $K_O$, (iii) every key installed against a commitment in $C_O$ lies in $K_O$, and (iv) neither a native account nor a contract arises at $a$. Two hops lead to a game in which the probability that $A$'s execution is not ideal is bounded directly. The first replaces the owner's proofs by simulated ones, at a loss of $\varepsilon_{\mathrm{ZK}}$. The second replaces the seeds of the owner's post-quantum keys, which the wallet derives with HMAC-SHA512 keyed with the seed over distinct labels (\cref{sec:wallet}), by uniform ones. Its loss $\varepsilon_{\mathrm{KDF}}$ is the advantage in telling these outputs from uniform strings given $\mathsf{mk}(s)$, in which the seed is the message of the same function; a random oracle in place of HMAC-SHA512 has this property, and A2 assumes it. The resulting game uses the seed only for $\mathsf{mk}(s)$, as Game~1 of \cref{thm:auth} does. In each game, $K_O$, $C_O$ and $\mathrm{Auth}$ are those of that game, and a repeated label yields the same key. In the last game, a migration outside $\mathrm{Auth}$ is an accepted carrier with a public input for $a$ of which no proof was simulated, an event of probability at most $\varepsilon_{\mathrm{SE}} + \varepsilon_{\mathrm{wit}}$ as in that proof; a failure of (ii) is a forgery against ML-DSA-44, whose advantage over all of the owner's keys is $\varepsilon_{\mathrm{sig}}$ (A3); a failure of (iii) is a second preimage of one of the owner's key commitments; and a failure of (iv) is a preimage of the 160-bit address $a$ under the hash that fixes the address of a native account or of a contract. An execution is therefore ideal except with probability $\varepsilon_{\mathrm{auth}} + \varepsilon_{\mathrm{sig}} + \varepsilon_{\mathrm{hash}}$, where $\varepsilon_{\mathrm{hash}}$ sums $\varepsilon_{\mathrm{KDF}}$ and the advantages against the two preimage properties (A2). In an ideal execution, $A$ uses the secp256k1 key freely, migrates $a$ only with public inputs in $\mathrm{Auth}$, and activates, uses and rotates only under keys outside $K_O$, unless it forwards a transaction of the owner.

\emph{Invariants.} Every state of an ideal execution satisfies (I1): $m = \text{BY-PROOF}$ implies $c \in C_O$; and (I2): $k \ne \bot$ implies that $c$ commits to $k$ and that $m \ne \text{LEGACY}$. Both hold at the start, where $m = \text{LEGACY}$ and $k = \bot$. A migration sets $c$ to a commitment in $C_O$, by (i), and $k$ to $\bot$. An enrollment leads to mode ENROLLED with $k = \bot$. An activation changes only $k$, to a key to which $c$ commits, in a mode other than LEGACY. A rotation installs $\mathit{pk}'$ with its commitment; in mode BY-PROOF, I1, I2 and (iii) put the signing key $k$ in $K_O$, so by (ii) the owner signed, and it rotates to a key in $K_O$. The other transitions change none of $m$, $c$ and $k$.

\emph{Steps.} By inspection of \cref{tab:machine}, every transition from $(m, c, n, k)$ at $t$ to $(m', c', n', k')$ at $t'$ satisfies the following.
(P1)~$m \ne \text{LEGACY}$ implies $m' \ne \text{LEGACY}$, $m = \text{BY-PROOF}$ implies $m' = \text{BY-PROOF}$, and $t' \ge t$.
(P2)~$n' = n + 1$ for an enrollment, a migration and a rotation, and $n' = n$ otherwise.
(P3)~If $m = \text{ENROLLED}$ and $t < T + W$, the transition is a migration, an activation, a use or time, since the others need mode LEGACY or a final account; an activation and a use leave $(m, c, n)$ unchanged.
(P4)~If $m = \text{LEGACY}$ and $t \ge T$, the transition is a migration or time, since the first two need $t < T$ and the others a mode other than LEGACY or, with I2, an installed key.

\emph{The four properties.} (a)~Only the first two transitions of \cref{tab:machine} act under the secp256k1 key. Both need $m = \text{LEGACY}$ and $t < T$, a condition that, once false, stays false by P1; rules R1--R3 enforce it at the three places of \cref{sec:retire}. (b)~A migration applies $x \in \mathrm{Auth}$ only while the nonce of $x$ is $n$, and it sets $n + 1$. By P2 the nonce never decreases, so $x$ applies at most once. The charge that the proof authorizes is that of R4 (\cref{sec:fees}): unsponsored, the gas limit times the maximum price, and sponsored, the reimbursement, each within the fee cap of $x$; the gas of a carrier is charged to the sponsor that signed it. In mode BY-PROOF, which no transition leaves (P1), $c \in C_O$ by I1, and an installed key is one to which $c$ commits by I2, so it lies in $K_O$ by (iii). (c)~While $m = \text{ENROLLED}$ and $t < T + W$, the migration of every $x \in \mathrm{Auth}$ with nonce $n$ passes its gate, and by P3 the transitions under the enrolled key leave $(m, c, n)$ unchanged. An override is a migration: it leads to mode BY-PROOF with a commitment in $C_O$ and $k = \bot$, and (b) holds from then on; the commitment then changes only by another migration in $\mathrm{Auth}$ or, once a key is installed, by a rotation that this key of the owner signs. An enrollment needs $t < T$, so more than $W$ remains until $T + W$. P4 is the statement on frozen accounts. (d)~A \texttt{0x7e} that is not valid at its position makes its block invalid, and an entry of a \texttt{0x7f} that is not valid is skipped without a write to its account or a charge of its own, while the other entries apply and the sponsor pays the carrier's gas (\cref{sec:tx}). A recovery is therefore the migration of \cref{tab:machine} with the charge of (b), or it makes no write to the record of $a$ and no charge of its own. By P2 the nonce changes only by an enrollment, a migration or a rotation; a rotation needs a final account, a migration a public input in $\mathrm{Auth}$, and an enrollment the mode LEGACY, to which no account returns (P1), and $t < T$.

A reorganization replaces the end of the chain by other valid blocks, so every chain is an execution, and the properties hold on each. \Cref{tab:machine} fixes $T$ and $W$; for releases that move $T$ or change $W$, the window lemma (\cref{sec:override}) keeps every deadline and the freeze.
For the authority record of an ideal execution, Apalache checks that a strengthened invariant, which contains I1 and I2, holds at the start and is kept by every step. It also checks P1--P4 on every step and that the nonces of the applied public inputs strictly increase, for unbounded integer time and nonce and all integer parameters with $T_0 < T$ and $W \ge 1$. This model keeps of a key and of a commitment only whether they are the owner's; it has no expiry and no batch, and it leaves fees to the TLC model (\cref{app:model}).
\end{proof}

\begin{proof}[Proof of \cref{cor:recovery}]
Let the owner send a migration $x$ whose nonce is $n$, and let a sponsor keep its other conditions valid for $\Delta$, so that $x$ fails only if the record of the account refuses it. If another migration applies first, the owner proved it, by (b), and that migration binds the account to a key of the owner; A8 bounds its finality from its own sending. Otherwise, for a frozen account, P4 leaves no other transition that changes the record, so $x$ stays valid until it is applied, and by A8 it is finalized within $\Delta$. For an enrolled account, P3 gives the same as long as $t < T + W$. For a LEGACY account before $T$, the only other transition before $x$ that changes the record is an enrollment (P2). It leaves the account enrolled, and a second proof, at the new nonce, succeeds as in the second case. If $T$ comes first, the account is frozen and $x$ still passes its gate.
\end{proof}

\begin{proof}[Proof of \cref{prop:noseed}]
(a)~$A$ knows the secp256k1 key of an account only from $Q_0$ on (\cref{sec:adversary}), so whatever that key signed in a block that is final before $Q_0$, the owner signed. Let $K_O$ be the post-quantum keys that the owner generates and $C_O$ their commitments; the owner signs only its own transactions and rotates only to keys in $K_O$. It has no witness and proves nothing, so $\mathrm{Auth}$ is empty. Let an enrollment by the owner be final. It sets a commitment in $C_O$, the mode ENROLLED and an account nonce above zero, and from then on no transition acts under the secp256k1 key (P1). The reduction of the proof of \cref{thm:account} applies without its hop for the key derivation, since the keys in $K_O$ come from fresh randomness. The seed proofs that $A$ sees are those of other accounts; replacing them by simulated ones costs $\varepsilon_{\mathrm{ZK}}$. An accepted carrier with a public input for $a$ then yields, by weak simulation extractability, a seed and a path that derive $a$, except with probability $\varepsilon_{\mathrm{SE}}$, and $A$ with the simulator and the extractor computes them with probability at most $\varepsilon_{\mathrm{wit}}$, here the probability that $A$, with all it knows of $a$, finds such a seed and path; up to these terms, no carrier for $a$ is accepted and no migration of $a$ applies, which is condition (i). Conditions (ii) and (iii) fail only by a forged signature or by a second preimage of a commitment in $C_O$, the only parts of $\varepsilon_{\mathrm{sig}}$ and $\varepsilon_{\mathrm{hash}}$ that are needed. Condition (iv) cannot fail: a native account is created only in mode LEGACY, and a contract only at an address whose account nonce is zero. In an ideal execution, $m = \text{ENROLLED}$ and $c \in C_O$ hold from the enrollment on, and every installed key lies in $K_O$: no migration applies; an activation installs the key to which $c$ commits (iii); and a rotation, possible once the account is final, is signed under an installed key, hence by the owner (ii), and leads to a key in $K_O$.
(b)~Let two copies of the owner's program generate the keys $K_0$ and $K_1$, and compare two runs from the same chain with the same randomness of the protocol: in run $i$, copy $i$ is the owner, and $A$ runs the other copy as an owner would, which it can because it holds the secp256k1 key and so produces whatever the protocol accepts from the owner. The chain records nothing that tells $K_0$ from $K_1$, and no builder's order of the two copies' messages depends on $i$, so the protocol sees the same in both runs, and its outcome has the same distribution. Let $S_i$ be the event that $a$ ends bound to keys of copy $i$ alone. $S_0$ and $S_1$ exclude each other, so their probabilities sum to at most $1$: in one of the two runs, $a$ ends bound to the owner's keys alone with probability at most $1/2$, where staying unbound and being bound to keys of both copies count as failures. A record of the owner's key that is on the chain but not final when $A$ obtains the secp256k1 key protects the owner only if no reorganization drops it.
\end{proof}

\begin{proof}[Proof of \cref{prop:minimal}]
(a)~If $A$ can reproduce every owner authentication that the protocol accepts, given all records final before $Q_0$, the two runs of the proof of \cref{prop:noseed}(b) apply, and in one of them the account ends bound to keys of the owner alone with probability at most $1/2$. The protocol must therefore check a secret that $A$ cannot compute. It tells the owner from $A$ only by the account's own records and wallet secrets, and no record final before $Q_0$ ties the account to a secret from which it does not derive; a check of such a secret, $A$ passes with a secret of its own, and the two runs apply again. The secret must thus be one from which the account derives. The seed is one even if $A$ holds every BIP-32 key of the wallet (A1, A2), though not the only one: the inner digest of the HMAC that yields the master key is another. G2 requires that the check reveal nothing about the seed: knowing it, $A$ could prove for every other account derived from it and compute the wallet's post-quantum keys (\cref{sec:wallet}). (b)~A key that replaces BLS counts only through a registration that the old chain finalizes, and the corrections of \cref{sec:design} count only once finalized, on the old chain or on a chain that the new set runs. Suppose that the old chain has not finalized the registrations that determine the new set when $A$ can suppress its finality until the corrections' deadlines pass. In an execution in which $A$ does so, no new set is authoritative and the old chain finalizes nothing, so no pending correction is finalized by its deadline; this uses neither A5 nor A8. $A$ gains this power, for instance, by forged equivocations: once the validators that it has had slashed hold more than a third of the active stake, the votes that count, those of unslashed validators, fall short of two thirds of the active stake, which still includes the slashed validators until they exit or the inactivity leak drains them~\cite{ConsensusSpecs}; since $A$ may break every BLS key, it can slash more as they drain.
\end{proof}

\emph{Validator transition (G5).} Let $V$ be the validator set that the checkpoint tool computes at $B_{\mathrm{ck}}$.

\begin{proposition}[Validator transition]\label{prop:n1}
(a)~For every order of the registrations and every position of $B_{\mathrm{ck}}$, a validator whose owner's valid registration is included up to $B_{\mathrm{ck}}$ enters $V$ with the owner's keys or is excluded, and a validator enters $V$ with other keys only if $A$ has broken its BLS key and no valid registration of its owner is included up to $B_{\mathrm{ck}}$.
(b)~Under A2, A3 and A5--A7, and with the owners' leanXMSS keys generated from fresh randomness, $A$ holds less than a third of the weight of $V$ at $S$, and the Lean chain that starts from $V$ at $B_{\mathrm{ck}}$ is safe and live while $A$'s share stays below a third.
(c)~In an ideal execution, from $T_0$ on a request that its owner did not authorize is accepted from the withdrawal address of a protected owner only through an enrollment of $A$'s that the owner left uncorrected until $T + W$.
\end{proposition}

\begin{proof}
(a)~Let $D_v$ be the set of distinct tuples of new keys in the valid registrations for validator $v$ up to $B_{\mathrm{ck}}$. For a validator that meets its other conditions, such as withdrawal credentials that name an execution address (\cref{sec:n1-late}), the checkpoint tool includes $v$ with the tuple of $D_v$ if $|D_v| = 1$, excludes it if $|D_v| \ge 2$ and counts it as missed if $D_v$ is empty (\cref{sec:n1-bound}); it excludes every other validator. The outcome thus depends on $D_v$ alone and not on the order of the registrations. A registration is valid only if $v$'s BLS key signed it, so its signer is the owner or, for a key that $A$ has broken, $A$. If the owner's tuple $\tau_v$ is in $D_v$, then $v$ enters $V$ with $\tau_v$ or is excluded. If $v$ enters $V$ with another tuple, then $\tau_v \notin D_v$, so no valid registration of the owner is included up to $B_{\mathrm{ck}}$, and $A$ signed, so it has broken $v$'s key. (b)~By (a), the validators of $V$ outside the set $U$ of \cref{sec:n1-bound} entered with their owners' keys. A5 bounds the weight of $U$, $f_A < 1/3$, and makes $B_{\mathrm{ck}}$, the last block finalized by $D_{\mathrm{reg}}$, a block of the honest chain. Nodes start the Lean chain from $B_{\mathrm{ck}}$ and $V$ (A6), so BLS keys that $A$ breaks later do not affect it. A broken BLS key does not yield an owner's leanXMSS keys, which come from fresh randomness, so by A2 and A3 only validators of $U$ sign as $A$ chooses. A7 then gives safety and liveness for as long as $A$'s share of the weight stays below a third. (c)~From $T_0$ on, the request contracts accept a request only from an account in mode BY-PROOF, a native account, an enrollment at or after $T + W$, or a contract (\cref{sec:n1-late}). The withdrawal address of a protected owner is an account derived from the seed: in an ideal execution it is neither native nor a contract, and with an EIP-7702 delegation it still counts by its mode. In mode BY-PROOF its installed key is the owner's (\cref{thm:account}), so by (ii) the owner signed the transaction that submits the request. An enrollment of $A$'s reaches $T + W$ only if the owner did not override it while the gate was open, for more than $W$.
\end{proof}

\emph{Zero knowledge of the final layer's IOP.} $\mathsf{F}$ proves with Plonky3's hiding FRI commitment~\cite{Plonky3}, whose masking follows~\cite{HK24}. We consider the interactive oracle proof (IOP) that underlies it: the prover's free random choices are uniform and independent, the verifier's challenges are uniform and independent of them, and the verifier sees opened values instead of commitments. Let $\mathbb{F}$ be the KoalaBear field of $p$ elements, $\mathbb{K}$ its extension of degree 8, $\Omega$ the trace domain of $2^{14}$ rows with generator $\omega$, and $\Lambda$ the coset of $2^{24}$ points over which the prover commits; $\Lambda$ is disjoint from $\Omega$, and $\omega\Lambda = \Lambda$. We write $v_\Omega$ and $v_{\Omega_i}$ for vanishing polynomials. The prover commits to polynomials over $\mathbb{F}$:
(i)~every column $f$ of the trace, and every coordinate of a column of the LogUp lookup argument~\cite{Habock22}, as $\hat f = f + v_\Omega r_f$ with a mask $r_f$ of degree below $2^{14}$;
(ii)~every coordinate $q$ of the quotient by $v_\Omega$ of the folded constraints, evaluated on the masked columns $\hat f$, in eight chunks: $q = \sum_i \ell_i q_i$ over eight cosets $\Omega_i \subset \Lambda$ of $2^{14}$ points, where $\ell_i = c_i \prod_{j \neq i} v_{\Omega_j}$ is one on $\Omega_i$ and $\deg q_i < 2^{14}$, each chunk committed as $\hat q_i = q_i + v_{\Omega_i} b_i$ with masks $b_i$ of degree below $2^{14}$, the eighth fixed by $\sum_i c_i b_i = 0$;
(iii)~a matrix of sixteen random polynomials of degree below $2^{15}$, the first eight of which we call $u_0, \dots, u_7$; and
(iv)~random polynomials whose joint distribution does not depend on the witness: the other eight of (iii), and eight in every committed matrix of columns and in every chunk, those of the chunks masked as in (ii).
The columns of the lookup argument exist unless one of its at most $2^{17}$ denominators vanishes, each a challenge minus a value that the witness and another challenge fix; the prover then stops. Given the out-of-domain point $\zeta \in \mathbb{K}$, the prover opens every committed polynomial at $\zeta$, and at $\omega\zeta$ the columns of (i) that a constraint reads at the next row, possibly with other polynomials of (i) and (iv); those of (ii) and (iii) are opened at $\zeta$ only. Given the batching challenge $\alpha \in \mathbb{K}$, FRI~\cite{BBHR18FRI} tests $\Phi = \sum_{(h, z)} \alpha^{e(h, z)} (h - h(z))/(X - z)$, summed over the committed polynomials $h$ and their opening points $z$, with distinct exponents, $e_u + k$ for $u_k$. The prover sends FRI's layers at the queried positions and its final polynomial, all functions of $\Phi$ and of FRI's challenges, and the values of every committed polynomial at the 29 query points $\lambda_j \in \Lambda$. The lookup argument's terminal is zero, because $\mathsf{F}$ has one table. These values are the \emph{openings}; the other fields of a proof are commitments, authentication data and constants.

\begin{proposition}[Honest-verifier zero knowledge of the IOP underlying $\mathsf{F}$]\label{prop:zk}
For every witness of a public input, the challenges and openings of an honest execution of this IOP are within statistical distance $2^{-122}$ of the output of a polynomial-time simulator that is given the public input only.
\end{proposition}

\begin{proof}
Let $E$ be the event that $\alpha$ or $\zeta$ lies in the subfield of $p^4$ elements, or that a denominator of the lookup argument vanishes for the witness at hand. Its probability is at most $2p^{-4} + 2^{17} p^{-8} < 2^{-122}$. The simulator draws the challenges as the verifier does; if $\alpha$ or $\zeta$ lies in that subfield, it outputs them alone, and otherwise it samples openings as below, whatever the denominators are. An execution in which the prover stops lies in $E$. We show that outside $E$ the openings, given the challenges, have a distribution that does not depend on the witness and that the simulator samples.

Outside $E$, the powers $1, \alpha, \dots, \alpha^7$ are a basis of $\mathbb{K}$ over $\mathbb{F}$. The minimal polynomials $\mu_\zeta$ and $\mu_{\omega\zeta}$ over $\mathbb{F}$ have degree 8 and differ, since $\zeta^{p^j} = \omega\zeta$ implies $\omega^m = 1$ for some $m \le 8$, and $\zeta$ lies neither in $\Omega$ nor in $\Lambda$. A polynomial over $\mathbb{F}$ takes conjugate values at conjugate points, so its residue modulo $\mu_\zeta$ and its value at $\zeta$ determine each other. Let $M = \mu_\zeta \prod_j (X - \lambda_j)$, the product over the distinct $\lambda_j$; $\deg M \le 37$.

\emph{Columns.} For a column, let $M_f$ be $M$, times $\mu_{\omega\zeta}$ if the column is opened at $\omega\zeta$, times the factors $X - \omega\lambda_j$ that $M$ lacks if a constraint reads the column at the next row; $\deg M_f \le 37 + 8 + 29 = 74 \le 2^{14}$. No root of $M_f$ lies in $\Omega$, so $r \mapsto v_\Omega r \bmod M_f$ maps the masks linearly onto the residues modulo $M_f$. The residue of $\hat f$ modulo $M_f$ is therefore uniform, whatever $f$ is, and independent of the other columns. It gives the column's values at its opening points, at the $\lambda_j$ and, if read at the next row, at the $\omega\lambda_j$.

\emph{Quotient.} A coordinate of the quotient at a point $y$ is a polynomial over $\mathbb{F}$, fixed by the challenges, in the values at $y$ of the columns and of public polynomials and in the values at $\omega y$ of the columns read at the next row, divided by $v_\Omega(y)$. The residues above thus determine $q$ modulo $M$. Fix the columns and a root $y$ of $M$. The residues of $b_1, \dots, b_7$ modulo $M$ are uniform and independent, so their values at $y$ are, and independently of their values at the roots that are not conjugate to $y$. If $y$ lies in no $\Omega_i$, the values $\hat q_i(y) = q_i(y) + v_{\Omega_i}(y) b_i(y)$, $i < 8$, are uniform, and $\hat q_8(y)$ follows from $q(y) = \sum_i \ell_i(y) \hat q_i(y)$, in which $\ell_8(y) \neq 0$. Otherwise $y$ is a $\lambda_j$ in some $\Omega_i$: there $\hat q_i(y) = q(y)$, and the other seven values are uniform: if $i \neq 8$, the eighth chunk's is so through $b_i(y)$, which enters it with a nonzero coefficient and no other chunk's value at $y$.

\emph{FRI.} Fix all committed polynomials but the $u_k$, and write $u_k = s_k + \mu_\zeta g_k$ with $\deg s_k < 8$. The $s_k$, that is, the values $u_k(\zeta)$, are uniform in $\mathbb{K}$, and the $g_k$ are uniform of degree below $2^{15} - 8$, all independent. Then $g = \sum_{k < 8} \alpha^k g_k$ is uniform among the polynomials over $\mathbb{K}$ of that degree, and $\Phi = \Phi_0 + \alpha^{e_u} g \mu_\zeta / (X - \zeta)$, where $\Phi_0$ does not involve the $g_k$. Hence, given also the $u_k(\zeta)$, $\Phi$ is uniform among the polynomials over $\mathbb{K}$ of degree below $2^{15} - 1$ with fixed values at the seven other conjugates $\xi$ of $\zeta$. In these values every $h(\xi)$ is the conjugate of $h(\zeta)$, so the openings at $\zeta$ and $\omega\zeta$ determine them, and the distribution of $\Phi$ depends on the fixed polynomials through these openings only. Finally, $\Phi(\lambda_j)$ and the other openings determine $\sum_k \alpha^k u_k(\lambda_j)$, hence each $u_k(\lambda_j) \in \mathbb{F}$.

\emph{Simulator.} It draws a uniform residue modulo $M_f$ for every column, computes the coordinates of the quotient at $\zeta$ and at the $\lambda_j$ from the constraints, draws the chunks' values as in the second step, samples the polynomials of (iv) as the prover does and the $u_k(\zeta)$ uniformly in $\mathbb{K}$, draws $\Phi$ as in the third step, solves for the $u_k(\lambda_j)$ and computes FRI's messages from $\Phi$.
\end{proof}

\emph{From the IOP to the proof.} \Cref{prop:zk} concerns uniform masks and opened values. $\mathsf{F}$'s proof also holds Merkle roots, authentication paths and the salts of the opened leaves; its challenges come from a Fiat--Shamir transcript; and its randomness comes from generators: every Merkle leaf is hashed with a salt of eight field elements, ChaCha20 keystream under a fresh key for each committed matrix, and the masks come from ChaCha12, seeded by the operating system for every proof. For salted leaves that hide statistically and random oracles that the simulator programs, the BCS transformation preserves zero knowledge, also against quantum adversaries~\cite{BCS16,COS20}. $\mathsf{F}$ does not meet these premises: its salts have the digest's size and are pseudorandom, and one Poseidon2 permutation serves its leaf hash, its tree compression and its transcript. Carrying \cref{prop:zk} over to $\mathsf{F}$'s proof thus needs a computational argument against quantum adversaries for the joint commitments, their partial openings and the programmed transcript, with generators that are pseudorandom against such adversaries. We do not give it: the zero-knowledge part of A4 still assumes it, which is more than A2 states. An aggregate is computed from four final proofs and public data, so it reveals no more than they do. Weak simulation extractability remains an assumption.

\fi
\end{document}

%% file: generated/bill_macros.tex
\newcommand{\simDaysBEnroll}{29.4}
\newcommand{\simDaysBActiveEnroll}{10.0}
\newcommand{\simDaysBRestBatched}{155}

\newcommand{\simDaysBBatched}{235}
\newcommand{\simDaysBSingle}{494}
\newcommand{\simDaysBSplitBatched}{165}

\newcommand{\simDaysBEnrollUpper}{14.7}

\newcommand{\simEnrollPerSlot}{21.1}
\newcommand{\simEnrollPerSlotCounted}{22.1}
\newcommand{\simEnrollRoomShare}{0}
\newcommand{\simEnrollMigShare}{40}

\newcommand{\simEnrollServed}{62}

\newcommand{\simEnrollFeeEth}{0.00057}

\newcommand{\simEnrollMigShareUpper}{80}

\newcommand{\simEnrollServedUpper}{25}

\newcommand{\simBatchedRoomShare}{22}

\newcommand{\simBatchedFeeEth}{0.00416}

\newcommand{\simNodeBaselineDownGB}{19.2}

\newcommand{\simNodeBaselineClDownGB}{18.87}
\newcommand{\simNodeBaselineGethCpu}{0.23}
\newcommand{\simNodeBaselineReamCpu}{3.31}

\newcommand{\simBaselineServed}{100}

\newcommand{\simLinHalf}{5.44}
\newcommand{\simLinFull}{5.60}
\newcommand{\simOwnerProofSeconds}{111}

\newcommand{\simOwnerCount}{996}
\newcommand{\simProveL}{23.3}
\newcommand{\simProveB}{28.0}
\newcommand{\simProveF}{46.5}
\newcommand{\simProveAll}{98}
\newcommand{\simBatchSeconds}{88}
\newcommand{\simBatchCount}{237}

\newcommand{\simValRegHours}{10.5}

\newcommand{\simValRegHoursUpper}{5.2}

\newcommand{\simBatchSecondsPortable}{342}

\newcommand{\simOwnerLaptopHoursBRound}{138,000}

\newcommand{\simValKeySeconds}{8.4}
\newcommand{\simValSecretMB}{16.8}
\newcommand{\simValKeyLaptopHours}{1,900}
\newcommand{\simValSecretTB}{13.7}

\newcommand{\simDaysTokenBatched}{306}
\newcommand{\simDaysTokenSingle}{641}

\newcommand{\simTokenEoasM}{5.78}

\newcommand{\simSingleSixFill}{88}

\newcommand{\simSingleSixProcessSeconds}{0.50}
\newcommand{\simSingleSixImportDelay}{2.3}
\newcommand{\simSponsorCheckBatchedTwelve}{2.67}
\newcommand{\simSponsorCheckBatchedFortyEight}{2.63}
\newcommand{\simSponsorCheckBatchedRoomTwelve}{34}
\newcommand{\simSponsorCheckSingleTwelve}{2.63}
\newcommand{\simSponsorCheckSingleFortyEight}{2.62}

\newcommand{\simBatchedRuns}{three}
\newcommand{\simBatchedPerSlotMin}{2.46}
\newcommand{\simBatchedPerSlotMax}{2.77}
\newcommand{\simNodeEnrollAddedDownGB}{4.9}

\newcommand{\simNodeBatchedAddedDownGB}{6.8}

\newcommand{\simNodeSingleAddedDownGB}{10.6}

\newcommand{\simMachineCheckRecord}{2.67}
\newcommand{\simMachineCheckOther}{2.77}
\newcommand{\simMachineCheckGap}{4.0}

\newcommand{\simBatchedBlockKBMax}{712}

\newcommand{\simSingleBlockKBMax}{1,227}

\newcommand{\simEnrollWaitMedian}{1,145}
\newcommand{\simEnrollWaitMedianMin}{19.1}
\newcommand{\simEnrollWaitPten}{1,049}
\newcommand{\simEnrollWaitPninety}{1,193}

\newcommand{\simEnrollBacklog}{1,988}
\newcommand{\simEnrollSealedPerSlot}{18.7}
\newcommand{\simEnrollLittleSealedPerSlot}{21.6}

\newcommand{\simEnrollLittleGap}{2}

\newcommand{\simOrdWaitMedian}{11.7}
\newcommand{\simOrdWaitPninetyNine}{26}

\newcommand{\simEnrollWaitMedianUpper}{557}

\newcommand{\simOverridesInWBatchedM}{6.9}

\newcommand{\simEoasAExposedM}{20.43}
\newcommand{\simDaysAExposedBatched}{1,080}

\newcommand{\simEoasBExposedM}{2.75}
\newcommand{\simDaysBExposedBatched}{146}

\newcommand{\simDaysBEnrollSeedTwo}{32.6}
\newcommand{\simDaysBBatchedSlowest}{251}
\newcommand{\simDaysBBatchedFastest}{223}

\newcommand{\simFeeEthBSplitBatchedRound}{13,100}

%% file: generated/bill_gpu_macros.tex
\newcommand{\gpuLaptopOwnerSeconds}{87}

\newcommand{\gpuLaptopProveL}{16.1}
\newcommand{\gpuLaptopProveB}{21.7}
\newcommand{\gpuLaptopProveF}{32.0}

\newcommand{\gpuLaptopOwnerCount}{20}

\newcommand{\gpuBestOwnersPerHour}{82}
\newcommand{\gpuBestOwners}{12}
\newcommand{\gpuLaptopDeviceOwnerGiB}{11.4}
\newcommand{\gpuLaptopDeviceBatchGiB}{14.5}
\newcommand{\gpuLaptopHostGiB}{31.6}
\newcommand{\gpuBestPipelines}{3}
\newcommand{\gpuBestCpus}{13.6}

\newcommand{\gpuBestDollarsPerThousand}{22}

\newcommand{\gpuReverifiedOwnerProofs}{120}
\newcommand{\gpuReverifiedCarriers}{31}

\newcommand{\gpuMachineHoursBRound}{54,000}
\newcommand{\gpuDollarsBk}{97}
\newcommand{\gpuCpuOnlyDollarsBk}{178}

\newcommand{\gpuMachinesToKeepUp}{10}
\newcommand{\gpuSponsorGpusToKeepUp}{2.3}

%% file: generated/exchange_macros.tex
\newcommand{\exNumExchanges}{357}
\newcommand{\exAddresses}{30{,}570}
\newcommand{\exEthM}{11.19}
\newcommand{\exEthContractM}{1.19}
\newcommand{\exShareEl}{14.6}
\newcommand{\exKeys}{1{,}869}

\newcommand{\exKeysEthM}{10.01}
\newcommand{\exExposedValueShare}{97.0}

\newcommand{\exEnrollMin}{17.7}
\newcommand{\exEnrollMinUpper}{8.9}
\newcommand{\exFeeEth}{1.06}

\newcommand{\exValidators}{112{,}292}
\newcommand{\exStakeM}{5.30}
\newcommand{\exStakeShare}{12.0}

\newcommand{\exTopExchanges}{39}
\newcommand{\exExamined}{25}
\newcommand{\exExaminedShare}{94}

%% file: generated/backlog_macros.tex
\newcommand{\backlogWaitMinHalfK}{4.3}

\newcommand{\backlogWaitMinOneK}{9.3}

\newcommand{\backlogOverNineOneK}{73}

\newcommand{\backlogWaitMinFourK}{38.1}

%% file: generated/sens_macros.tex
\newcommand{\sensDemandMaxPct}{14}

\newcommand{\sensDaysEnrollRepriceHigh}{58.1}
\newcommand{\sensDaysBatchedRepriceHigh}{446}
\newcommand{\sensDaysControl}{248}
\newcommand{\sensDaysRecordSeedOne}{251}
\newcommand{\sensShareNone}{80}
\newcommand{\sensShareHalfLo}{36}
\newcommand{\sensShareHalfHi}{40}
\newcommand{\sensShareThreeQ}{20}
\newcommand{\sensShareCapLo}{36}
\newcommand{\sensShareCapHi}{43}

\newcommand{\sensDaysSingleHighRate}{765}

%% file: figures/fig1.tex
\begin{tikzpicture}[
  x=1pt,y=1pt,
  font=\fontsize{8}{9}\selectfont,
  >={Latex[length=4pt,width=4pt]},
  every node/.style={inner sep=0pt,outer sep=0pt},
  guide/.style={black!35,densely dotted,line width=.45pt},
  ico/.style={font=\fontsize{9.5}{9.5}\selectfont},
  intext/.style={anchor=base west,font=\fontsize{8}{9}\selectfont},
  chain/.style={rounded corners=1.5pt,line width=.6pt,minimum height=11pt,inner xsep=3pt,anchor=center},
  tick/.style={line width=.9pt,-{Latex[length=2.6pt,width=3.4pt]}},
  arrowlabel/.style={fill=white,inner xsep=1.5pt,text=risktext,font=\fontsize{8}{9}\selectfont}
]
  \definecolor{proofblue}{HTML}{2166AC}
  \definecolor{riskorange}{HTML}{B36A15}
  \colorlet{risktext}{riskorange!75!black}
  \def\figsub#1{_{\mbox{\fontsize{7}{7}\selectfont$#1$}}}
  \def\aico#1{\textcolor{riskorange}{\fontsize{8.5}{8.5}\selectfont#1}}
  \def\xi{10}\def\xs{20}\def\tz{88}\def\tf{160}\def\tw{206}\def\xe{243}
  \def\ychain{126}\def\yback{113.5}\def\yband{100}\def\yecd{83}\def\yenr{67}\def\yovr{51}\def\ymig{37}\def\yax{24}\def\h{5}
  \path[use as bounding box] (0,0) rectangle (243.9,132.5);

  \node[chain,draw=proofblue,fill=proofblue!8,text=proofblue] (seed) at (30,\ychain) {\faSeedling\ \ seed phrase};
  \node[chain,draw=black!55,fill=black!6] (key) at (123.5,\ychain) {\faKey\ \ private key};
  \node[chain,draw=black!55,fill=black!6] (pub) at (222,\ychain) {public key};
  \draw[black!40,line width=.5pt,->] (seed.east) -- (key.west);
  \draw[black!40,line width=.5pt,->] (key.east) -- (pub.west);
  \node[anchor=base,text=proofblue,font=\fontsize{7.5}{8}\selectfont] at (17,\yback-2.84) {secret};
  \draw[riskorange,line width=.8pt,->] (pub.south) |- ($(key.south)+(8,-6.5)$) -- ($(key.south)+(8,-1)$);
  \node[arrowlabel] at (176.75,\yback) {\aico{\faAtom\,\faCheck}\ quantum attack};
  \draw[riskorange,line width=.8pt,densely dashed,->] ($(key.south)+(-8,-1)$) |- ($(seed.south)+(4,-6.5)$) -- ($(seed.south)+(4,-1)$);
  \node[arrowlabel] at (74.75,\yback) {\aico{\faAtom\,\faBan}\ no way back};

  \draw[guide] (\tz,\yax) -- (\tz,\yecd+\h);
  \draw[guide] (\tf,\yax) -- (\tf,\yecd+\h);
  \draw[guide] (\tw,\yax) -- (\tw,\yenr+\h+2);

  \shade[left color=white,right color=riskorange!24] (\xs-6,\yband-7.5) rectangle (\xs+10,\yband+7.5);
  \fill[riskorange!24] (\xs+10,\yband-7.5) rectangle (243.9,\yband+7.5);
  \node[ico,text=riskorange] at (5.2,\yband) {\faUserSecret};
  \node[text=riskorange,font=\fontsize{8}{8}\selectfont] at (15.2,\yband) {\faAtom};
  \node[anchor=base west] at (\xs+12,\yband-2.8)
    {\textcolor{risktext}{\faEyeSlash\ \ unobservable $Q\figsub{0}\le Q\figsub{p}$:} $A$ may hold every exposed key};

  \node[ico,text=black!60] at (\xi,\yecd) {\faKey};
  \filldraw[fill=black!20,draw=black!55,line width=.5pt,rounded corners=1.5pt] (\xs,\yecd-\h) rectangle (\tf,\yecd+\h);
  \node[intext,text=black] at (\xs+4,\yecd-2.8) {ECDSA, until enrolled or migrated};
  \node[circle,fill=white,draw=black!60,line width=.6pt,minimum size=12.5pt] at (\tf,\yecd) {};
  \node[ico,text=black!70] at (\tf,\yecd) {\faSnowflake};
  \node[anchor=base west] at (\tf+9,\yecd-2.8) {frozen, recoverable};

  \node[anchor=base,text=black] at (\xi,\yenr+2.4) {owner};
  \node[anchor=base,text=risktext] at (\xi,\yenr-5.8) {or $A$};
  \draw[black!55,densely dotted,line width=.8pt] (\tf,\yenr) -- (\tw,\yenr);
  \draw[black!55,fill=white,line width=.9pt,dash pattern=on 2.5pt off 1.2pt,rounded corners=1.5pt]
    (\tz,\yenr-\h) rectangle (\tf,\yenr+\h);
  \draw[tick,black!45] (100,\yecd-\h) -- (100,\yenr+\h+.6);
  \node[intext,text=black] at (\tz+4,\yenr-2.8) {enroll PQ key};
  \node[text=risktext,anchor=east,font=\fontsize{8.5}{8.5}\selectfont] at (\tf-3.5,\yenr) {\faUserSecret\,?};
  \node[circle,fill=white,minimum size=11pt] at (\tw,\yenr) {};
  \node[ico,text=black!70] at (\tw,\yenr) {\faLock};
  \node[anchor=base west] at (\tw+7,\yenr-2.8) {final};

  \node[ico,text=proofblue] at (\xi,\yovr-1) {\faSeedling};
  \node[anchor=base,text=proofblue] at (\xi,\yovr-13.5) {seed};
  \node[anchor=base,text=proofblue] at (\xi,\yovr-21.5) {proof};
  \fill[proofblue,rounded corners=1.5pt] (\tz,\yovr-\h) rectangle (\tw,\yovr+\h);
  \node[intext,text=white] at (\tz+4,\yovr-2.8) {\faUndo\ \ override};
  \node[anchor=base east,text=white,font=\fontsize{8}{9}\itshape] at (\tw-4,\yovr-2.8) {unforgeable by $A$};
  \draw[tick,proofblue] (140,\yovr+\h) -- (140,\yenr-\h-1.2);
  \fill[proofblue,rounded corners=1.5pt] (\tz,\ymig-\h) -- (\xe-8,\ymig-\h) -- (\xe-8,\ymig-\h-2)
    -- (\xe,\ymig) -- (\xe-8,\ymig+\h+2) -- (\xe-8,\ymig+\h) -- (\tz,\ymig+\h) -- cycle;
  \node[intext,text=white] at (\tz+4,\ymig-2.8) {owner migrates if not enrolled, even if frozen};

  \draw[black!70,line width=.5pt,->] (\xs,\yax) -- (\xe,\yax);
  \foreach \x in {\tz,\tf,\tw} \draw[black!70,line width=.5pt] (\x,\yax-2) -- (\x,\yax+2);
  \node[anchor=base] at (\tz,\yax-11) {$T\figsub{0}$};
  \node[anchor=base] at (\tz,\yax-21) {PQ-ready fork};
  \node[anchor=base] at (\tf,\yax-11) {$T=Q\figsub{p}$};
  \node[anchor=base] at (\tf,\yax-21) {migration fork};
  \node[anchor=base] at (\tw,\yax-11) {$T+W$};
  \node[anchor=base] at (\tw,\yax-21) {deadline};
\end{tikzpicture}%

%% file: figures/fig_overview.tex
\begin{tikzpicture}[
  x=1pt,y=1pt,
  font=\fontsize{8}{9}\selectfont,
  >={Latex[length=3.6pt,width=3.4pt]},
  every node/.style={inner sep=0pt,outer sep=0pt},
  word/.style={anchor=base west},
  wordr/.style={anchor=base east},
  lab/.style={anchor=base},
  head/.style={anchor=base west,font=\fontsize{8}{9}\bfseries\selectfont},
  small/.style={font=\fontsize{7}{7}\selectfont},
  note/.style={anchor=base east,font=\fontsize{7}{8}\selectfont},
  ico/.style={font=\fontsize{9}{9}\selectfont,text=black!65},
  stepk/.style={circle,fill=black!60,draw=white,line width=.4pt,text=white,minimum size=7.8pt,
    font=\fontsize{7}{7}\bfseries\selectfont},
  stepb/.style={stepk,fill=proofblue,minimum size=10pt,font=\fontsize{8}{8}\bfseries\selectfont},
  dotk/.style={circle,fill=black!55,draw=white,line width=.4pt,minimum size=4.6pt},
  dotb/.style={circle,fill=proofblue,minimum size=6.8pt},
  flow/.style={line width=.45pt,black!60,->},
  proof flow/.style={line width=1pt,proofblue,->}
]
  \definecolor{proofblue}{HTML}{2166AC}
  \definecolor{riskorange}{HTML}{B36A15}
  \colorlet{oldgrey}{black!50}
  \colorlet{proofband}{proofblue!8}
  \def\figsub#1{_{\mbox{\fontsize{7}{7}\selectfont$#1$}}}
  \def\adv{\textcolor{riskorange}{\faUserSecret\,?}}
  \def\W{505.89}\def\H{141}
  \path[use as bounding box] (0,0) rectangle (\W,\H);

  \def\ra{95}\def\rb{125}\def\rc{157}
  \def\lw{170}
  \def\rowh{9.45}\def\ytop{121}
  \def\rowm{4.6}\def\rowt{7.2}
  \fill[proofband] (112,26.5) rectangle (138,\H);
  \node[head] at (1.5,\H-7.5) {System layers};
  \node[font=\fontsize{8}{8}\selectfont,text=black!65] at (\ra,\H-5) {\faKey};
  \node[font=\fontsize{10}{10}\selectfont,text=proofblue] at (\rb,\H-5) {\faSeedling};
  \node[font=\fontsize{8}{8}\selectfont,text=black!65] at (\rc,\H-5) {\faUsers};
  \node[lab,small,text=black!75] at (\ra,\H-16.3) {Account};
  \node[lab,font=\fontsize{7}{7}\bfseries\selectfont,text=proofblue] at (\rb,\H-16.3) {Proof};
  \node[lab,small,text=black!75] at (\rc,\H-16.3) {Validator};
  \draw[black!60,line width=.4pt] (0,\ytop) -- (\lw,\ytop);
  \def\layer#1#2#3#4{%
    \pgfmathsetmacro\yy{\ytop-#1*\rowh}%
    \draw[black!15,line width=.3pt] (0,\yy-\rowh) -- (\lw,\yy-\rowh);
    \ifcase#2 \fill[black!80] (1.6,\yy-\rowm-2.1) rectangle (5.8,\yy-\rowm+2.1);
    \or \draw[black!80,line width=.6pt] (3.7,\yy-\rowm) circle [radius=2.1];
    \else \draw[black!80,line width=.7pt] (1.6,\yy-\rowm) -- (5.8,\yy-\rowm);\fi
    \node[word,small] at (10.5,\yy-\rowt) {#3};
    \node[word] at (22,\yy-\rowt) {#4};}
  \layer{0}{1}{L0}{Wallets, keys}
  \layer{1}{0}{L1}{Transactions}
  \layer{2}{1}{L2}{Network}
  \layer{3}{1}{L3}{Mempool}
  \layer{4}{0}{L4}{Building, import}
  \layer{5}{0}{L5}{Execution}
  \layer{6}{0}{L6}{State, history}
  \layer{7}{0}{L7}{Consensus}
  \layer{8}{2}{L8}{Data availability}
  \layer{9}{1}{L9}{RPC, tools}
  \def\dy#1{\ytop-#1*\rowh-\rowm}
  \foreach \i in {0,1,3,4,5,6,9} {\node[dotk] at (\ra,\dy{\i}) {};}
  \foreach \i in {0,1,2,3,4,5,6,8,9} {\node[dotb] at (\rb,\dy{\i}) {};}
  \foreach \i in {0,1,5,6,7,9} {\node[dotk] at (\rc,\dy{\i}) {};}
  \node[dotk] at (3.7,19.9) {};
  \node[word,small] at (9,17.6) {affected layer};
  \fill[black!80] (1.6,10) rectangle (5.8,14.2);
  \node[word,small] at (9,9.7) {consensus rules change at a fork};
  \draw[black!80,line width=.6pt] (3.7,4.2) circle [radius=2.1];
  \node[word,small] (lg) at (9,1.8) {wallet or client software};
  \draw[black!80,line width=.7pt] ($(lg.east |- 0,4.2)+(6,0)$) -- ++(4.2,0);
  \node[word,small,xshift=13pt] at (lg.east |- 0,1.8) {no rule change};
  \draw[black!25,line width=.4pt] (173,0) -- (173,\H);

  \def\na{184.5}\def\ia{198.5}\def\la{179}
  \def\cb{316}\def\cc{396}\def\cd{467}
  \def\tb{141}\def\ta{83.5}\def\tc{41.75}\def\fb{86.5}

  \fill[proofband] (176,\fb) rectangle (\W,\H);
  \fill[proofblue] (176,\fb) rectangle (177.4,\H);
  \node[font=\fontsize{11}{11}\selectfont,text=proofblue] at (185,\tb-6.5) {\faSeedling};
  \node[head,font=\fontsize{9.5}{10}\bfseries\selectfont,text=proofblue]
    at (196,\tb-8) {Seed proof (zero knowledge)};
  \node[note] at (\W-2,\tb-8)
    {\textcolor{proofblue}{\faSnowflake}\ \,frozen: no deadline\quad
     \textcolor{proofblue}{\faUndo}\ \,enrolled: override before $T+W$};
  \def\ys{22}\def\yla{40.8}\def\ylb{49.3}
  \def\proofstep#1#2#3#4#5{%
    \node[stepb] at (#1-10.5,\tb-\ys) {#2};
    \node[font=\fontsize{13}{13}\selectfont,text=proofblue] at (#1+4,\tb-\ys) {#3};
    \node[lab] at (#1,\tb-\yla) {#4};
    \node[lab] at (#1,\tb-\ylb) {#5};}
  \node[stepb] at (\na,\tb-\ys) {1};
  \draw[proofblue!65,line width=.65pt,rounded corners=2pt,fill=white]
    (193,\tb-\ys-8) rectangle (250,\tb-\ys+8);
  \node[text=black!60,font=\fontsize{7}{7}\selectfont] at (199.5,\tb-\ys) {\faLaptop};
  \node[font=\fontsize{7.5}{8}\selectfont] at (227,\tb-\ys) {Local proving};
  \node[fill=proofblue,text=white,rounded corners=1pt,minimum width=30pt,minimum height=19pt,
    align=center,font=\fontsize{7}{7.5}\selectfont] (finalproof) at (273,\tb-\ys) {Final ZK\\proof};
  \draw[proof flow] (250,\tb-\ys) -- (finalproof.west);
  \node[word] at (\la,\tb-\yla) {only the final proof leaves.};
  \proofstep{\cb}{2}{\faPaperPlane}{\textbf{Migrate} (\texttt{0x7e}), from $T\figsub{0}$}{or \textbf{migrate four} (\texttt{0x7f})}
  \proofstep{\cc}{3}{\faIcon{shield-alt}}{\textbf{Verify} (all nodes)}{190\,KB, 38\,ms}
  \node[stepb] at (\cd-10.5,\tb-\ys) {4};
  \node[font=\fontsize{13}{13}\selectfont,text=proofblue] at (\cd+4,\tb-\ys) {\faLink};
  \node[wordr] at (\W-1,\tb-\yla) {\textbf{Bind} owner's key;};
  \node[wordr,font=\fontsize{7.5}{8}\selectfont] at (\W-1,\tb-\ylb) {then activate and transact};
  \draw[proof flow] (finalproof.east) -- (\cb-18,\tb-\ys);
  \draw[proof flow] (\cb+17,\tb-\ys) -- (\cc-18,\tb-\ys);
  \draw[proof flow] (\cc+17,\tb-\ys) -- (\cd-18,\tb-\ys);

  \def\yt{7.2}\def\yi{4.9}\def\ys{17.6}\def\yla{30.2}\def\ylb{38.4}
  \def\first#1#2#3#4{%
    \node[stepk] at (\na,#1-\ys) {1};
    \node[ico] at (\ia,#1-\ys) {#2};
    \node[word,font=\fontsize{7.5}{8}\selectfont] at (\la,#1-\yla) {#3};
    \node[word,small] at (\la,#1-\ylb) {#4};}
  \def\step#1#2#3#4#5#6{%
    \node[stepk] at (#1-9.5,#2-\ys) {#3};
    \node[ico] at (#1+3,#2-\ys) {#4};
    \node[lab,font=\fontsize{7.5}{8}\selectfont] at (#1,#2-\yla) {#5};
    \node[lab,small] at (#1,#2-\ylb) {#6};}
  \def\last#1#2#3{%
    \node[wordr,font=\fontsize{7.5}{8}\selectfont] at (\W-1,#1-\yla) {#2};
    \node[wordr,small] at (\W-1,#1-\ylb) {#3};}
  \begin{scope}[text=black!85]
    \node[font=\fontsize{8}{8}\selectfont,text=black!65] at (183,\ta-\yi) {\faKey};
    \node[head,font=\fontsize{7.5}{8}\bfseries\selectfont] at (191,\ta-\yt)
      {Account key: \textcolor{oldgrey}{\mdseries ECDSA}\,$\to$\,ML-DSA-44};
    \node[note] at (\W-2,\ta-\yt) {\adv\ \,may be the adversary's};
    \first{\ta}{\faFileSignature}{\textbf{Enroll} (\texttt{0x7d}) in $[T\figsub{0},T)$,}{signed with ECDSA \adv}
    \step{\cb}{\ta}{2}{\faPowerOff}{\textbf{Activate} (\texttt{0x7c})}{installs the PQ key}
    \step{\cc}{\ta}{3}{\faPenNib}{\textbf{Transact} (\texttt{0x7c})}{sign with ML-DSA-44}
    \node[ico] at (\cd,\ta-\ys) {\faBan};
    \last{\ta}{from $T$: no ECDSA,}{no \texttt{ecrecover}}
    \draw[flow] (\ia+10,\ta-\ys) -- (\cb-17,\ta-\ys);
    \draw[flow] (\cb+13,\ta-\ys) -- (\cc-17,\ta-\ys);

    \draw[black!20,line width=.35pt] (176,41.6) -- (\W,41.6);
    \node[font=\fontsize{8}{8}\selectfont,text=black!65] at (183,\tc-\yi) {\faUsers};
    \node[head,font=\fontsize{7.5}{8}\bfseries\selectfont] at (191,\tc-\yt)
      {Validator key: \textcolor{oldgrey}{\mdseries BLS}\,$\to$\,leanXMSS};
    \node[note] at (\W-2,\tc-\yt) {from $T\figsub{0}$, exit requests follow \S\ref{sec:n1-late}};
    \first{\tc}{\faUserPlus}{\textbf{Register} two leanXMSS keys,}{BLS-signed, before $T\figsub{0}$ \adv}
    \step{\cb}{\tc}{2}{\faFlagCheckered}{\textbf{Checkpoint:}}{conflicts excluded}
    \step{\cc}{\tc}{3}{\faIcon{exchange-alt}}{\textbf{Switch} at $T\figsub{0}$:}{a Lean chain}
    \node[ico,text=proofblue] at (\cd,\tc-\ys) {\faSeedling};
    \last{\tc}{Late: seed proof for}{withdrawal account; exit}
    \draw[flow] (\ia+10,\tc-\ys) -- (\cb-17,\tc-\ys);
    \draw[flow] (\cb+13,\tc-\ys) -- (\cc-17,\tc-\ys);
  \end{scope}
\end{tikzpicture}%

%% file: generated/bill_table.tex
\begin{table*}[t]
\caption{The migration on mainnet, simulated. In each run, ten geth/Ream nodes on one Amazon EC2 m7i.4xlarge instance (16 vCPU, 64\,GiB) run the protocol on a population with mainnet's composition under mainnet's ordinary demand (60\,M gas limit, EIP-1559 unchanged, migration bids at most 1\,gwei); a time on mainnet is a population divided by the steady-state rate (migration gas per slot over the gas of one account), at 7,200 slots a day. The demand of record has half of the ordinary users re-price to the base fee; at today's bids all keep their bids, an upper bound on the migration's speed. The batched recovery of record pools three runs with other random draws (2.46--2.77 accounts per slot each); its node rows are those of the first. Without the migration a node downloads 19.2\,GB a day and uses 0.23\,s (geth) and 3.31\,s (Ream) of CPU per slot. Ream's CPU per slot depends on the instance: the run without migration and the single recovery of record ran on one instance (3.31 and 3.38\,s), the node rows of the other five columns on another (2.77--2.92\,s), and the batched recovery of record used 2.83\,s there and 3.34\,s on a third.}
\label{tab:sim}
\centering\footnotesize
\setlength{\tabcolsep}{4pt}
\begin{tabular}{@{}lrrrrrr@{}}
\toprule
 & \multicolumn{2}{c}{enroll} & \multicolumn{4}{c}{recover after $T$} \\
\cmidrule(lr){4-7}
 & \multicolumn{2}{c}{before $T$} & \multicolumn{2}{c}{batches of four} & \multicolumn{2}{c}{one by one} \\
\cmidrule(lr){2-3}\cmidrule(lr){4-5}\cmidrule(lr){6-7}
 & of record & today's bids & of record & today's bids & of record & today's bids \\
\midrule
accounts per slot & 21.1 & 42.2 & 2.6 & 5.6 & 1.3 & 2.6 \\
migration gas per slot, share of the target & 40\,\% & 80\,\% & 36\,\% & 78\,\% & 39\,\% & 80\,\% \\
base fee, median (gwei) & 1.04 & 1.01 & 1.03 & 0.97 & 1.06 & 1.01 \\
ordinary demand included & 62\,\% & 25\,\% & 64\,\% & 26\,\% & 64\,\% & 25\,\% \\
gas per account, activation included & 567,878 & 567,878 & 4,161,560 & 4,161,560 & 9,219,679 & 9,219,679 \\
fee per account (ETH) & 0.00057 & 0.00057 & 0.00416 & 0.00416 & 0.00922 & 0.00922 \\
\midrule
days, EOAs $\geq$ 0.01\,ETH (26.28\,M) & 173 & 86.5 & 1,389 & 652 & 2,914 & 1,395 \\
days, EOAs $\geq$ 0.1\,ETH (4.45\,M) & 29.4 & 14.7 & 235 & 110 & 494 & 236 \\
days, EOAs $\geq$ 1\,ETH (1.45\,M) & 9.5 & 4.8 & 76.5 & 35.9 & 160 & 76.8 \\
days, EOAs $\geq$ 10\,ETH (0.28\,M) & 1.9 & 0.9 & 15.1 & 7.1 & 31.6 & 15.1 \\
\midrule
P2P download per node per day (GB) & 24.1 & 29.3 & 26.0 & 27.7 & 29.8 & 35.9 \\
CPU per slot, geth / Ream (s) & 0.73 / 2.77 & 0.82 / 2.87 & 0.59 / 2.83 & 0.94 / 2.84 & 0.62 / 3.38 & 0.87 / 2.92 \\
\bottomrule
\end{tabular}
\end{table*}

%% file: generated/exchange_table.tex
\begin{table}[t]
\caption{What migration at~$T_0$ costs each centralized exchange: the nine largest at block 26,344,089. ETH and stake in millions; keys: EOAs with at least 0.1\,ETH; custody as documented ($\dagger$: for its infrastructure or custody arm); Enroll: minutes of block space to enroll and activate all keys at the demand of record; fee: ETH at 1\,gwei.}
\label{tab:exchanges}
\centering
\footnotesize
\setlength{\tabcolsep}{1.5pt}
\begin{tabular}{@{}lrrrlrr@{}}
\toprule
Exchange & ETH & Keys & Stake & Custody & Enroll & Fee \\
\midrule
Binance & 3.21 & 173 & 3.59 & multisig, MPC~\cite{CustodyBinance}$^\dagger$ & 1.6 & 0.10 \\
Bitfinex & 1.60 & 21 & 0.05 & HSM, multisig~\cite{CustodyBitfinex} & 0.2 & 0.01 \\
Robinhood & 1.24 & 10 & -- & MPC~\cite{CustodyRobinhood} & 0.1 & 0.01 \\
OKX & 1.18 & 185 & 0.64 & multisig~\cite{CustodyOKX} & 1.8 & 0.11 \\
Kraken & 0.60 & 21 & 0.05 & undisclosed & 0.2 & 0.01 \\
Bybit & 0.39 & 65 & -- & MPC, HSM~\cite{CustodyBybitOverhaul} & 0.6 & 0.04 \\
Gemini & 0.32 & 5 & -- & HSM, multisig~\cite{CustodyGemini2019} & <0.1 & <0.01 \\
Crypto.com & 0.23 & 16 & -- & MPC~\cite{CustodyCryptoComCustody}$^\dagger$ & 0.2 & 0.01 \\
Bitstamp & 0.22 & 6 & 0.23 & multisig contract~\cite{CustodyBitstampWalletSimple} & 0.1 & <0.01 \\
\midrule
All 357 firms & 11.19 & 1{,}869 & 5.30 & & 17.7 & 1.06 \\
\bottomrule
\end{tabular}
\end{table}

%% file: generated/sens_table.tex
\begin{table}[t]
\caption{The simulated times with one parameter changed at a time, for the 4.45\,M EOAs with at least 0.1\,ETH. Enrollment: 10,000 accounts on the laptop, against the enrollment of record; recovery in batches: 500 owners on the laptop, against a control with the settings of record; one by one: against the single recovery of record. Share: migration gas in per cent of the gas target; base fee: steady-state mean in gwei.}
\label{tab:sens}
\centering\footnotesize
\setlength{\tabcolsep}{3pt}
\begin{tabular}{@{}lrrrrr@{}}
\toprule
 & Accounts & & & & Base \\
Setting & per slot & Days & Change & Share & fee \\
\midrule
\multicolumn{6}{@{}l}{\emph{Enrollment}} \\
Of record (half re-price, 1\,gwei) & 21.07 & 29.4 & -- & 39.9 & 1.04 \\
Other random draws & 18.96 & 32.6 & +11\,\% & 35.9 & 1.05 \\
Demand of February 2025 & 24.43 & 25.3 & $-$14\,\% & 46.2 & 1.03 \\
Migration bids at most 2\,gwei & 22.84 & 27.1 & $-$8\,\% & 43.2 & 2.08 \\
Migration bids at most 0.5\,gwei & 18.99 & 32.6 & +11\,\% & 36.0 & 0.52 \\
Three quarters re-price & 10.65 & 58.1 & +98\,\% & 20.2 & 1.07 \\
\midrule
\multicolumn{6}{@{}l}{\emph{Recovery in batches of four}} \\
Control (settings of record) & 2.50 & 248 & -- & 34.6 & 1.02 \\
Verification at 0.104\,gas/ns & 3.20 & 193 & $-$22\,\% & 35.9 & 1.02 \\
Demand of February 2025 & 2.80 & 221 & $-$11\,\% & 38.9 & 1.00 \\
Three quarters re-price & 1.39 & 446 & +80\,\% & 19.2 & 1.05 \\
\midrule
\multicolumn{6}{@{}l}{\emph{Recovery one by one}} \\
Of record & 1.25 & 494 & -- & 38.5 & 1.05 \\
Verification at 0.261\,gas/ns & 0.81 & 765 & +55\,\% & 37.4 & 1.05 \\
\bottomrule
\end{tabular}
\end{table}

%% file: generated/backlog_table.tex
\begin{table}[t]
\caption{An enrollment's wait in the mempool against the number of enrollments pending at once, at the demand of record: median and 10th to 90th percentile over the $n$ enrollments sent in the steady state, the share in per cent that waits longer than a fast-clock CRQC needs per key~\cite{BZG26}, and the mean wait that Little's law predicts beside the measured one. Blocks seal 20.6 to 22.2 enrollments per slot in every run. Laptop runs; the row marked EC2 is the enrollment of record.}
\label{tab:backlog}
\centering\footnotesize
\setlength{\tabcolsep}{2.5pt}
\begin{tabular}{@{}lrrrrrr@{}}
\toprule
Pending & & Median wait & \multicolumn{2}{c}{longer than} & \multicolumn{2}{c}{mean (min)} \\
\cmidrule(lr){4-5}\cmidrule(l){6-7}
at once & $n$ & (10--90\,\%), min & 9\,min & 12\,min & Little & measured \\
\midrule
500 & 6,783 & 4.3 (3.1--5.9) & 0 & 0 & 4.6 & 4.5 \\
1,000 & 5,824 & 9.3 (8.3--10.5) & 73 & 0.5 & 9.6 & 9.4 \\
2,000 & 4,049 & 19.9 (17.3--21.5) & 100 & 100 & 18.8 & 19.7 \\
2,000 (EC2) & 4,137 & 19.1 (17.5--19.9) & 100 & 100 & 18.4 & 18.7 \\
4,000 & 766 & 38.1 (37.9--38.3) & 100 & 100 & 35.9 & 38.1 \\
\bottomrule
\end{tabular}
\end{table}

%% file: generated/bill_gpu_table.tex
\begin{table}[t]
\caption{Proving with a GPU: owners an hour (owners/h), each proved ($\mathsf{L}$, $\mathsf{B}$, $\mathsf{F}$) and aggregated in fours on the machine, with the same machine and pipelines without the GPU in parentheses; median seconds per owner pipeline and per batch with the GPU; dollars per 1,000 owners at the rental price; $n$, the owners proved, without the GPU in parentheses where measured. Both A100 machines rent at \$1.80 an hour.}
\label{tab:gpu}
\centering
\footnotesize
\setlength{\tabcolsep}{3pt}
\begin{tabular}{@{}lrrrrr@{}}
\toprule
Machine & Pipelines & Owners/h & Owner, batch & \$/1,000 & $n$ \\
\midrule
Laptop, A5500 & 1 & 35.8 (26.4) & 87, 49\,s & -- & 20 (8) \\
A100, 20.4 CPUs & 1 & 48.6 (31.1) & 64, 41\,s & 37 (58) & 48 (8) \\
 & 2 & 75.0 (45.1) & 74, 42\,s & 24 (40) & 8 (8) \\
A100, 13.6 CPUs & 1 & 43.4 (25.1) & 72, 42\,s & 41 (72) & 4 (4) \\
 & 2 & 65.7 & 88, 44\,s & 27 & 8 \\
 & 3 & 82.3 & 104, 52\,s & 22 & 12 \\
 & 4 & 80.6 & 131, 68\,s & 22 & 16 \\
\bottomrule
\end{tabular}
\end{table}